\documentclass[reprint,aps,prx,groupedaddress,amsmath,amssymb,superscriptaddress]{revtex4-2}
\usepackage{graphicx}
\usepackage{dcolumn}
\usepackage{bm}
\usepackage{xcolor}
\usepackage[normalem]{ulem}
\usepackage[USenglish]{babel}
\usepackage{float}
\usepackage{times}
\usepackage{physics}
\usepackage{comment}
\usepackage{xr}
\usepackage[colorlinks]{hyperref}

\global\long\def\abs#1{\left\vert#1\right\vert}

\newcommand{\Eq}[1]{Eq.\,(\ref{#1})}

\newcommand{\Fig}[1]{Fig.\,\ref{#1}}

\makeatletter
\let\ORIbbl@fixname\bbl@fixname
\def\bbl@fixname#1{%
  \@ifundefined{languagealias@\expandafter\string#1}
    {\ORIbbl@fixname#1}
    {\edef\languagename{\@nameuse{languagealias@#1}}}%
}
\newcommand{\definelanguagealias}[2]{%
  \@namedef{languagealias@#1}{#2}%
}
\makeatother

\definelanguagealias{en}{english}
\definelanguagealias{EN}{english}

\usepackage{ifthen}
\usepackage{tcolorbox}
\newboolean{showcontent}
\setboolean{showcontent}{false} 
\newcommand{\itemizedparag}[1]{%
    \ifthenelse{\boolean{showcontent}}{%
        \begin{tcolorbox}[colframe=red, colback=white, sharp corners, width=\linewidth]
        {\bfseries #1} 
        \end{tcolorbox}
    }{}
}

\begin{document}



\title[]{Long-time soliton dynamics via a coarse-grained space-time method}

\makeatletter

\date{\today}

\author{Dung N. Pham}
\affiliation{Department of Electrical and Computer Engineering, Princeton University, Princeton, NJ 08544, USA}
\author{Zoe Zager}
\affiliation{Department of Electrical and Computer Engineering, Princeton University, Princeton, NJ 08544, USA}
\author{Wentao Fan}
\affiliation{Department of Electrical and Computer Engineering, Princeton University, Princeton, NJ 08544, USA}
\author{Hakan E. T\"ureci}
\affiliation{Department of Electrical and Computer Engineering, Princeton University, Princeton, NJ 08544, USA}
\begin{abstract}
We investigate the long-time dynamics of the Sine-Gordon (SG) model under a class of perturbations whose quantum field theoretic analog -- via bosonization -- corresponds to the massive Schwinger model describing 1+1D relativistic QED of Dirac fermions. Classical SG solutions offer critical insight into non-perturbative effects in this quantum theory, but capturing their long-time behavior poses significant numerical challenges. To address this, we extend a coarse-graining method to spacetime using a dual-mesh construction based on the Minkowski-metric. We first validate the approach against the well-studied variant of the SG model describing magnetic fluxon dynamics in Josephson transmission lines (JTLs), where analytical and numerical benchmarks exist. We then apply the method to the Schwinger-inspired SG model and uncover long-lived bound states -- “Schwinger atoms” -- in which a soliton is trapped by a fixed central charge. In certain regimes, the system exhibits limit cycles that give rise to positronium-like states of oppositely charged solitons, while in others such formation is suppressed. Accessing such long-time solutions requires a rigorous implementation of outgoing boundary conditions on a finite computational domain that provide radiative dissipation to allow relaxation toward states that exist only in an infinite domain. Here we provide such a construction. Our results also suggest the possibility of analog quantum simulation of relativistic quantum field theories with JTLs. These results demonstrate the utility of spatio-temporal coarse-graining methodology for probing non-perturbative structure formation in non-linear field theories.
\end{abstract}

\maketitle


\section{Introduction}

\itemizedparag{Why is it important to accurately integrate over long time periods the Sine-Gordon Model? 
\begin{itemize}
    \item The role of the perturbed SG equation in studying formation of coherent electromagnetic structures in JTLs
    \item The connection to the soliton solutions of the SG model
    \item Technological importance of the past work on fluxons viewed from today
    \item The technical challenges to and importance of capturing long-time evolution, the challenge we address in our paper. 
    \item State our main goal, why we felt we had to address this challenge in a way that no one went out of their way to do before: understanding stabilization of coherent structures in relativistic ED. 
    \item Couch it as an opportunity: The similarities between the models describing the two processes suggest that the rich literature on Josephson junction physics can provide valuable insights into the dynamics of the Schwinger process, while the differences in their perturbing terms guarantee that there is room for possible discoveries of new phenomena. (I like this very much). 
\end{itemize}
}

The Sine-Gordon (SG) equation has emerged as a fundamental model for investigating the formation and evolution of coherent electromagnetic structures in long Josephson junctions or an array of short Josephson junctions, commonly known as Josephson Transmission Lines (JTLs)~\cite{mclaughlin1978_SG,scott1976JJSG, ustinov1998solitons, berdiyorov2014parametric, fujimori2016field, sheikhzada2019instability, stern2019fractional, krasnov2020josephson}. The SG model in particular describes the dynamics of supercurrent vortices, known as fluxons, which can be accelerated by bias current up to the Swihart velocity~\cite{swihart1961} — the characteristic speed at which electromagnetic waves propagate along the tunneling region. These particle-like nonlinear wave packets, along with their bound states - formed by soliton-antisoliton pairs oscillating around a common center of mass (breathers) - and their interactions - establish a bridge between the realms of condensed matter physics, quantum information science, and relativistic field theory~\cite{berdiyorov2014parametric, fujimori2016field, sheikhzada2019instability, stern2019fractional, krasnov2020josephson, ustinov1992dynamics, wallraff2003singlevortex, fedorov2013josephson, sato2022quasiparticle, wildermuth2022fluxons,laub1995lorentz, gulevich2008timedilation}. Beyond their well-characterized classical relativistic dynamics, it was shown that individual fluxons can exhibit intrinsically quantum behaviors~\cite{wallraff2003singlevortex}, further adding to the rich physics that these solitons have in store. The remarkable stability and robustness of these particle-like, macroscopic quantum entities have spurred proposals for their deployment in a variety of advanced technologies. Proposed applications range from analog quantum simulation~\cite{petrescu2018qsim} and cryogenic memory~\cite{kalashnikov2024memory} to single-shot detectors~\cite{soloviev2014detector}, quantum flux shuttles~\cite{fulton1973fluxshuttle}, logic circuits~\cite{nakajima1976logic_circuit}, and amplifiers~\cite{nordman1995amplifier}, among many others~\cite{devoret2013sc_review, welp2013emitters_review, golod2021phaseshifter, ghosh2024diode}.


The purpose of this paper is to (1) highlight a fundamentally different connection between the SG equation and relativistic physics, specifically its relation to the relativistic electrodynamics of electrons and positrons in 1+1 dimensions (1+1D), (2) propose and implement a numerical methodology that ensures stable long-time evolution of its solutions; and (3) armed with that investigate a particular class of long-lived coherent structures that manifest as stable, atom-like solutions. We will then put forward the conjecture that these classical solutions can serve as fundamental components in the corresponding second-quantized theory, the Schwinger Model. This perspective will be presented at the end of the paper, with a more detailed investigation of the full quantum theory deferred to future work.


The connection we wish to build on -- between the SG model and the relativistic electrodynamics of electrons and positrons -- is well-known in high-energy physics: the zero-charge sector of the massive Schwinger Model describing the quantum electrodynamics of charged fermions in 1+1D has been shown~\cite{coleman1975, coleman1976} to be equivalent to the Quantum SG Model, but with important additional terms. This connection has played a pivotal role in exploring phenomena in quantum gauge theories that lie beyond the scope of field-theoretical perturbation methods, such as confinement~\cite{wilson1974confinement}, chiral symmetry breaking, axial anomaly~\cite{manton1985schwinger}, and fermion condensation~\cite{coleman1975}. One of these phenomena of interest here, the Sauter–Schwinger effect (also called the Schwinger effect)~\cite{sauter1931,Schwinger1951}, is a conjecture in Quantum Electrodynamics (QED) that suggests the possibility of spontaneous creation of electron-positron pairs out of vacuum when subjected to an electric field above a certain threshold. This prediction is significant and has been intensely investigated~\cite{schutzhold2008dynamically, chu2010capacitordischarge, semenoff2011holographic, sonner2013holographic, otto2015dynamical, otto2017afterglow} because it suggests a non-perturbative phenomenon that indicates the scale of electromagnetic field strength at which nonlinear effects in vacuum become evident. This problem has originally been studied in the context of the massive Schwinger Model \cite{schwinger1962model}. Here, we are specifically interested in the long-time evolution that follows spontaneous electron-positron (e-p) creation and relaxation towards coherent, stable structures in the long-time limit of this model.

A second point we wish to make in this paper is the possibility to analyze non-perturbative quantum electrodynamics with an analog simulator based on JTLs. The conditions under which such analog simulation will be interesting and relevant to phenomenology of relativistic QED are subtle, and we leave a full analysis to a future paper. In this manuscript, we focus on the {\it Schwinger process} viewed from the lens of a semiclassical analysis of the bosonized Schwinger model, and how that semiclassical model, under conditions investigated here can be realized in JTL. We are particularly interested in the formation and the electromagnetic stability of particle-like wavepackets as well as their interactions and backreaction in the presence of an ``external" electric field created by an immobile and structureless charged particle at the origin (representing a proton). In the bosonized Schwinger model, the Dirac field is thought to correspond to soliton solutions of the associated sine-Gordon (SG) model~\cite{coleman1975,coleman1976}. This correspondence has been firmly established only in the context of a closely related quantum field theory — the Thirring model~\cite{colemanthirring}, which crucially does not have the mass term generated by the gauge field. In this work, we explore time-dependent solutions of the exact form of the bosonized Schwinger model and in the presence of an external source, and discuss how this model can be created using a judiciously applied bias current distribution in a JTL.

On the numerical modeling front, to fully exploit the full `design space' (since we are talking about a 1+1D model, we have the freedom to study this model for different parameter regimes) and not to be restricted only to dynamics that are within or near analytically tractable evolutions, an efficient and reliable numerical method is needed. In particular, numerical stability over very long time periods is crucial for the precise study of the formation and stabilization of coherent electromagnetic structures. As we shall see, the issue is exacerbated when there are multiple perturbations to the system that disturb the balance between the nonlinearity and the dispersion, and when the dynamics to be simulated are far from the stable solitary wave state. Overcoming these modeling challenges can allow the detailed and reliable analysis of a wide class of strongly perturbed SG models that describe far-from-equilibrium soliton dynamics.

Radiative dissipation is another critical ingredient in modeling relaxation toward stable structures. Accurately capturing this radiative relaxation places significant demands on the size of the computational domain. An alternative technique to go around this difficulty is an open quantum system approach, where the infinite domain is divided into a system and a bath, with stringent assumptions on the bath subsystem and its coupling to the system. Here we take an alternative, third route, introducing a space-time approach to implement outgoing boundary conditions at the surface of an appropriately chosen finite computational domain. A variant of this approach has recently been shown to be compatible with the corresponding second quantized theory in the case of scalar non-relativistic electrodynamics~\cite{malekakhlagh2016non,kanupaper}. 

\itemizedparag{
Discuss the connection between the Schwinger model and the perturbed SG model (also state what is the specific perturbation without going into equations). 
\begin{itemize}
    \item Describe the connection
    \item Cite earlier work on relativistic physics in JTLs (what is that exactly, I do not think it is related to the Schwinger model). 
    \item The importance of the Schwinger model is that it enables studying non-perturbative quantum electrodynamics, provided a lot of insight (list phenomena), and led to the development of Lattice Gauge Theory which in turn provided for the first time numerical tools to study and model QCD.  
    \item Our goal here is to set the stage for a longer-term study: Structure formation in non-perturbative QED.
    \item Clearly state the caveat that we are studying semi-classical dynamics of fermionic four-current and exactly under what conditions this is accurate.    
\end{itemize}
}

\itemizedparag{
Lay out the program of the paper 
\begin{itemize}
    \item First, we present our numerical method
    \item Discuss DEC-QED and earlier work
    \item We first validate our model, discuss numerical accuracy and efficiency, by reproducing earlier work on fluxons under various settings
    \item Then we come to the main subject matter, semiclassical ED of electrons and positrons, the Schwinger process, atom formation, and stabilization by dissipation
    \item Emphasize that dissipation due to radiation is hard to describe, unlike other types of dissipation, and our solution to this problem in the context of DEC-QED
    \item Also emphasize the multi-scale nature of the dynamics we encounter, and how coarse-graining is an effective solution (cite and connect to earlier work on coarse-graining, in space and in time and that here we extend to space-time in a relativistically covariant fashion).  
\end{itemize}
}

\section{Summary of Results}

To guide the reader through the paper, this section presents a summary of the results.

We study the long-time evolution and structure formation in SG models. To do so, we first present our numerical method that builds upon DEC-QED - a coarse-graining approach that was recently developed for charged quantum fluids in multiscale heterogeneous environments~\cite{dec-qed, pham2023spectral}. By incorporating the Minkowski metric into the formulation of the nonlinear wave equation, the spatial and temporal dimensions can be consistently discretized together. Combining with the structure-preserving framework of DEC-QED, this allows for an enhanced stability for long-time simulations that is also resource-efficient.


With the numerical technique in place, we first validate our method by reproducing both the transient and the long-time dynamics of fluxons and breathers in Josephson transmission lines subject to various types of perturbation that are typically met in realistic junctions. These include the effects of resistive loss, external bias current, microshorts, and the dynamics of vortex-antivortex collisions. These numerical calculations serves a number of purposes: (1) first, they verify the results expected from previous asymptotic and perturbative analysis but have not been clearly demonstrated by direct long-time simulations. (2) Secondly, these calculations validate the accuracy of our proposed method and serve as benchmark examples to illustrate the advantages of our approach compared to existing methods for time-evolution of nonlinear dynamics. (3) Thirdly, the behavior of single fluxons or fluxon-antifluxon interaction are useful prototypical dynamics for our study of solitary waves in the massive Schwinger model. 

To showcase the versatility of the proposed numerical tool in practical modeling of Josephson junctions, we also examine fluxon dynamics in a junction featuring narrow constrictions. In this setting, we observe signatures of Cherenkov radiation — a phenomenon previously thought to occur in long Josephson junctions only when multiple junctions are coupled together~\cite{goldobin1998cherenkov}. 
We also study dynamical creation of fluxons and breathers in a junction using external electromagnetic pulses. The efficiency is illustrated through the finding of pulse parameters that would set off the desired soliton dynamics. 

To demonstrate the effectiveness of the proposed method for capturing multi-scale dynamics, we simulate the dynamical screening of a charged capacitor caused by the backreaction of e-p pairs generated via the Schwinger process~\cite{Schwinger1951} from vacuum. By comparing our numerical simulations with the analytical results in Ref.~\cite{chu2010capacitordischarge} for the massless case, we observe that the full evolution of charge and current distributions — including rapid oscillations, intermediate-time interference, and long-time asymptotic decay — is accurately reproduced. Furthermore, because the method is grounded in the rigorous framework of discrete differential forms~\cite{hirani2003discrete}, applying DEC-QED to the full spacetime grid enables precise enforcement of conservation laws, thereby improving numerical stability. In particular, the local charge-current continuity condition is exactly captured through its corresponding geometric identity, achieving machine-precision accuracy. Total energy is approximately conserved over long simulations, with numerical artifacts such as artificial micro-oscillations significantly reduced with respect to other numerical integration techniques we compare to.

We next turn to the dynamics in the bosonized Schwinger model, which is the central focus of this paper. Specifically, we investigate the behavior of a unit-charged soliton subject to the influence of a fixed external charge that neutralizes it. Building on semiclassical analysis of the bosonized model, we numerically demonstrate the formation of a one-dimensional analog of the Hydrogen atom in the asymptotic long time limit. We henceforth refer to this structure as the ``Schwinger atom". The full evolution — from initially separated components to a stable bound state — is captured. We also investigate the dynamics of oppositely charged soliton pairs, showing their potential to form a bound state we call ``Schwinger positronium". Depending on the parameter regime, the soliton pair either evolves into a stable, breather-like 1D positronium or scatters apart. In both the hydrogen-like and positronium-like cases, the bosonized massive Schwinger model appears to exhibit limit cycle behavior over a broad parameter range, driving the system toward stable atomic configurations in the long-time limit by radiating energy. Our results show that this process is stabilized by the dynamical mass and radiative boundary conditions, indicating that the dynamical mass term is a relevant perturbation shaping the nature of the solutions. These solutions are not only intrinsically interesting, but the computational framework and analysis developed here provide a foundation for broader investigations into structure formation in non-perturbative QED.


The rest of the paper is organized as follows: 
In Section~\ref{sec:background} we review the background on the SG model in Josephson transmission lines and the massive Schwinger process.
In Section~\ref{sec:dec_algo} we discuss the discretization of the nonlinear wave equation in the spacetime plane using DEC-QED. In Section~\ref{sec:bareJJ} we present the results for fluxons and breathers in JJs without perturbations, as well as compare the accuracy of our numerical method with the Euler's method. In Section~\ref{sec:JJwithperturbations} we discuss the dynamics of these solitons in junctions that contains resistive loss, bias currents, microshorts, and vortex-antivortex annihilation. 
In Section~\ref{ref:masslessSchwinger} we present the detailed 
backreaction dynamics arising from the massless Schwinger process during the discharging of a capacitor and compare our results with the Euler's method and the Crank-Nicholson method. We then focus on the massive Schwinger model in Section~\ref{sec:massiveSchwinger}, specifically delving into the dynamics of the Schwinger atom and Schwinger positronium. 
The technical details on how to derive the SG model for long Josephson junctions from the electrodydrodynamical description of the superconducting condensate is shown in Appendix~\ref{append:SG_from_EHDS}. Details on how to implement boundary conditions in our method are presented in Appendix~\ref{append:BCs}, while in Appendix~\ref{append:runtimes} we compare the runtime of this method with the Euler approach. 
More results on solitons in JJs are also discussed in Appendix~\ref{append:verylongtime_biasedBC} for the effect of boundary currents and magentic fields, Appendix~\ref{append:constrictions} for the dynamics of solitons in narrow constrictions, and Appendix~\ref{sec:pulse} for the dynamical creations of solitons in junctions. In Appendix~\ref{append:capdischarge_massive} we provide results on the capacitor discharge problem in the massive model.

\section{Background}\label{sec:background}
\begin{figure}[t]
    \centering
    \includegraphics[scale=0.39]{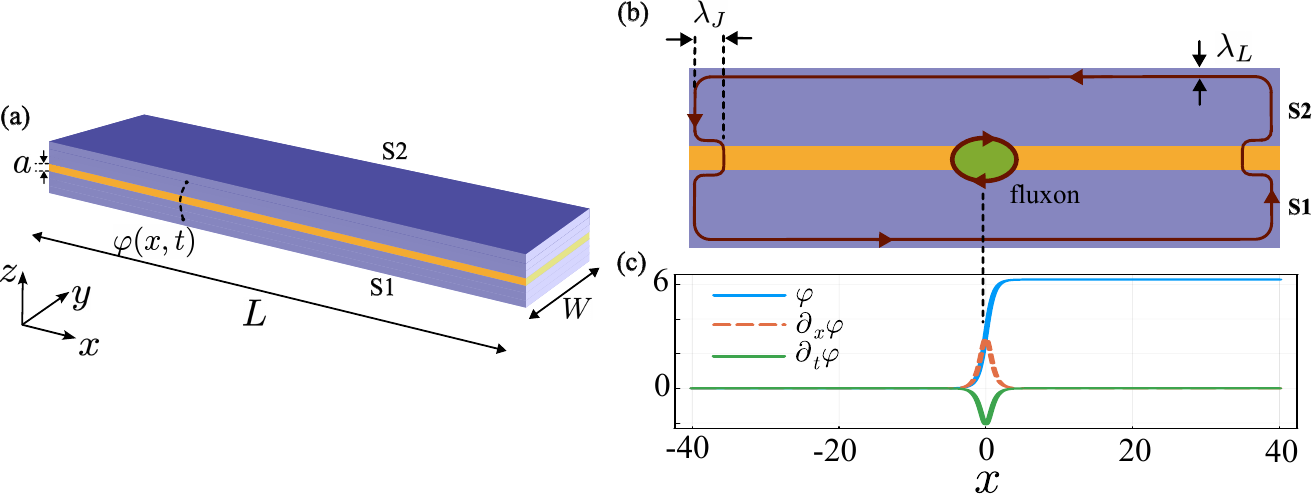}
    \caption{(a) 3D Schematic of a long junction in which the generalized flux variable $\varphi(x,t)$ varies along the length of the junction. (b) A view at an $xz$-slice of the junction. $\lambda_L$ is the London penetration depth and $\lambda_J$ is the Josephson penetration depth. (c) $\varphi(x)$, $\partial_x\varphi$ and $\partial_t\varphi$ of a $2\pi-$kink solution of the pure SG equation.}
    \label{fig:JJschem}
\end{figure}

\subsection{The Sine-Gordon Equation for Fluxons}

In this section, we provide some background information on how the Sine-Gordon equation arises in long Josephson junctions.  A more simple model, the short Josephson junction, is one whose lateral dimensions are much shorter than the Josephson penetration depth ($L\!\ll\!\lambda_J$ in Fig.\,\ref{fig:JJschem})~\cite{buckel2008superconductivity}. For a junction made of an insulator with thickness $a$ sandwiched between two superconducting electrodes having London penetration depth $\lambda_L$, $\lambda_J$ is given by $\lambda_J\approx\sqrt{\Phi_0/[2\pi\mu_0j_c\,(a+2\lambda_L)]}$~\cite{weihnacht1969influence}, where $\Phi_0$ is the flux quantum, and $j_c$ is the critical current density. For junctions typically used in superconducting microwave circuits, $\lambda_J$ is in the order of a few $\mu m$, while $\lambda_L \approx $ few $nm$.
The dynamics of Josephson junctions in an electromagnetic environment are encoded in the gauge-invariant, generalized flux variable $\varphi$ given by
\begin{equation}\label{eq:varphi_def}
    \varphi = -\frac{2\pi}{\Phi_0}\int_{1}^{2}{\mathbf d\ell}\cdot\bm{{\mathcal A}},
\end{equation}
where the line integral is carried out from one superconducting island to the other, while \mbox{$\bm{\mathcal{A}} = {\bf A} - \frac{\hbar}{q}\nabla\theta$}, with ${\bf A}$ being the magnetic vector potential and $\theta$ is the phase variable in the Madelung representation of the superconducting order parameter $\psi$ ($\psi=\rho e^{i\theta}$) ~\cite{madelung}.  In this case, the generalized flux variable $\varphi(t)$ will be described by
\begin{equation}
    \partial_t^2 \varphi + \alpha \partial_t \varphi + \sin \varphi = -\beta
\end{equation}
where the dissipative term $\alpha$ quantifies the resistive loss in the junction, and the term $\beta$ describes the amplitude of an external bias current injected the junction.

If one of the two lateral dimensions of the junction is sufficiently long such that the externally applied magnetic field only penetrates into a thin layer at the boundary of the junction ($L\gg\lambda_J$), it is called a long Josephson junction and the generalized flux variable $\varphi = \varphi(x,t)$ will vary in $x$ along the junction (see Fig.\,\ref{fig:JJschem}) and its dynamics are described by the perturbed SG equation~\cite{mclaughlin1978_SG, ustinov1998solitons}
 \begin{equation}\label{eq:perturbedSG}
     \partial^2_{t}\varphi - \partial^2_{x}\varphi + \alpha\partial_t\varphi + \sin\varphi = -\beta. 
 \end{equation}
In Eq.\,\ref{eq:perturbedSG} above, $x$ labels the coordinate (in units of Josephson penetration depth $\lambda_J$) along the long dimension of the junction, and $t$ is the time variable normalized by the inverse of the Josephson plasma frequency $\omega_p/2\pi=c\sqrt{\mu_0j_c/\Phi_0}$.  For the remainder of the text, we will specifically focus on the long junction. For a long junction with overlap geometry, $\beta$ models the bias current density across its length and can be spatiotemporally varying. For the inline configuration, the bias is injected at the junction endpoints and enters the model via boundary conditions on the sine-Gordon equation\,\cite{parmentier1993solitons}.

It is well known that supercurrent vortices can be created and stabilized within the nonlinear medium centered around the nanometer-thick insulating region that separates the two superconducting islands of a Josephson junction. These vortices - also called {\it fluxons} - are circulating supercurrents that carry a $2\pi-$ jump (a ``kink") in the value of $\varphi(x)$ (see \Fig{fig:JJschem}(b)).

If there are no perturbations ($\alpha=\beta=0$), a fluxon corresponds to the analytical solution of Eq.\,\ref{eq:perturbedSG} that describes a single soliton~\cite{mclaughlin1978_SG}. In general, the unperturbed SG equation permits solutions containing an arbitrary number of fluxons, antifluxons (i.e. antivortices), as well as radiation. While analytical solutions exist for the pure SG case~\cite{mclaughlin1978_SG} and for SG model with small perturbations~\cite{rubinstein1970SG, fogel1977perturbedSG} (we discuss some analytically known solutions in Section\,\ref{sec:LJJ}), the precise dynamics of solitons in the presence of significant losses and external biases can only be studied numerically.  Note that although throughout this paper we refer to any additional terms added on to the Sine-Gordon model as ``perturbations," we do not use any perturbation theory to study the systems discussed in this text.

\subsection{The Sine-Gordon Equation for the Bosonized Schwinger Model}
\label{ref:massiveSchwinger}

The Schwinger model~\footnote{Strictly speaking, the term Schwinger model traditionally refers to 1+1D QED with {\it massless} fermions, first introduced as a toy model for illustrating confinement in Ref.~\cite{schwinger1962}. In this work, however, we will use the term to refer to its massive counterpart.} is a 1+1D model of relativistic quantum electrodynamics of Dirac fermions coupled to the electromagnetic field.  The Lagrangian is given by
\begin{equation}
    \mathcal{L} = \Bar{\psi} ( i \gamma^\mu D_\mu - m ) \psi - \frac{1}{4} F_{\mu \nu} F^{\mu \nu},
    \label{eq:mschwingerlag}
\end{equation}
where $\psi$ is the fermion spinor field, $m$ is the mass of the fermion,  $D_\mu=\partial_\mu-i e A_\mu$ is the gauge covariant derivative, and $F_{\mu\nu}=\partial_\mu A_\nu-\partial_\nu A_\mu$ is the electromagnetic field strength tensor, which in 1+1D has only one independent component, $F^{\mu \nu} = \epsilon^{\mu \nu} F^{01}$ (where $\epsilon^{\mu\nu}$ is the 1+1D Levi-Civita tensor, so that $\epsilon^{01} = 1$), identified as the electric field $F^{01} = - F^{10} = E(x,t)$. The Schwinger model can be equivalently rewritten as a theory of bosons where components of the Dirac field $\psi$ are mapped to exponentials of the corresponding boson fields, i.e. the vertex operators in the bosonic theory~\cite{coleman1975, coleman1976, mandelstam1975}. In particular, the bosonized equivalence of the Schwinger model is described by the following Lagrangian
\begin{equation}
    \mathcal{L} = \frac{1}{2} \partial_\mu \varphi \partial^\mu \varphi + \frac{\kappa}{2} \cos (2 \sqrt{\pi} \varphi ) + \frac{g}{2} \varphi \epsilon^{\mu\nu} F_{\mu\nu} - \frac{1}{4} F_{\mu\nu} F^{\mu\nu}
    \label{eq:qSGmodel}
\end{equation}
where $\varphi(x,t)$ is a Hermitian scalar field satisfying bosonic commutation relations, $g=\frac{e}{\sqrt{\pi}}$, and $\kappa = \frac{m}{\pi \epsilon}$ where $\epsilon$ is a UV cutoff scale in energy. The term proportional to $\kappa$ originates from the fermionic mass term in \Eq{eq:mschwingerlag}, while the term proportional to $g$ — as will become clear once the gauge field is integrated out (see \Eq{eq:massiveQSG}) — leads to a dynamically generated mass for the scalar field $\varphi$. Intuitively, this can be traced back to earlier analyses~\cite{schwinger1962, lowenstein1971, coleman1975} of the exactly solvable massless fermion case,  which revealed that the fundamental excitations are not isolated charged particles but rather neutral, dipole-like mesonic states. These mesons acquire mass through their hybridization with the gauge field. A concise summary of the relevant bosonization framework is provided in Appendix~\ref{append:bosonization}.

The Euler-Lagrange equations for \Eq{eq:qSGmodel} yield
\begin{equation}\label{eq:skg_full}
    (\partial_t^2 - \partial_x^2) \varphi - \frac{g}{2} \epsilon_{\mu\nu} F^{\mu\nu} + \sqrt{\pi} \kappa \sin(2 \sqrt{\pi} \varphi) = 0,
\end{equation}
\begin{equation}\label{eq:emfield_eom}
    \partial_\mu F^{\mu\nu} = g \epsilon^{\mu\nu}\partial_\mu \varphi = j_\varphi^\nu
\end{equation}
where $j_\varphi^\nu$ represents the conserved current in the bosonic picture.  From this, Eq.\,\ref{eq:emfield_eom} can be integrated to show that $E=g\varphi$ and plugged into Eq.\,\ref{eq:skg_full} to get a single equation of motion for the system:
\begin{equation}
    \label{eq:massiveQSG}
    (\partial_t^2 - \partial_x^2) \varphi + g^2 \varphi + \sqrt{\pi} \kappa \sin(2\sqrt{\pi}\varphi) = 0
\end{equation}
where we can now explicitly see the role of $g$ as a mass-gap term for the scalar field.

From Eq.\,\ref{eq:emfield_eom}, one can see that the conserved {\it total charge } is given by the difference in the values of $\varphi$ at the boundaries of the system.
\begin{equation}
    Q = -g \int_{-L}^L dx \, \partial_x \varphi = -g\Big[ \varphi(L)-\varphi(-L) \Big]
\end{equation}
Therefore, the boundary conditions on $\varphi$ restrict us to a subset of the state space in which the total fermion number (fermions minus antifermions) is fixed. The system evolves in time by creating and annihilating electron-positron pairs. An external electric field applied to a system in the vacuum state creates electron-positron pairs through the Schwinger process~\cite{Schwinger1951}, and the electric fields between the pairs will have a backreaction effect on the original external field~\cite{chu2010capacitordischarge}. In Sections \ref{ref:masslessSchwinger}, \ref{sec: schwinger atom}, we consider situations where the electric field is created by a central point source ~\footnote{The original Schwinger process was proposed as a non-perturbative effect in 3+1D QED. Throughout this work, references to the Schwinger process and to the creation or annihilation of electron-positron pairs should be understood as their analogous counterparts in the context of the 1+1D model.}.

In the limit where $\kappa \to 0$ (massless fermions), the equation of motion is linear and can be solved exactly.  The assignment of the variable name $\varphi$ to both the generalized flux in a Josephson junction and to the bosonic field in the Schwinger model is an intentional choice here. By comparing Eq.\,\ref{eq:perturbedSG} to Eq.\,\ref{eq:skg_full}, one can see that there are similarities: both are wave equations that have a non-linear sinusoidal term. In this manuscript, whenever the variable $\varphi$ is discussed, the context will make it clear whether the Josephson flux or the Schwinger bosonic field is being referred to.

Key insight into the non-perturbative aspects of quantum field theories of the type in \Eq{eq:qSGmodel} can be gained through a semiclassical approach~\cite{dashen1974, dashen1975}  which involves analyzing solutions of the corresponding classical field theory. In the remainder of this paper, we carry out such an analysis in detail.

In addition to having the same pure SG terms, Eqs.\,\ref{eq:perturbedSG} and \ref{eq:skg_full} differ in the additional perturbing terms, which greatly affect how the systems evolve over time and their steady states. Nevertheless, as we will see, the solutions of \Eq{eq:skg_full} still feature solitary waves. In the bosonized picture of the Schwinger model, the corresponding kink solution in $\varphi$ represents a transition between two discrete vacua, which we will constrain to be equal to a single electron charge in the case of the central charge problem we analyze in Section \ref{sec: schwinger atom}. Therefore, a SG soliton here will be a finite-size field packet that carries exactly one unit of charge \cite{colemanthirring}.  

Based on their solitary wave solutions, we can also establish a correspondence between a macroscopic Josephson vortex (antivortex) and an electron (positron) as in Table\,\ref{table:mapping}: in addition to the mapping of the single-kink solution to a flux quantum carried by the Josephson vortex and to an electron charge in the Schwinger model that we have already discussed, the spatial derivative $\partial_x\varphi$ maps to the magnetic field $H_y(x)$ that pierces the junction and corresponds to the charge density $\rho$ in the Schwinger model. The temporal derivative $\partial_t\varphi$, on the other hand, represents the voltage $V$ in the long JJ, whereas in the Schwinger model it is proportional to the current density $J$.

\begin{table}[h!]
		\centering
		\vspace{0.1in}
		\begin{tabular}{c|c|c}
			\hline
			\hline
			  & Josephson vortex & Schwinger electron \\
			\hline
			$\varphi_{+\infty}-\varphi_{-\infty}$ & \hspace{0.05in} flux quantum $\Phi_0$ \hspace{0.05in} & \hspace{0.05in} charge $e$ \hspace{0.05in} \\
			$\partial_x\varphi$ & magnetic field $H_y$ & charge density $\rho$ \\
			$\partial_t\varphi$ & voltage $V$ & current density $J$\\
			\hline
			\hline
		\end{tabular}
            \caption{\label{table:mapping} Mappings from the kink solution $\varphi$ for SG equation and its derivatives to the quantities that they are proportional to. $\varphi_{-\infty}$ and $\varphi_{+\infty}$ are the asymptotic values of $\varphi$ at $x\!=\!\pm\infty$.} 
	\end{table}

\section{Discretization of 1+1D field theory with DEC-QED }\label{sec:dec_algo}
 \begin{figure}[t]
    \centering
    \includegraphics[scale=0.19]{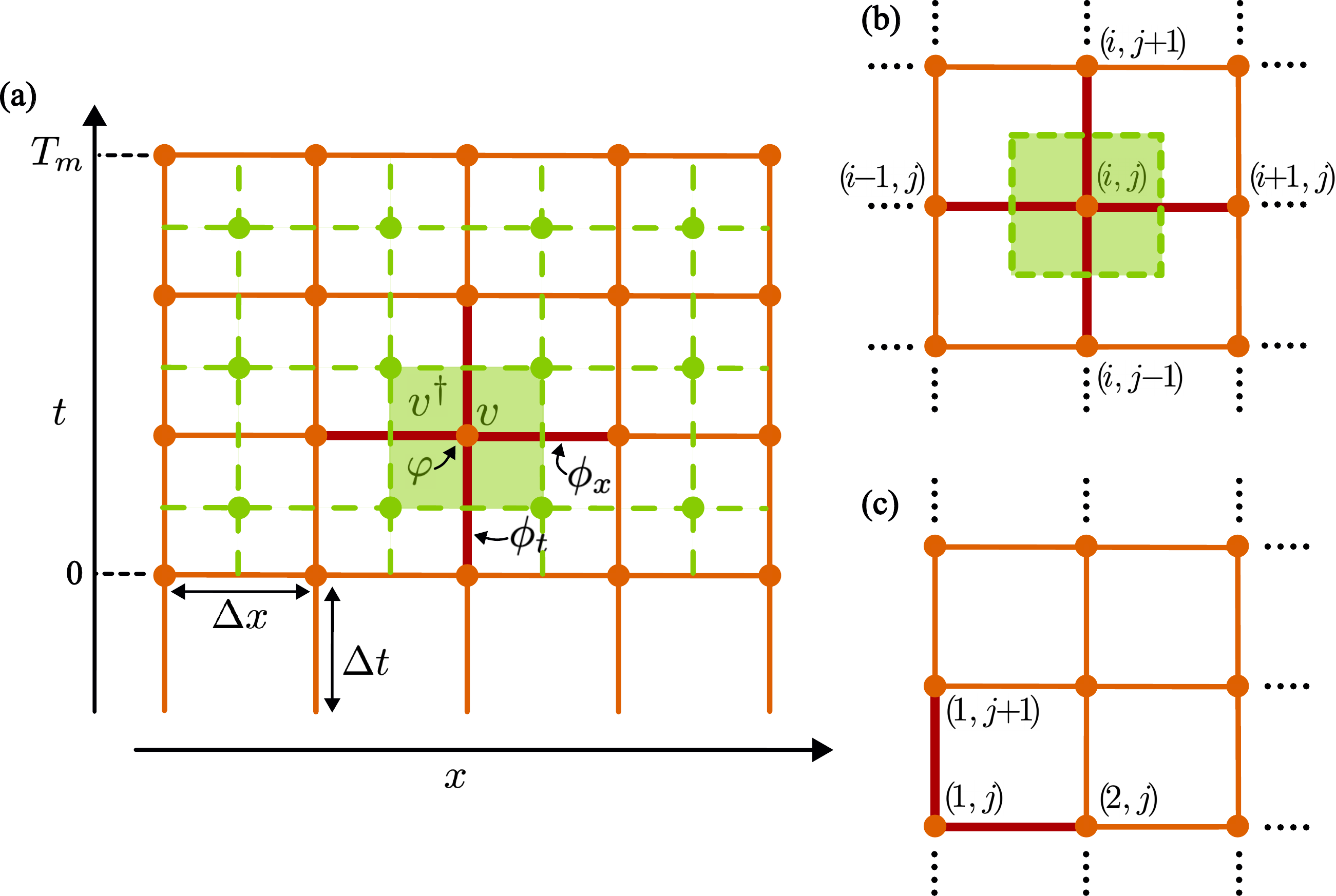}
    \caption{(a) A schematic of the 1+1D spacetime grid used for the DEC algorithm. The vertices and edges on the primal mesh are colored orange, while the dual vertices and dual edges are in green. An example cell $v^\dagger$ that is the dual of a primal vertex $v$ is also shaded in green. The primal edges connected to $v$ are bolded in red. (b) A closed-up view of an internal primal vertex $v(i,j)$ along with its single dual face $v^\dagger$, the neighboring vertices, and the primal edges associated with this vertex. Dual faces of neighboring vertices are not shown here. (c) A closed-up view of a boundary vertex $v(i,j)$ with the boundary primal edges associated with it.}
    \label{fig:dec_grid}
\end{figure}

In this section, we discuss a discretization procedure for the 1+1D field equations to enable their symplectic integration. Although in this section we will focus on the derivation of the coarse-grained version of the perturbed sine-Gordon equation in space-time, the process for coarse-graining other nonlinear field equations follows a similar procedure (see e.g. Ref.~\cite{dec-qed}).  We consider a primal rectangular grid $M$ that discretizes the space-time computational domain, in which $\Delta x$ and $\Delta t$ are spacings between the neighboring grid points along $x$ and $t$, respectively. We label the vertices, edges, and elemental faces of this mesh as $v$, $e$ and $f$, respectively.
 Additionally, consider a dual mesh whose vertices are the circumcenters of the rectangular cells $f$ of the primal grid. Each edge of this dual mesh is constructed by connecting the two dual nodes corresponding to two neighboring primal faces (faces that share a primal edge). The dual faces, edges, and nodes are labeled $v^\dagger$, $e^\dagger$, and $f^\dagger$, respectively. Fig.\,\ref{fig:dec_grid}a presents a schematic of the spacetime grid, where the primal mesh is colored orange and the dual mesh is in green.

 By introducing the 1+1D Minkowski metric tensor $g_{\mu\nu}$~\cite{lee2006riemannian, dirac1996GR}, Eq.\,\ref{eq:perturbedSG} can be conveniently rewritten as
 \begin{equation}\label{eq:perturbedSG_imt}
     g^{\mu\nu}\partial_\mu\partial_\nu\varphi + \alpha\partial_{t}\varphi + \sin\varphi = -\beta, 
 \end{equation}
 where $t$ is now treated on equal footing with the spatial coordinate $x$. 
 We now introduce the coarse-grained variables $\phi_x(e_x)$ and $\phi_{t}(e_t)$ living on the edges $e_x$ and $e_{t}$ along $x$ and $t$ axes, respectively  
 \begin{align}
     \phi_x(e_x) &= \int_{\Delta x}\nabla_x\varphi dx, \label{eq:phix} \\
     \phi_{t}(e_t) &= \int_{\Delta t}\nabla_{t}\varphi dt. \label{eq:phit}
 \end{align}
Integrating Eq.\,\ref{eq:perturbedSG_imt} over a dual face $v^\dagger$ surrounding a primal node $v$ (see Fig.~\ref{fig:dec_grid}(b)) and applying Gauss's law gives
 \begin{align}\label{eq:discretegauss_SG}
     0 &= \int_{v^\dagger}( g^{\mu\nu}\partial_\mu\partial_\nu\varphi + \alpha\partial_{t}\varphi + \sin\varphi + \beta)dA \\
     &= \!\int_{\partial v^\dagger}\!\!\hat{n}_\mu\!\cdot\!\left(\nabla\varphi\right)^\mu ds +  \left(\alpha\partial_{t}\varphi \!+\! \sin\varphi \!+\! \beta\right)\!\Delta A \nonumber\\
     &= \!\left(\!-\sum_{\partial_t v^\dagger}\!\frac{\Delta t}{\Delta x}\phi_x \!+\! \sum_{\partial_x v^\dagger}\!\frac{\Delta x}{\Delta t}\phi_t \right) \!\! + \! \left(\alpha\partial_{t}\varphi \!+\! \sin\varphi \!+\! \beta\right)\!\Delta A.  \nonumber
\end{align}
On the last line of Eq.\,\ref{eq:discretegauss_SG} above, the summations are done over all four boundary edges of the dual face $v^\dagger$, where $\partial_tv^\dagger$ denotes the two boundary edges along $t$  and  $\partial_xv^\dagger$ denotes the two boundary edges along $x$. By dividing both sides of Eq.\,\ref{eq:discretegauss_SG} by $\Delta A$, we obtain the discrete version of the perturbed SG equation
\begin{align}\label{eq:SG_dec}
     -\!\sum_{\partial_t v^\dagger}\!\frac{\phi_x}{\Delta x^2} \!+\! \sum_{\partial_x v^\dagger}\!\frac{\phi_t}{\Delta t^2} + \alpha\frac{\bar{\phi_t}}{\Delta t} + \sin\varphi + \beta = 0,
\end{align}
where $\bar{\phi_t}$ is the average of the field values on the two $\phi_t$ edges connected to $v$.
Given initial conditions on either $\phi_t$ and $\varphi$ everywhere in space or $\phi_t$ and $\phi_x$ everywhere combined with the value for $\varphi$ at the boundary, Eq.\,\ref{eq:SG_dec} can be used to propagate the system in time to study the evolution of $\phi_t$, $\phi_x$. The procedure is as follows: by labeling each primal vertex using their discrete spatial and temporal indices $i$ and $j$ and each primal edge by the two vertices connected to it, Eq.\,\ref{eq:SG_dec} when applied to a vertex $v(i,j)$ can be rewritten as

\begin{widetext}
\begin{align}\label{eq:SG_dec_updaterule}
    &\phi_t[(i,j),(i,j\!+\!1)] =\nonumber\\
    &\frac{\Delta t^2}{1+\alpha\Delta t/2}\left\{ \frac{\phi_x[(i,j),(i\!+\!1,j)]-\phi_x[(i\!-\!1,j),(i,j)]}{\Delta x^2} + \left(\frac{1}{\Delta t^2} - \frac{\alpha}{2\Delta t} \right)\phi_t[(i,j\!-\!1),(i,j)] - \sin\varphi(i,j) - \beta \right\}.
\end{align}
\end{widetext}
From Eq.\ref{eq:SG_dec_updaterule} one can see that at each $(j+1)^{\text{th}}$ time step the field living on each internal edge $\phi_t[(i,j),(i,j\!+\!1)]$ can be computed given the knowledge of the edge fields $\phi_x$, $\phi_t$ and the scalar field $\varphi$ at the previous time step. The value of $\phi_t$ at the boundary edges are determined by boundary conditions (BCs). Depending on the type of BCs, they can be imposed either on the boundary edges or the boundary vertices (Fig.\,\ref{fig:dec_grid}c).
Details on how to impose boundary conditions (both closed and outgoing BCs) are discussed in Appendix \ref{append:BCs}.
Once $\phi_t$ at the $(j+1)^{\text{th}}$ time step is determined everywhere in space, $\varphi$ can be updated by applying Eq.\,\ref{eq:phit} and then $\phi_x$ can be found by applying Eq.\,\ref{eq:phix}. On a discrete grid, these equations translate {\it exactly} to the following update rules
\begin{align}
    &\hspace{-0.12in}\varphi(i,j\!+\!1) = \phi_t[(i,j),(i,j\!+\!1)] \!+\! \varphi(i,j),\label{eq:update_varphi}\\
    &\hspace{-0.12in}\phi_x [(i,j\!+\!1),\!(i\!+\!1,j\!+\!1)] = \varphi(i\!+\!1,j\!+\!1) \!-\! \varphi(i,j\!+\!1).\label{eq:update_phix}
\end{align}
Eqs.\,\ref{eq:SG_dec_updaterule}, \ref{eq:update_varphi}, and \ref{eq:update_phix}, along with initial and boundary conditions allow us to determine the fields $\phi_t, \phi_x,$ and $\varphi$ everywhere on the spacetime grid.
Note that Eq.\,\ref{eq:SG_dec} is effectively a first-order equation in terms of the edge fields $\phi_x$ and $\phi_t$, making numerical simulations more time-efficient than directly simulating the second-order SG equation in Eq.\,\ref{eq:perturbedSG}. In the next sections, we demonstrate the stability and effectiveness of this scheme for studying the long-time nonlinear dynamics of the SG equation under various perturbations.

\section{Dynamics of solitons in perturbed long Josephson junctions}\label{sec:LJJ}

\subsection{The bare sine-Gordon equation}\label{sec:bareJJ}

\subsubsection{Single fluxon dynamics}\label{sec:singlefluxon_bareJJ}

We first apply the numerical scheme developed above to compute the dynamics of a single fluxon in an ideal long JJ, a case that has been well studied using other numerical techniques~\cite{nakajima1974, krasnov2020josephson}. Here, the dynamics of the generalized flux variable $\varphi$ are governed by the unperturbed SG equation ($\alpha=\beta=0$). The boundary conditions \cite{ustinov1998solitons}, which are typically applied to inline junctions, are given by 
\begin{align} 
    \partial_x\varphi(0,t) &= \eta + \xi \label{eq:longjj_BC1} \\
    \partial_x\varphi(L,t) &= \eta - \xi \label{eq:longjj_BC2}
\end{align}
where $\eta$ is the normalized value of the $y$ component of external magnetic field, and $\xi$ is the normalized value of the external current injected into the sides of the junction. In an ideal and infinitely long junction, the analytical soliton solution describing a single fluxon is given by \cite{mclaughlin1978_SG}
\begin{equation}\label{eq:SG_singlefluxon}
    \varphi(x,t) = 4\tan^{-1}\left[\exp\left(\frac{x-x_0-ut}{\sqrt{1-u^2}}\right)\right] + 2\pi n,
\end{equation}
where $n$ is an integer, $x_0$ is the location of the fluxon at $t=0$, and $u$ is its velocity that is normalized by the Swihart velocity~\cite{swihart1961}. In Eq.\,\ref{eq:SG_singlefluxon}, the Swihart velocity is normalized, $0\leq u \leq 1$. The solution in Eq.\,\ref{eq:SG_singlefluxon} corresponds to a $2\pi-$kink in $\varphi$ that travels at constant speed $u$ and is the 1+1D description of a single (2+1D) vortex trapped between the superconducting islands of the junction. One can show that the 1+1D dynamics of $\varphi$ that is given by the sine-Gordon equation can be derived from the full 2+1D electro-hydrodynamical equations that couple the superconducting order parameter and the EM field.
For completeness, in Appendix \ref{append:SG_from_EHDS} we present a rigorous derivation of this reduction. 

\begin{figure*}[t]
    \centering
    \includegraphics[scale=0.41]{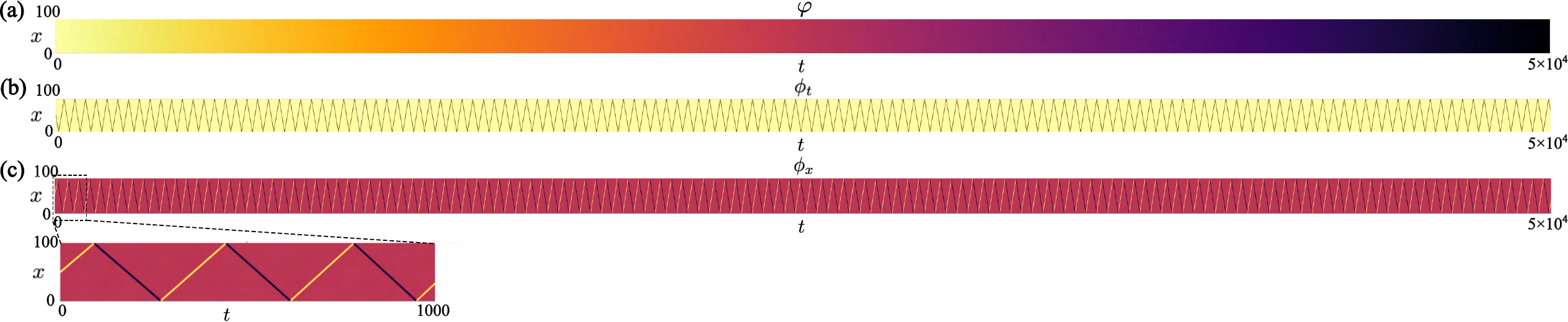}
    \caption{Long-time dynamics of a single fluxon trapped inside a Josephson junction. The junction length is $L=100$, the total time of the dynamics is $T_m=50000$, and the initial velocity of the fluxon is $u=0.55$, with no external bias on the boundary ($\eta=\xi=0$). (a), (b) and (c) show the value of $\varphi$, $\phi_t$, and $\phi_t$ respectively.}
    \label{fig:verylongtime_bareJJ}
\end{figure*}

\begin{figure}
    \centering
    \includegraphics[scale=0.18]{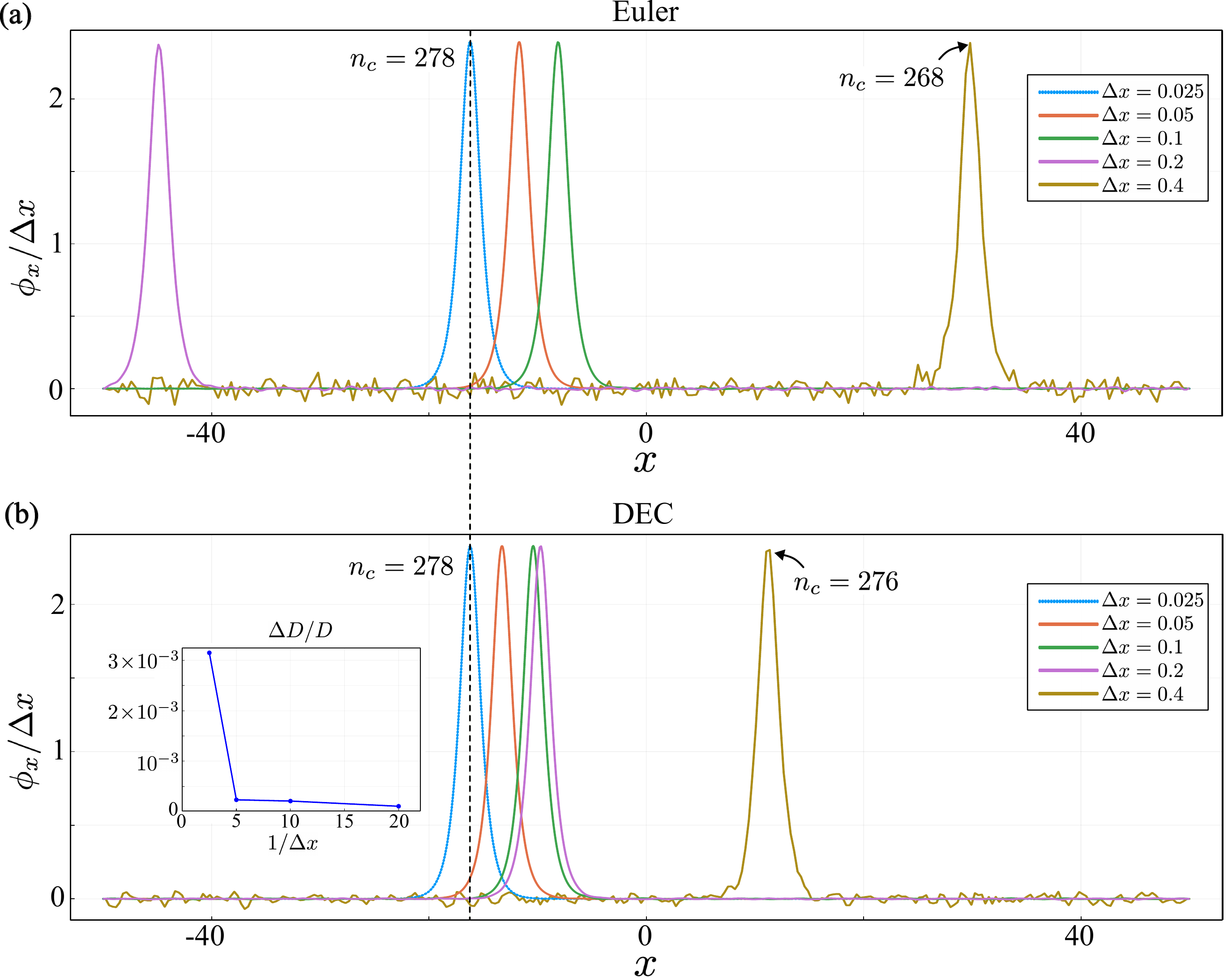}
    \caption{Comparisons on the final positions of the fluxon in the junction at the last time step \mbox{($t=T_m=50000$)} obtained with different discretization steps $\Delta x$ and $\Delta t$. (a) shows the results from using Euler method, and (b) shows the results from DEC-QED. In all the plots from both figures, $\Delta t=0.8\Delta x$, $u=0.55$, and $L=100$. The inset in (b) shows the difference $\Delta D$ in the distance traveled by the soliton calculated with different values of $\Delta x$ to the distance $D$ covered by the soliton simulated with $\Delta x = 0.025$.}
    \label{fig:verylongtime_laststep_DEC_euler_CN}
\end{figure}

Using the expression in Eq.\,\ref{eq:SG_singlefluxon} at $t=0$ as the initial condition, we perform long-time simulations of a single fluxon in a long JJ. Figs.\,\ref{fig:verylongtime_bareJJ}a, \ref{fig:verylongtime_bareJJ}b, and \ref{fig:verylongtime_bareJJ}c show the plots for $\varphi$, $\phi_t$, and $\phi_x$ respectively when there is no external bias on the boundary ($\eta\!=\!\xi\!=\!0$), implemented by a Neumann-type boundary condition for $\varphi$ at $x=0,L$. This translates to Dirichlet conditions on the edge field $\phi_x$. The length of the junction is $L\!=\!100$, with a total time $T_m\!=\!50000$, and the initial condition is given by Eq.\,\ref{eq:SG_singlefluxon} in which $u\!=\!0.55$, $x_0\!=\!0$, and $n\!=\!0$. 
The Lorentz factor, which determines the size of the soliton, is therefore \mbox{$1/\sqrt{1-u^2}\approx 1.2$}. 
The spacings in the space-time grid are kept at $\Delta x\!=\!0.05,$ and $\Delta t\!=\!0.04$ in order to resolve the kink.  With no external bias on the boundary, the fluxon simply collides with the junction boundary, is reflected back, and becomes an anti-fluxon that travels in the opposite direction with the same velocity. Without external force or perturbations, this process repeats indefinitely - as shown in Figs.\,\ref{fig:verylongtime_bareJJ}a-\ref{fig:verylongtime_bareJJ}c. 
Within the simulated time $(T_m\!=\!50000)$ the fluxon collides with the two boundaries a total of 278 times, while still maintaining its original shape and speed. By the end of the simulation, the soliton has traveled a distance that is 11583 times its size, providing an early hint for the stability of the numerical scheme even in the long-time regime. To also study the effect of boundary currents and magnetic fields, in Appendix \ref{append:verylongtime_biasedBC} we present simulations for the long-time dynamics with nonzero $\eta$ and $\xi$. 

For a more quantitative insight into the stability of DEC-based integration, we now provide a comparative analysis to the Euler method. Although the Euler method is generally first-order accurate for linear wave equations and can be unstable at long times due to dispersion~\cite{shlager1993relative, petropoulos1994phase, hyatt2025multiple}, the soliton solutions to the {\it unperturbed} sine-Gordon equation are the steady-state solutions in which the nonlinear term perfectly balances out the dispersion. Therefore, the Euler method can be used here to compare with our method. In Figs.\,\ref{fig:verylongtime_laststep_DEC_euler_CN}a and \ref{fig:verylongtime_laststep_DEC_euler_CN}b we compare the final positions of the fluxon in the JJ at the last time step $T_m\!=\!50000$ computed using the two methods with various discretization steps $\Delta x$ and $\Delta t$. With sufficiently small steps ($\Delta x\!=\!0.025$), the Euler method and DEC converge to the the same location for the fluxon at $t=T_m$ that we use as the benchmark (the blue curves in Fig.\,\ref{fig:verylongtime_laststep_DEC_euler_CN}a and Fig.\,\ref{fig:verylongtime_laststep_DEC_euler_CN}b) for more more coarse discretizations. As we increase $\Delta x$ (and $\Delta t$ accordingly by setting $\Delta t\!=\!0.8\Delta x$), the final location of the fluxon gradually drifts away from the benchmark, but solutions produced by DEC are noticeably closer to the benchmark than those obtained from the Euler method with the same dicretization. Overall, as $\Delta x$ increases, Euler's method diverges from the accurate solution faster than DEC does. In Fig.\,\ref{fig:verylongtime_laststep_DEC_euler_CN} the discrepancies are most significant with $\Delta x\! =\! 0.4$, when the soliton in both methods falls behind in the number of collisions with the boundary; the soliton in the Euler method only completes $n_c=268$ collisions, 10 less than 278. The soliton in DEC, on the other hand, has $n_c=276$, only two collisions shy of the correct number. In the inset in Fig.\,\ref{fig:verylongtime_laststep_DEC_euler_CN}b we show as a function of $1/\Delta x$ the difference $\Delta D$ in the distance traveled by the soliton with respect to the distance $D$ that the soliton traveled when a sufficiently small $\Delta x = 0.025$ was used; even for $\Delta x = 0.4$ -- a discretization step 16 times larger than the benchmark step size -- the relative error in the final position of the soliton is only $\Delta D/D \approx 3\times 10^{-3}$ after having traveled a distance that is 11583 times its size.

We also recorded the run times of our spacetime DEC-QED method as well as the Euler method in these simulations and found that although the two methods scale at the same rate, DEC-QED is always faster because it deals directly with the first-order equation for the edge fields rather than a second order equation for $\varphi$ as in the Euler method. The quantitative comparison is presented in Appendix \ref{append:runtimes}.

\subsubsection{Fluxon-antifluxon dynamics}\label{sec:vav_bareJJ}

\begin{figure}[t]
    \centering
    \includegraphics[scale=0.105]{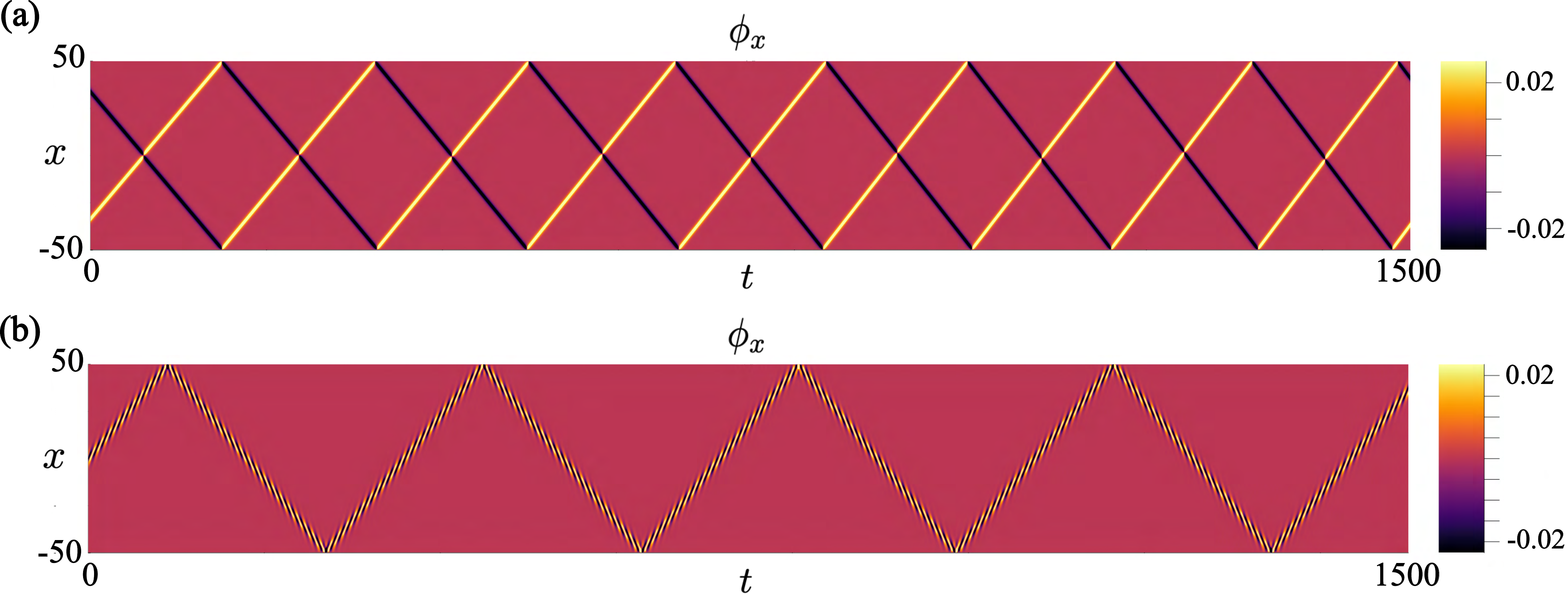}
    \caption{(a) Simulated dynamics of the field $\phi_x(x,t)$ that corresponds to a vortex-antivortex pair in a long Josephson junction. The initial distance between the pair is $d=66$, with each soliton having an initial speed of $u=0.55$ and set to travel towards each other. (b) The field $\phi_x(x,t)$ that corresponds to a breather traveling inside a long JJ also with speed $u=0.55$.}
    \label{fig:pairs_bareJJ}
\end{figure}

In general, the unperturbed sine-Gordon equation admits solutions that contain an arbitrary number of fluxons, antifluxons, and non-soliton radiation. A case that is typically investigated is the vortex-antivortex solution whose analytical form is known \cite{mclaughlin1978_SG}
\begin{equation}\label{eq:vav_pair}
    \varphi_{vav}(x,t) = -4\tan^{-1}\left(\frac{\sinh\left[(ut-d/2)/\sqrt{1-u^2}\right]}{u\cosh\left[(x-x_0)/\sqrt{1-u^2}\right]}\right),
\end{equation}
where $x_0$ is the initial position of the center of mass of the fluxon-antifluxon pair, $u$ is the velocity of the fluxon in the center-of-mass frame of reference, and $d$ sets the initial distance between the fluxon and the antifluxon. As the name suggests, the fluxon-antifluxon solution in Eq.\,\ref{eq:vav_pair} describes the generalized flux variable $\varphi (x,t)$ in a long JJ containing a vortex-antivortex pair. Using Eq.\,\ref{eq:vav_pair} as the initial condition, the dynamics of a vortex-antivortex pair in an unperturbed junction is simulated using our method and shown in Fig.\,\ref{fig:pairs_bareJJ}. Without any loss mechanism or perturbations and with sufficient kinetic energy, the two-soliton initial condition is sustained by the system and the two solitons pass through each other without losing speed. In this case, shown in Fig.\,\ref{fig:pairs_bareJJ}, the fluxon and antifluxon are not bound together; after colliding with each other, each continues to travel with the same speed towards the opposite boundaries before being reflected back. 

\subsubsection{Breathers}\label{sec:breather_bareJJ}
In certain cases, it is more energetically favorable for a pair to be bound together rather than being two separate solitons. These are called breathers, and their solutions in the rest frame of the center of mass are given by
\begin{equation}\label{eq:breathers}
    \varphi_B(x,t) = 4\tan^{-1}\!\!\left(\frac{\tan\nu\sin\left[(\cos\nu)(t-t_0)\right]}{\cosh\left[(\sin\nu)(x-x_0)\right]}\right).
\end{equation}
Eq.\,\ref{eq:breathers} describes a fluxon and an antifluxon oscillating around the pair's center of mass with a frequency given by $\cos\nu$.
If the breather is moving inside the junction at velocity $u$, a Lorentz transformation $x\rightarrow (x-ut)/\sqrt{1-u^2}, t\rightarrow (t-ux)/\sqrt{1-u^2}$ will boost the solution in Eq.\,\ref{eq:breathers} to one that describes a traveling breather. A numerical calculation of a breather bouncing between the two boundaries of a junction is shown in Fig.\,\ref{fig:pairs_bareJJ}.


\subsection{Vortices under perturbations}\label{sec:JJwithperturbations}
So far, we have presented numerical results for the pure sine-Gordon equation where the form of some families of solutions are analytically known to verify the accuracy and stability of DEC-QED based integration. We now apply the method to study the sine-Gordon model with perturbations. 

\begin{figure}[t]
    \centering
    \includegraphics[scale=0.05]{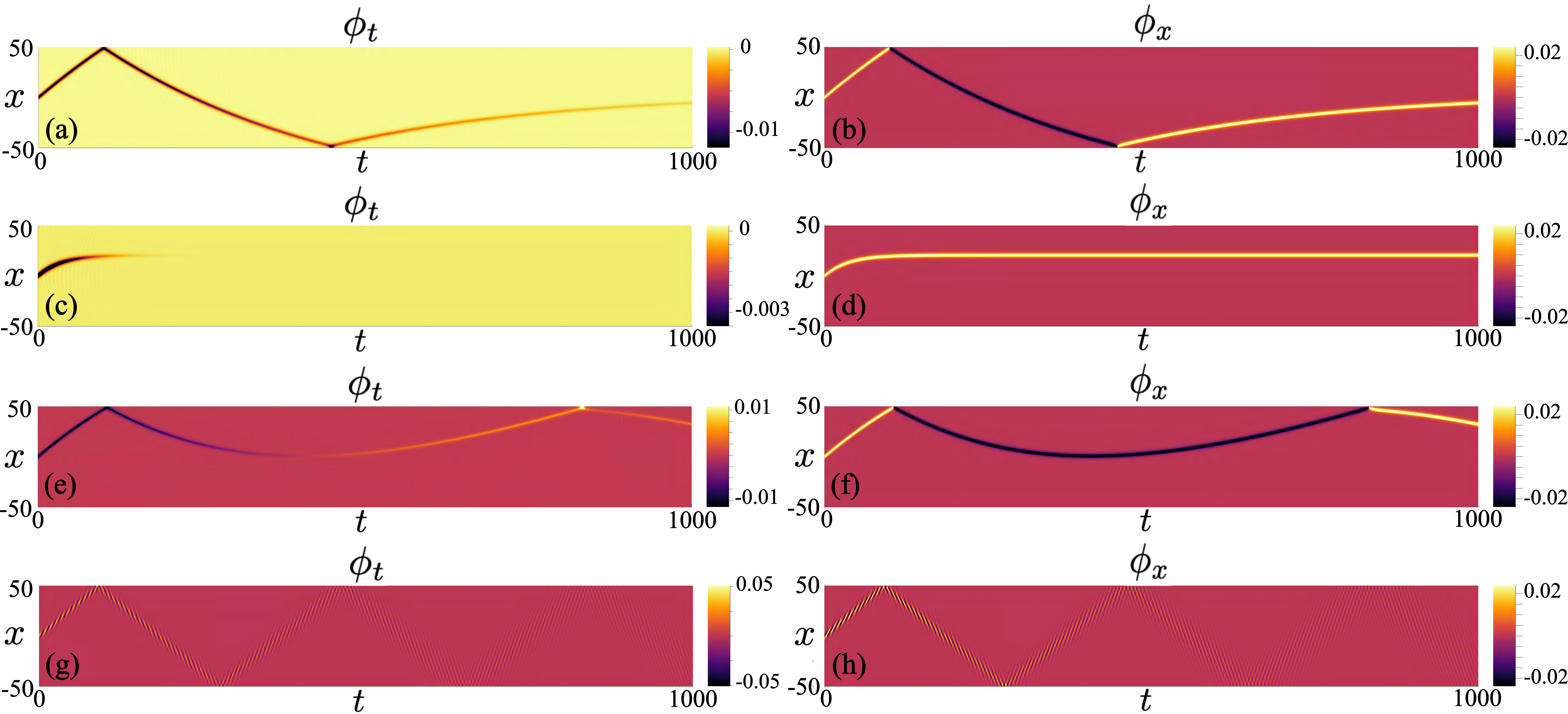}
    \caption{Dynamics of a SG soliton under dissipation and external bias current. The total simulated time is $T_m=1000$, the initial velocity of the fluxon is $u=0.55$, and the junction length is $L=100$ with boundary coefficients $\eta=0.002$ and $\xi=0.006$.
    In parts (a)-(f), the initial soliton is a fluxon. The coefficients for the resistive loss and bias current (a) and (b) are [$\alpha=0.003$, $\beta=0$], (c) and (d) are [$\alpha=0.03$, $\beta=0$], and in (e) and (f) are [$\alpha=0.003$, $\beta=0.001$]. The initial soliton in parts (g) and (h) is a breather traveling in a medium with coefficients [$\alpha=0.003$, $\beta=0$]. }
    \label{fig:soliton_w_loss_n_bias}
\end{figure}

\subsubsection{Solitons under resistive loss and external bias}

Consider the perturbed sine-Gordon model given in Eq.\,\ref{eq:perturbedSG} with finite resistive loss $\alpha$ and bias current $\beta$. Although analytical solutions for Eq.\,\ref{eq:perturbedSG} do not exist, numerical simulations provide a clear understanding of how these perturbations affect the SG solitons. We shall see that soliton solutions appear to be stable to perturbations of the class studied here, at least when such perturbations are small. 

In Figs.\,\ref{fig:soliton_w_loss_n_bias}a-b, we plot the dynamics of a fluxon that has initial velocity $u\!=\!0.55$ and travels in a lossy junction with $\alpha\!=\!0.003$ and without a bias current ($\beta=0$). The fluxon slows down over time due to friction, as evidently seen from the change in the slope of the trace that the soliton makes in space-time. The fading intensity of the trace in Fig.\,\ref{fig:soliton_w_loss_n_bias}a, where we plot $\phi_t$, also signifies the slowing down of the soliton. This is more apparent in Figs.\,\ref{fig:soliton_w_loss_n_bias}c-d, where the dissipation rate is stronger with $\alpha\!=\!0.03$, forcing the soliton to quickly come to a stop without ever reaching the junction boundary. In Figs.\,\ref{fig:soliton_w_loss_n_bias}e-f the dissipation rate is set to $\alpha\!=\!0.003$ as in Figs.\,\ref{fig:soliton_w_loss_n_bias}a-b, but now we also include a bias current by setting $\beta\!=\!0.001$. This bias current exerts a Lorentz force on the soliton and pulls it towards the \mbox{$-x$} direction if the soliton is a vortex and towards the \mbox{$+x$} direction if it is an antivortex. We can see this force in action in Figs.\,\ref{fig:soliton_w_loss_n_bias}e-f; the vortex is initially set to travel towards the $+x$ direction until it hits the boundary and becomes an antivortex that travels in the $-x$ direction. The antivortex slows down due to both the friction coming from the $\alpha$ term and the Lorentz force pulling it in the $+x$ direction and comes to a complete stop. It then accelerates in the opposite direction ($+x$) due to the Lorentz force until it hits the boundary and turns into a vortex again. 

Note that when traveling in a lossy medium and experiencing a bias current, the fluxon remains a soliton the whole time while changing its velocity. Depending on the velocity of the soliton, its shape varies according to the Lorentz contraction rule but it never losses its structural integrity (as shown in Figs.\,\ref{fig:soliton_w_loss_n_bias}b, \ref{fig:soliton_w_loss_n_bias}d, and \ref{fig:soliton_w_loss_n_bias}f). This is, however, not the case for breathers, as seen in Figs.\,\ref{fig:soliton_w_loss_n_bias}g-h, where the dynamics of a breather in a lossy medium is plotted. Unlike (anti)fluxons, breathers spread out and decrease in amplitude over time when dissipation is present. This is because a breather contains a fluxon-antifluxon pair oscillating around the center of mass, causing it to lose amplitude as it dissipates energy into the lossy medium. 

\subsubsection{Vortex-antivortex annihilation}\label{sec:vav_annhiliation}
As we saw in Section \ref{sec:bareJJ}, without any perturbation in a pure JJ, a fluxon-antifluxon pair can either exist as two separate solitons or as a bound breather that oscillates indefinitely. In the presence of dissipation (for example, through a finite loss term $\alpha$), the system loses energy and a vortex and antivortex can annihilate each other by binding together, forming a breather, and then the oscillations of the breather dissipating its energy to the lossy medium.  

\begin{figure}[t]
    \centering
    \includegraphics[scale=0.1]{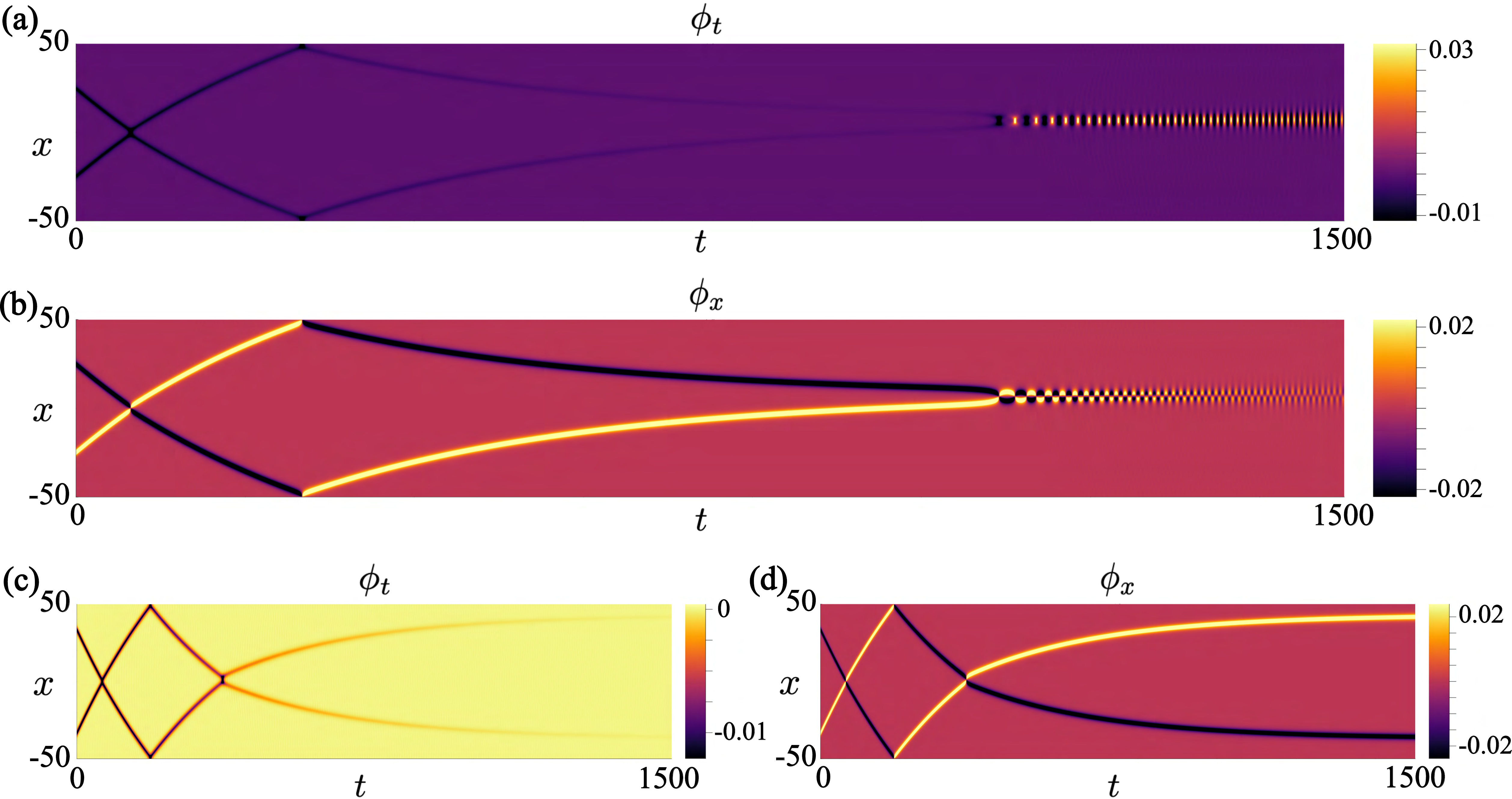}
    \caption{Vortex-antivortex dynamics in a lossy Josephson junction. The total simulated time is $T_m=1500$, the junction length is $L=100$, and the resistive loss coefficient is $\alpha=0.003$. In (a) and (b), the initial speeds of the two solitons are both $u=0.4$. In (c) and (d) their initial speeds are $u=0.55$.}
    \label{fig:vav_annihilate}
\end{figure}

The dynamics shown in Figs.\,\ref{fig:vav_annihilate}a-b is an example of such a scenario; here, the initial speeds of the two solitons are $u=0.4$, large enough so that they pass through each other at first. Then, they each reflect off their respective boundaries, and in the second time around they meet each other, the resistive medium has slowed down the solitons sufficiently such that they are bound together. The kinetic energy of each of them is no longer large enough to escape the other's attraction, and they oscillate together as a breather, losing energy in the process. 

Total annihilation of a pair does not always happen, however, as the pair has to eventually form a breather for this to happen. In Figs\,\ref{fig:vav_annihilate}c-d, we present a different scenario. The initial speed of the pair is $u=0.55$, slightly higher than in the previous case, while all other parameters are kept the same. We see that upon their second collision with each other, the solitons now still have enough kinetic energy to escape from each other. They are eventually slowed down to a complete stop at a finite distance away from each other and become two stationary solitons. Here we do not see a destruction of the pair because it is never bound into a breather.

\subsubsection{Interaction between a vortex and a microshort}\label{subsec:microshort}
\begin{figure}[t]
    \centering
    \includegraphics[scale=0.1]{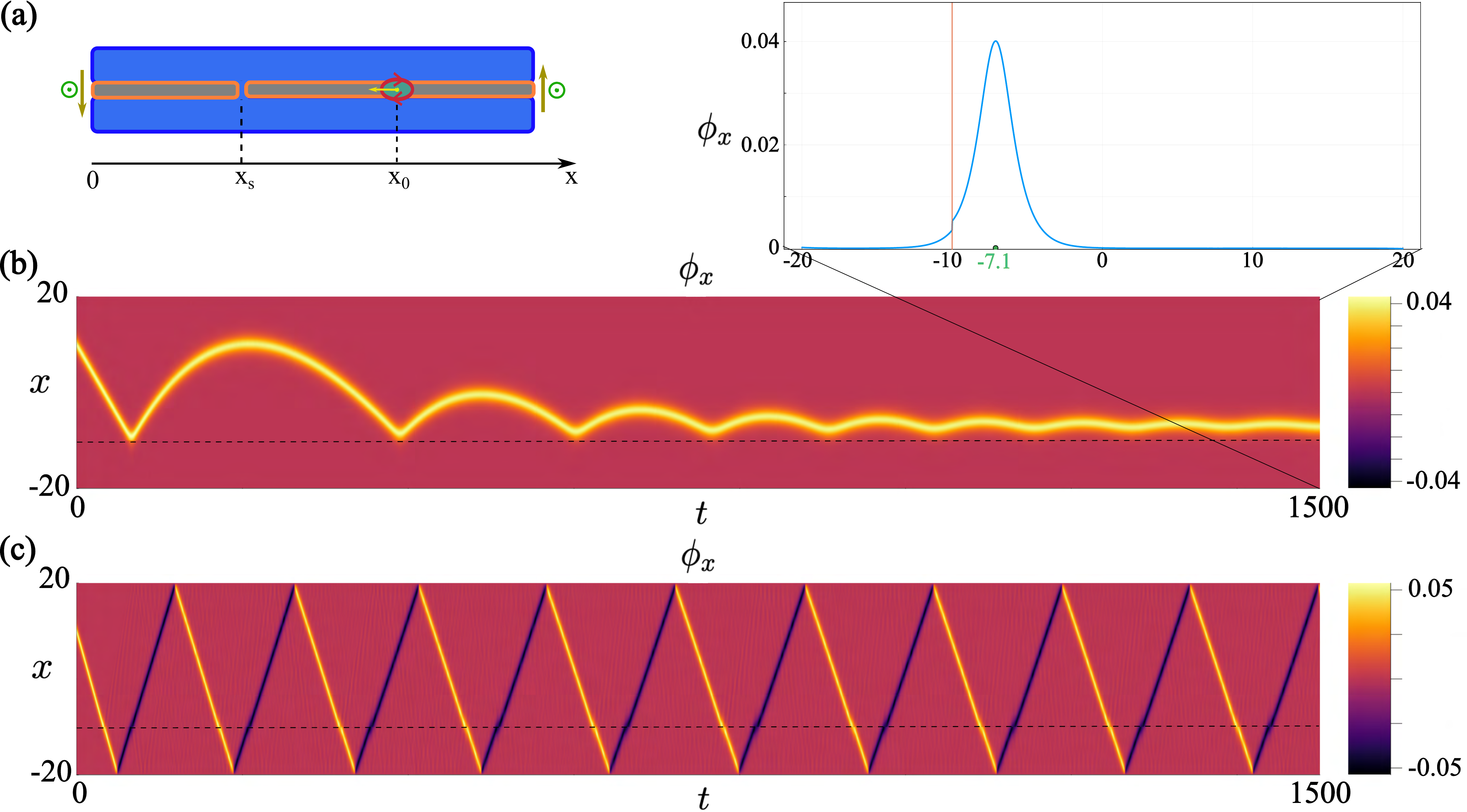}
    \caption{(a) Schematic of a fluxon initially located at $x_0$ and is moving towards a microshort at location $x_s$. (b) The dynamics of a fluxon repelled by the microshort, whose location is indicated by the horizontal dashed line. The initial speed of the fluxon is $u_i=0.3$. The inset shows the fluxon at the equilibrium point $x_e$ where the repulsion from the short is balanced by the Lorentz force. (c) Dynamics of a fluxon that has enough kinetic energy to pass through the short. The initial speed of the fluxon is $u_i=0.618$. }
    \label{fig:microshort}
\end{figure}

Mircoshorts are micro-scale (or sometimes nano-scale) short-circuits that can appear in long Josephson junctions. These short circuits happen when the insulating region becomes too narrow at some locations in the junction, causing current density to approach infinity there. They can either be unwanted defects that occur due to fabrication imperfections or intentional design decisions~\cite{price2010vortexqbit}. Therefore, precise modeling of vortex interactions near microshorts is important. The modified sine-Gordon equation that takes into account the presence of microshorts is given by

\begin{equation}
    \varphi_{tt} -\varphi_{xx} + \alpha\varphi_t + \sin\varphi + \sum_s\mu_s\delta(x-x_s)\sin\varphi = -\beta,
\end{equation}
where $x_s$ are the locations of the short circuits, and $\mu_s$ is the critical current at the shorts. The contribution of a short to the Hamiltonian is therefore given by
\begin{align}\label{eq:short_energy}
    H_s = \mu_s(1-\cos\varphi)\delta(x-x_s).
\end{align}
We therefore see that for $\varphi= 2\pi n$ with $n$ being an arbitrary integer, Eq.\,\ref{eq:short_energy} vanishes and there is no energy stored in the short. If a fluxon is in the vicinity of the microshort, $\varphi$ deviates from the bulk value and now $H_s>0$. The short draws energy directly from the vortex and causes it to slow down, or equivalently the vortex experiences a repulsion as it approaches the short. The dynamics of this process is shown in Fig.\,\ref{fig:microshort}. We consider a junction whose length is $L=40$, has a resistive loss rate of $\alpha=0.005$ and is subjected to a bias current $\beta$. A fluxon is initially prepared at terminal velocity $u_t$ (velocity at which friction and Lorentz force cancel each other) at $x_0=10$, sufficiently far away from and moving towards a microshort at $x_s=-10$. In Fig.\,\ref{fig:microshort}b, where we set $u_t=0.3$ (which corresponds to $\beta=0.002$), the fluxon is repelled by the short. After the first repulsion, the fluxon travels in the opposite direction until it is slowed down to a complete stop by both resistance and the Lorentz force. The Lorentz force, which always pulls the fluxon in the $-x$ direction, then accelerates the fluxon towards the short again. As seen in Fig.\,\ref{fig:microshort}b, the fluxon bounces off the mircoshort multiple times, losing energy in the process due to resistance, and eventually oscillates inwards into the pinning position, which is an equilibrium point $x_e$ where the Lorentz force and the repulsion from the short balance out each other (see inset of Fig.\,\ref{fig:microshort}b). Note that when the fluxon is repelled by the microshort, it does not turn into an antifluxon, unlike what happens when it collides with the boundary. 

If the initial kinetic energy of the vortex is sufficiently large, it can escape pinning~\cite{mclaughlin1978_SG}. This scenario is shown in Fig.\,\ref{fig:microshort}c,  in which the initial speed of the soliton is $u_t=0.618$ and corresponds to a bias current $\beta=0.005$. The fluxon simply passes through the microshort with negligible changes in kinetic energy. 

Microshorts are the extreme examples of when Josephson junctions have varying critical current across their length. More generally, the thickness of the insulating region separating the superconducting islands of a long junction may be nonuniform, which may affect the structure and the dynamics of fluxons as they travel along the junction. In Appendix \ref{append:constrictions} we investigate such a situation, where we consider a fluxon traveling in junctions that have one or a few constricted regions that force the fluxon to slow down by emitting radiation. In particular, we observe signatures of Cherenkov radiation~\cite{akhmediev1995cherenkov, goldobin1998cherenkov} when a fluxon arrives at a constriction with an incident velocity larger than the Swihart velocity in that constricted region.

\section{The bosonized Schwinger model}

We now turn to the analysis of the bosonized Schwinger model and its classical solutions. 


\subsection{Massless case ($\kappa=0$)}\label{ref:masslessSchwinger}
\begin{figure*}[t]
    \centering
    \includegraphics[scale=0.17]{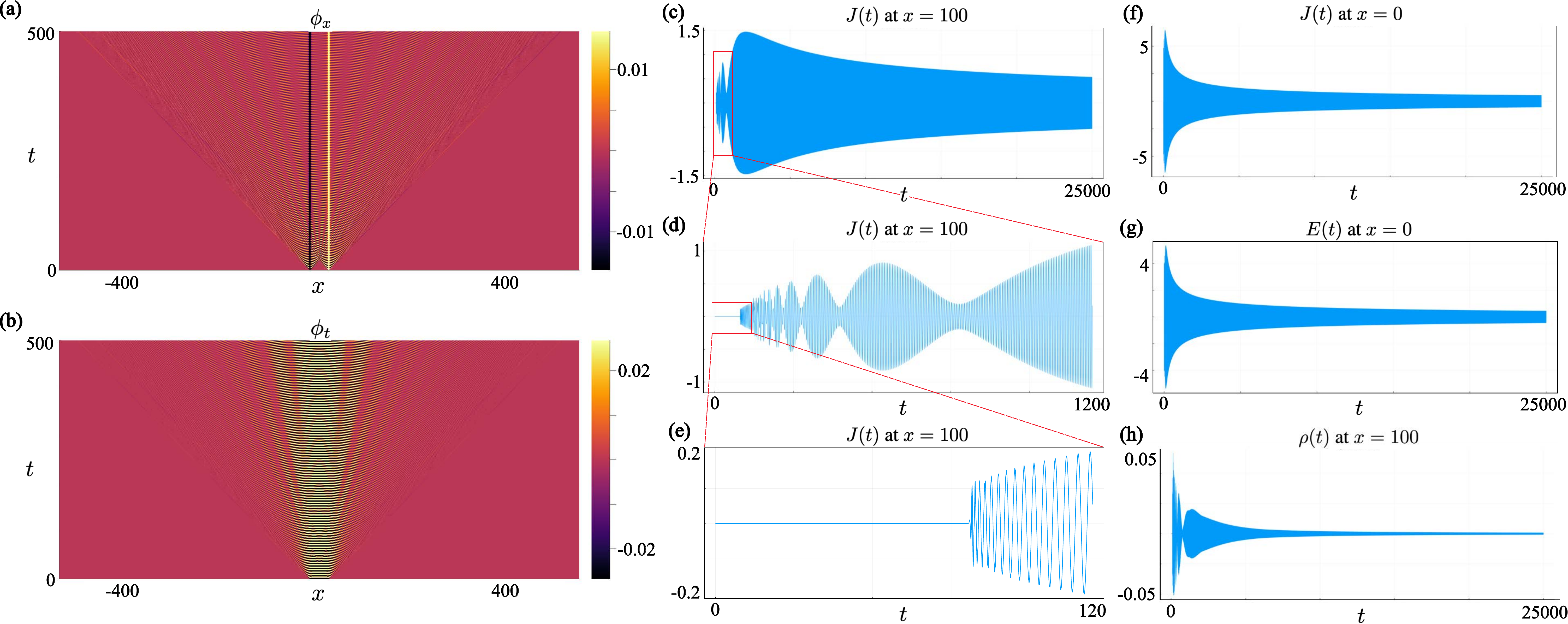}
    \caption{Backreaction dynamics of the massless fermionic fluid in response to a capacitor with fixed charges $\pm Q$ at $\mp L/2$, with $Q=4$, $L=40$, and $g=1.2$. The dynamics is captured up to $T_m=25000$. The $x-t$ heatmaps at early times $0\leq t\leq 500$ for $\phi_x(x,t)$ and $\phi_t(x,t)$ are shown in (a) and (b), respectively. (c) shows the evolution of the current $J(t)$ at $x=100$ over the entire simulated time $0\leq t \leq T_m$, while (d) shows the current over an intermediate time $(t<1200)$, and (e) zooms into the current at early times $(t<120)$. The current at the center of the capacitor is shown in (f), while the electric field $E(t)$ there is plotted in (g). The field $\phi_x$ at $x=100$ is plotted in (h).}
    \label{fig:massless_capdischarge}
\end{figure*}

Through the bosonization of the Dirac field, the Schwinger process for massless fermions is described by the Klein-Gordon equation~\cite{chu2010capacitordischarge} 
\begin{equation}\label{eq:kleingordon}
    \partial^2_t\varphi - \partial^2_x\varphi + g^2\varphi = -gF,
\end{equation}
which is simply Eq.\,\ref{eq:skg_full} with $\kappa$ = 0. Recall that $\varphi$ is the real-valued scalar field and $F$ is the classical background field. The term proportional to $g$ engenders a mass gap in the bosonic sector that reflects the fact that the low-energy excitations of the fermionic theory are mesons, bound particle-antiparticle pairs of mass $e/\sqrt{\pi}$. This term was the first manifestation of quark trapping in a low-dimensional gauge theory, requiring a non-perturbative approach~\cite{coleman1975}. In modern parlance, the low-energy excitations of the massless Schwinger model are hybridized longitudinal excitations of the gauge field and fermionic dipoles. 

 As discussed in Eqs.\,\ref{eq:emfield_eom}, the current and charge densities of the fermionic fluid relates to the scalar field by $\rho = -g\partial_x\varphi$ and $J=g \partial_t \varphi$. Therefore, the edge fields $\phi_x$ and $\phi_t$, as defined in Eqs.\,\ref{eq:phix} and \ref{eq:phit} take on physical meanings as the average charge and average current integrated over a finite resolution $\Delta x$ and $\Delta t$, respectively.  
Hence, a discrete version of
Eq.\,\ref{eq:kleingordon} written in terms of the fields $\phi_x$ and $\phi_t$ can be used to study the backreaction dynamics of the electron-positron fluid in the presence of the source.
Before exploring the dynamics of massive fields, we benchmark our numerical method using this well-studied linear model of the Schwinger process~\cite{chu2010capacitordischarge}.

The setup under consideration is the backreaction dynamics of massless fermions to a pair of fixed charges $\pm Q$ stationed at $\mp L/2$ that form a 1D capacitor~\footnote{This setup has originally been studied using an axillary $\theta$ field to implement a non-zero external electrical field~\cite{coleman1975, coleman1976}. We follow the source-field procedure used in Ref.~\cite{chu2010capacitordischarge} which is more general and flexible.}. This setup creates an external electric field
\begin{equation}\label{eq:externalE}
    F = Q\left[\Theta\left(x+\frac{L}{2}\right) - \Theta\left(x-\frac{L}{2}\right) \right]
\end{equation}
that sources the Klein-Gordon equation given in Eq.\,\ref{eq:kleingordon}, where $\Theta(x)$ represents a Heaviside step function. The initial conditions are chosen to be $\varphi(x,0)=\partial_t\varphi(x,0)=0$ so that at $t=0$ there are no charges. We employ radiative boundary conditions - the details of which are discussed in Appendix \ref{append:BCs} - which allow us to simulate long-time dynamics while simultaneously limiting the computational domain to a finite spatial region. It is also interesting to note that while we have no reference to topological BCs~\cite{chu2010capacitordischarge} the total charge within the system is still conserved at all times. This is due to the fact that the total initial charge is $Q=0$ and that DEC enforces exactly local charge conservation (to be discussed in details in Sec. \ref{sec:charge_conserve}).
The DEC-QED simulations are shown in Fig.\,\ref{fig:massless_capdischarge}. The distance between the fixed charges is $L=40$, while $Q=4$ and $g=1.2$. In Fig.\,\ref{fig:massless_capdischarge}a, where the heatmap for $\phi_x(x,t)$ is plotted, we see that pairs of fermions are immediately created near the capacitor plates that screen the fixed charges $\pm Q$. Each capacitor is a source from which matter waves originate. These waves then travel within the light cone and create particle-antiparticle pairs on their path. 

\begin{figure}
    \centering
    \includegraphics[scale=0.28]{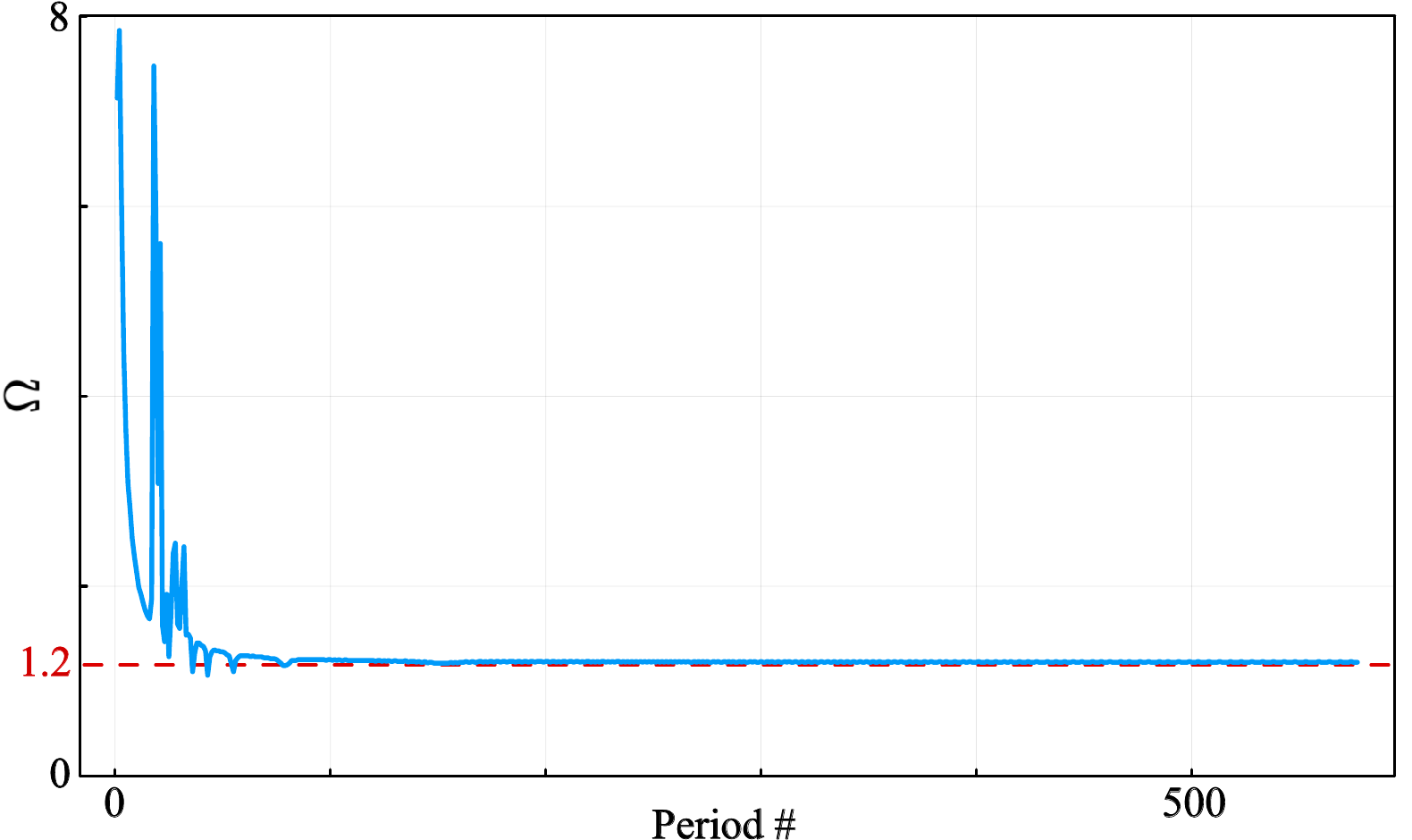}
    \caption{Oscillation frequency of the current $J$ at $x=100$ in the capacitor setup is plotted as a function of the period number. The parameters in the simulations are $\pm Q$ at $\mp L/2$, with $Q=4$, $L=40$, and $g=1.2$.}
    \label{fig:freq_vs_t}
\end{figure}

The ability of DEC-QED to capture evolution over very long times while simultaneously probing the detailed dynamics that happen within much shorter time scales is critical to to the phenomenology of this problem.
Figs.\,\ref{fig:massless_capdischarge}c, \ref{fig:massless_capdischarge}d, and \ref{fig:massless_capdischarge}e provide a snapshot of the current density at $x\!=\!100$ - a finite distance to the right of the capacitor - in long, intermediate, and early times, respectively. From Fig.\,\ref{fig:massless_capdischarge}e, we can see that it takes a finite amount of time for the waves to reach this location. The first wave to arrive is the one that originates from the source at $+L/2$. Note that when it first appears at $x\!=\!100$, the oscillation has very high frequency which slows down while the amplitude increases. After some time, the wave sourced by the other charge at $-L/2$ arrives and interferes with the first wave. As just mentioned,  because the frequencies of the waves also vary when they travel, the mismatch in frequencies of the two waves at $x\!=\!100$ results in a beating pattern seen at Fig.\,\ref{fig:massless_capdischarge}d. The widths of the beats stretch because both frequencies are decreasing over time. As seen in Fig.\,\ref{fig:massless_capdischarge}c, eventually at very long times both frequencies approach the asymptotic value and the beating pattern vanishes. The asymptotic carrier frequency of both waves, and hence of the overall oscillation frequency scale at every location, is precisely the effective mass $g$ (Fig.\,\ref{fig:freq_vs_t}). Note that the rate at which the frequency approaches this asymptotic value sets the intermediate time scale of the backreaction dynamics, i.e. the time scale over which the beating pattern occurs. 

The envelope switches from beats to decay over longer time scales. This is expected because at asymptotically long times, it is known from other considerations~\cite{coleman1975} that the fixed charges will be totally screened and the leakage of current out of the capacitor will have to cease. The charge density also decays, as the plot for $\phi_x(t)$ in Fig.\,\ref{fig:massless_capdischarge}c indicates. The envelope decays at an asymptotic long-time rate of $1/\sqrt{t}$~\cite{chu2010capacitordischarge}, though this behavior emerges only well after the oscillations have settled into their asymptotic frequency.

Within the capacitor region ($-L/2 < x < L/2$), electron-positron pairs are rapidly created to screen the electric field generated by the fixed charges. Since the fermions are massless in this model, pair production occurs almost instantaneously, leading to fast, underdamped oscillations in the polarization current. Over time, these charge oscillations dissipate as the system radiates energy and approaches equilibrium. Both the current (Fig.,\ref{fig:massless_capdischarge}f) and the electric field $E$ (Fig.,\ref{fig:massless_capdischarge}g) gradually decay as the generated pairs effectively screen the sources.

\begin{figure}
    \centering
    \includegraphics[scale=0.155]{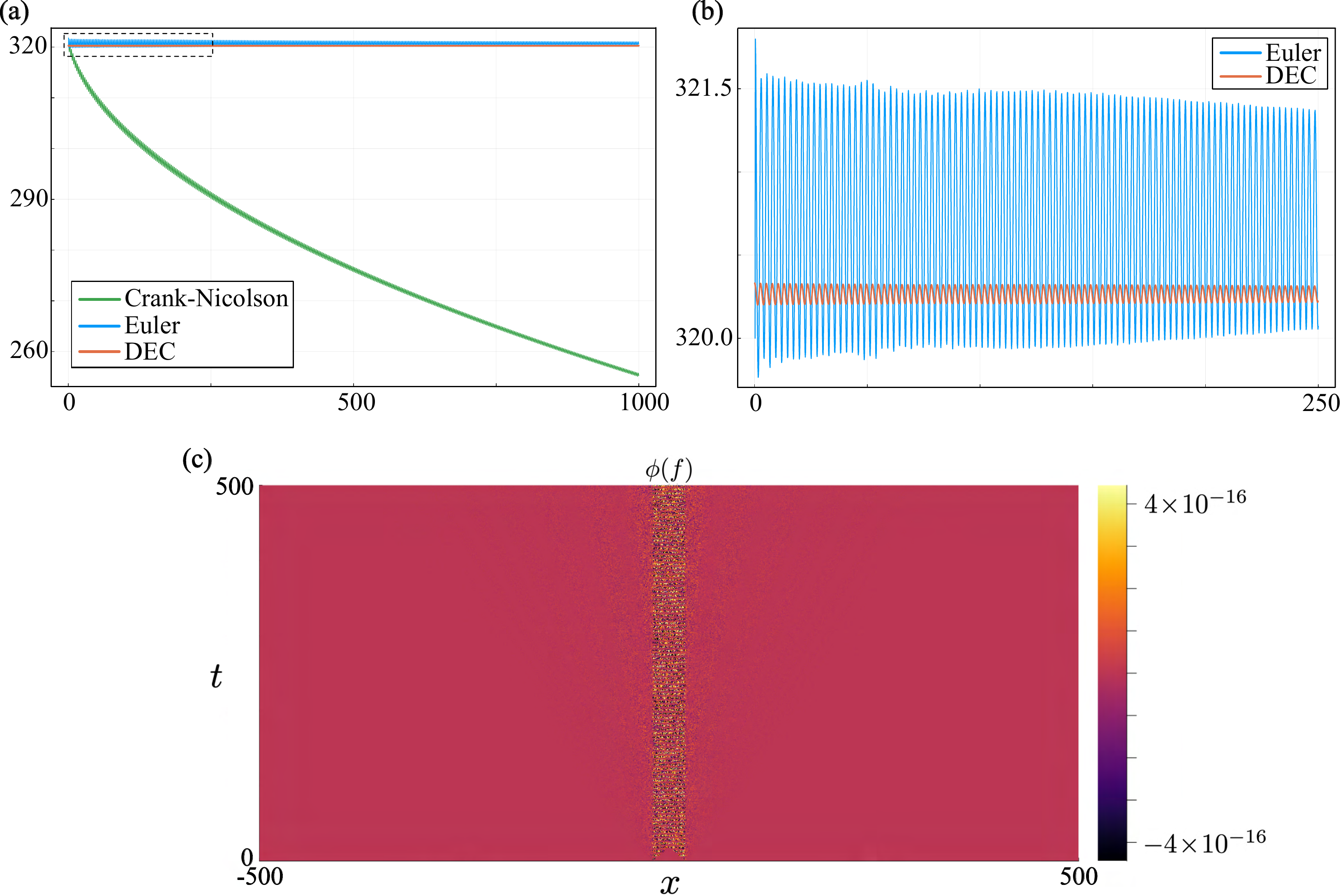}
Eneg    \caption{Energy and charge conservation in the backreaction dynamics of the massless fluid in response to a capacitor with fixed charges. (a) The energy computed from the simulation results obtained from DEC is compared to those calculated using the Euler method and the Crank-Nicolson method. For all three methods, $\Delta x\!=\! 0.333, \ \Delta t \!=\! 0.125, T_m\!=\!1000$. (b) A closed-up view of the energy computed using the Euler method and using DEC. (c) The heat map of $\varphi(f)$, evaluated at each primal face $f$ by summing up the edge fields $\phi_x$ and $\phi_t$ on the boundary of that face.}
    \label{fig:masslessCap_energies}
\end{figure}

\subsubsection{Conservation of energy}
We also computed the energies associated with the backreaction dynamics and used it as a metric to compare DEC-QED with other approaches for simulating time dynamics. Fig.\,\ref{fig:masslessCap_energies}(a) shows the energy components as a function of time obtained from simulations using the Crank-Nicolson method, the Euler method, and DEC. Due to the underlying property of the Crank-Nicolson (CN) method in the way it takes time-averages of spatial derivatives, there is no consistent way to compute the energies from this method. As a result, the total energy is not conserved in the Crank-Nicolson method and decays over time (see Fig.\,\ref{fig:masslessCap_energies}a). Instead of the traditional time propagation, this method also requires matrix inversions to obtain the field at all locations at once, making it less efficient and more memory intensive when applied to large problems.
One should note, however, that the CN method can be further optimized to achieve near $O(N)-O(N\text{log}N)$ scaling. For very long-time simulations, though, the runtime differences between CN, Euler’s method, and our approach eventually become significant. 
Euler's method and our spacetime-DEC approach also ensure that energy conservation is respected. Euler's method, however, introduces unphysical micro-oscillations in the total energy, as seen in Fig.\,\ref{fig:masslessCap_energies}b. These oscillations originate from the evaluation of the spatial and temporal derivatives using the discrete values of $\varphi$ obtained from solving the second-order Eq.\,\ref{eq:kleingordon}. The errors are then carried over to the calculation of the Hamiltonian
\begin{equation}
    H(t) = \int_{-L/2}^{L/2}dx\frac{1}{2}\left[\left(\partial_x\varphi\right)^2 + \left(\partial_t\varphi\right)^2 + \left(g\varphi + F\right)^2\right],
\end{equation}
where the terms in the bracket are the quadratic potential, kinetic energy, and the energy stored in the electric field, respectively. In DEC-QED, the derivatives are encoded on the edges of the spacetime grid and are directly solved for from the equation of motion. This allows for the suppression of unphysical oscillations in calculating the discretized Hamiltonian, which is given by
\begin{align}
    H_{dec}&(j) = \!\!\sum_{i=1}^{N_x-1}\frac{\phi_x^2[(i,j),(i+1,j)]}{2\Delta x}\nonumber\\
    +& \sum_{i=1}^{N_x}\frac{\Delta x}{4\Delta t^2}\left\{\phi_t[(i,j\!-\!1),(i,j)]+\phi_t[(i,j),(i,j\!+\!1)]\right\} \nonumber\\
    +& \sum_{i=1}^{N_x}\frac{\Delta x}{2}\left[g\varphi(i,j) + F(i,j)\right]^2,
\end{align}
 where $N_x$ is the number of grid points in the spatial dimension. As seen in Fig.\,\ref{fig:masslessCap_energies}b, the total energy computed using DEC is more stable with noticeably smaller oscillations than those obtained from the Euler method.

\subsubsection{Local conservation of charge}\label{sec:charge_conserve}
An outstanding feature of DEC is that it enforces exactly local current-charge conservation on the discrete grid. In the bosonic picture, local charge conservation in the Schwinger process is given by 
\begin{equation}\label{eq:chargeconserve}
\frac{\partial\rho}{\partial t} + \frac{\partial J}{\partial x}=0.
\end{equation}
In the spacetime plane, the left-hand side of Eq.\,\ref{eq:chargeconserve} is equivalent to $\nabla\!\times\!\nabla\varphi$.
The discrete version of local charge conservation rule is obtained by integrating Eq.\,\ref{eq:chargeconserve} over a primal face $f$ with dimensions $\Delta x$-by-$\Delta t$ and applying Stokes' theorem
\begin{subequations}
\begin{align}\label{eq:dec_chargeconserve_a}
    \int_f\! {\bf da}\!\cdot\! (\nabla\!\times\!\nabla\varphi) &=  \int_{\partial f}\!{\bf d\ell}\!\cdot\!\nabla\varphi \\
    &= \sum_{e_t\in \partial f}\!\phi_t(e_t) +\! \sum_{e_x\in \partial f}\!\phi_x(e_x) \label{eq:dec_chargeconserve_b} \\
    &= \sum_{e[v_i,v_j]\in\partial f}\!\!\varphi(v_j)-\varphi(v_i) \label{eq:dec_chargeconserve_c}\\
    &= 0.
\end{align}
\end{subequations}
In Eq.\,\ref{eq:dec_chargeconserve_b} above, the sums are over the four boundary edges of $f$.
From Eq.\,\ref{eq:dec_chargeconserve_b} to Eq.\,\ref{eq:dec_chargeconserve_c} we have used the definitions given in Eq.\,\ref{eq:phix} and Eq.\,\ref{eq:phit} for the edge fields.  The sum in Eq.\,\ref{eq:dec_chargeconserve_c} is zero because each vertex of $f$ appears twice in the sum but with opposite signs. We have therefore proven that the structure-preserving nature of DEC allows for local charge conservation to be transferred exactly from the continuous EoM to the discrete one. This exactness is demonstrated in Fig.\,\ref{fig:masslessCap_energies}c, where we show that the sum of $\phi_t$ and $\phi_x$ for every face $f$ (i.e. Eq.\,\ref{eq:dec_chargeconserve_b}) in the grid is zero up to machine precision.

\begin{figure}[t]
    \centering
    \includegraphics[scale=0.11]{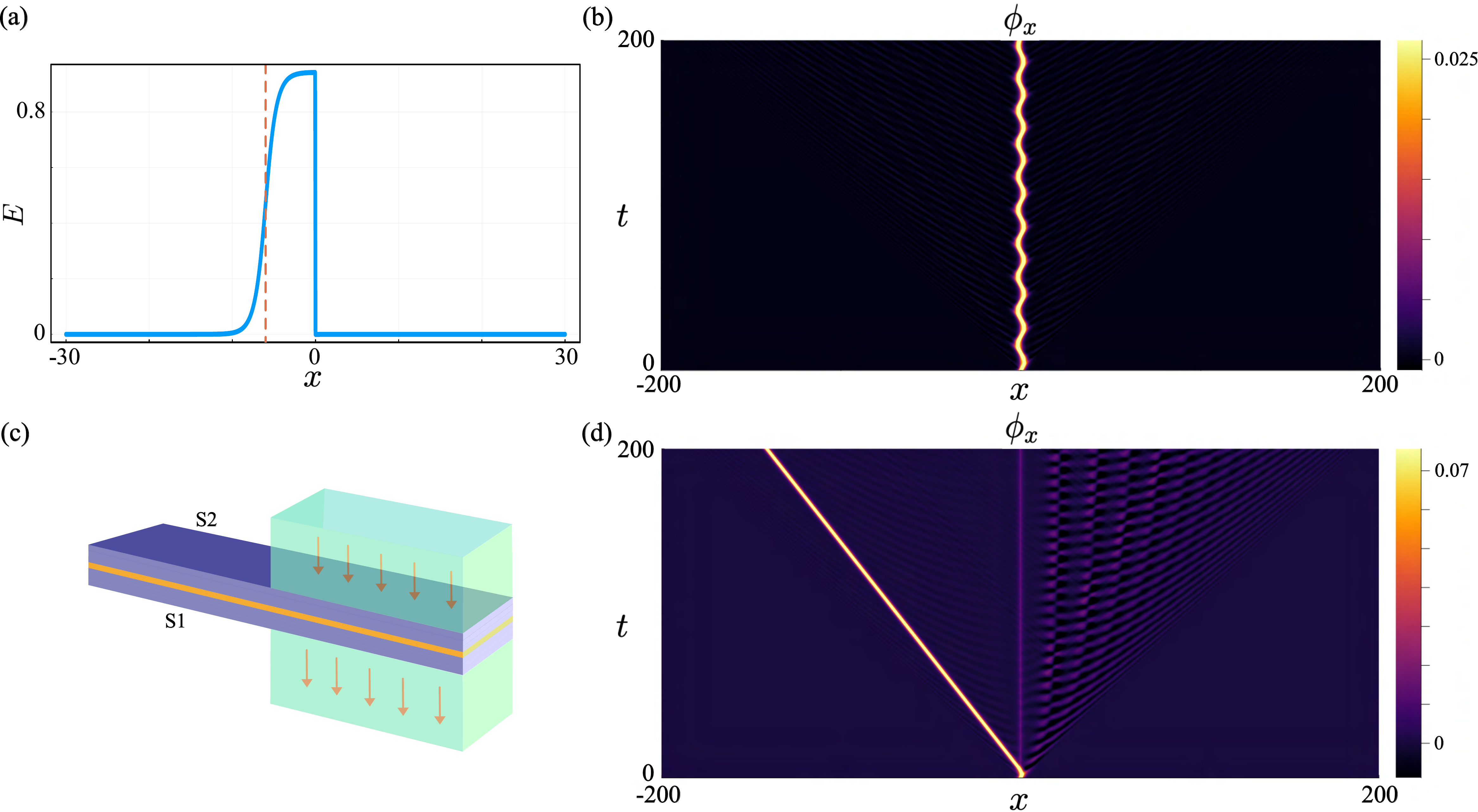}
    \caption{(a) The total electric field $E(x)$ created by a classical charge at $x=0$ and an oppositely charged soliton at a finite distance away from the classical source. The center of the soliton is indicated by the vertical orange dashed line. (b) The field $\phi_x$ that corresponds to the massive Schwinger dynamics of a soliton oscillating around the classical source. (c) Schematic of a long Josephson junction with half of its length (i.e. the region where $x\geq 0$) being biased by an external current. (d) The field $\phi_x$ that corresponds to the dynamics of a fluxon in a Josephson junction with a biased current influencing half the junction, where $x\geq0$. }
    \label{fig:schwinger_atom_x0_0_no_g^2}
\end{figure}

\begin{figure*}[t]
    \centering
    \includegraphics[scale=0.2]{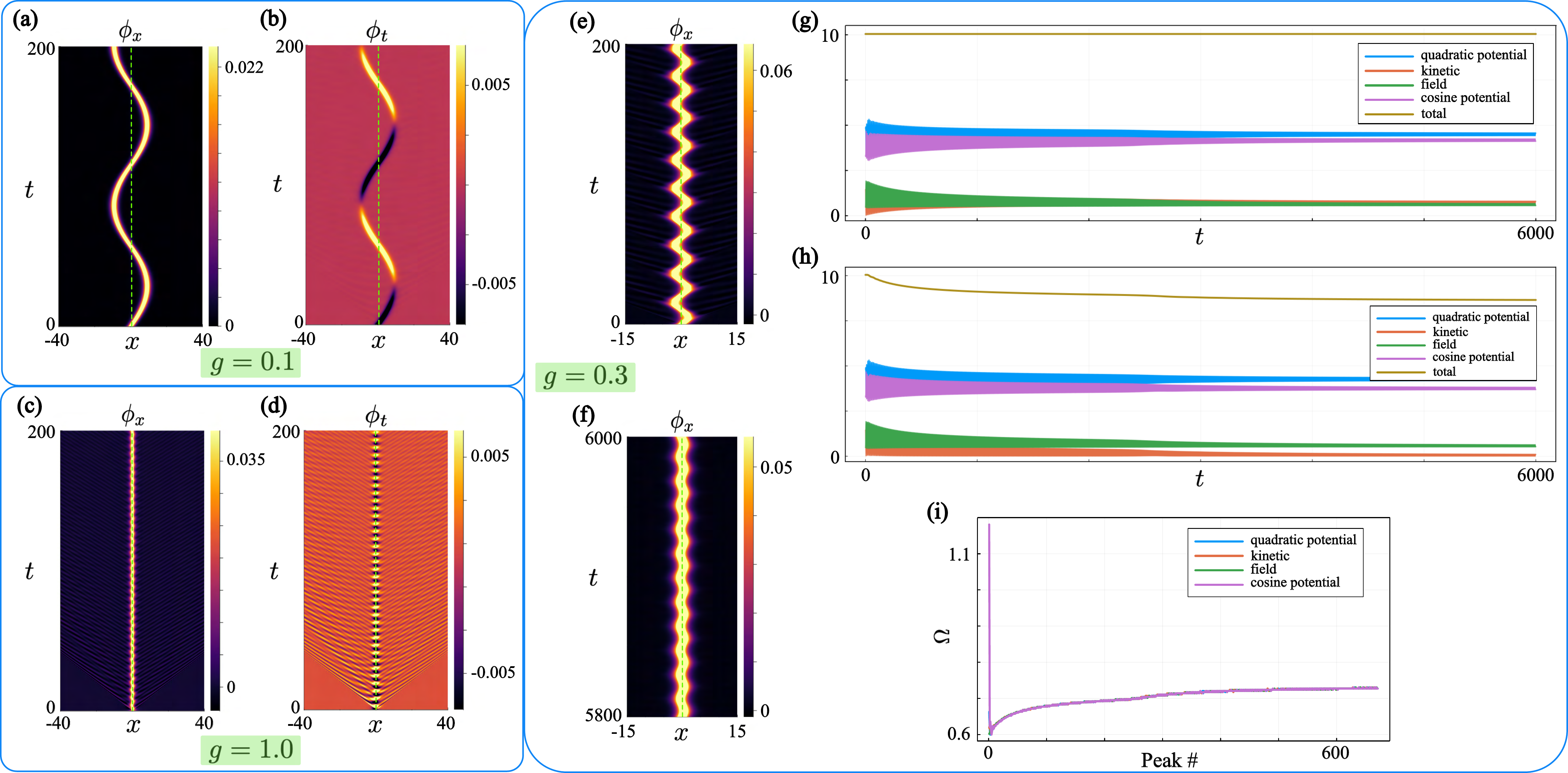}
    \caption{Interaction between a fixed classical charge and an oppositely charged soliton in the context of the massive Schwinger process. The fixed charge is located at $x=0$. The initial condition for the scalar field $\varphi$ is the SG fluxon solution that is also initially placed at the origin with an initial velocity $u=0.55$. (a) and (b) show the resulting fields $\phi_x$ and $\phi_t$ for $g=0.1$ at early times ($0<t<200$). The vertical green dashed line in each plot indicates the fixed location of the classical source. (c) and (d) also show the fields $\phi_x$ and $\phi_t$ at early times, but for $g=1.0$. (e) plots the field $\phi_x$ for $g=0.3$ at early times, while (f) plots the same field but at late times $5800\leq t\leq6000$. (g) plots each of the terms in the Hamiltonian of the system and the total energy as functions of time. (h) plots the dependence on time of the energies within a finite region $-16\leq x \leq 16$. (i) shows the frequencies in the oscillations of each term in the Hamiltonian as a function of the period number.}
    \label{fig:schwinger_atom_x0_0}
\end{figure*}

\subsection{Massive case}\label{sec:massiveSchwinger}
We now turn to the Schwinger process involving massive fermions, described by Eq.~\ref{eq:skg_full}. To simplify the algebra and to draw connections to the standard form of the Sine-Gordon equation, we rescale the variables as follows: \mbox{$\varphi\rightarrow2\sqrt{\pi}\varphi$}, \mbox{$t\rightarrow \sqrt{2\pi\kappa}t$}, \mbox{$x\rightarrow \sqrt{2\pi\kappa}x$}, and \mbox{$g\rightarrow g/\sqrt{2\pi\kappa}$}.
The nomalized equation of motion for the scalar field in the massive case is given by
\begin{equation}\label{eq:normalized_skg}
    \partial_t^2\varphi - \partial_x^2\varphi + g^2\varphi + \sin\varphi= -gF.
\end{equation}
Eq.\,\ref{eq:normalized_skg} is a sine-Gordon equation with a source term on the right-hand side and an additional term $g^2\varphi$ on the left. As was seen in the massless dynamics, the term $g^2\varphi$ is crucial to the creation of mesons and to mediating the interaction between classical charges and/or charged solitons. This is because the total electric field generated by the mesons is proportional to $g\varphi$. We will be especially interested in the effect of this additional term on the SG equation in the following simulations on the 1D atom and positronium. Also, one should keep in mind that after the rescaling of variables to obtain Eq.\,\ref{eq:normalized_skg}, $g$ is now also rescaled to be $g/\sqrt{2\pi\kappa}$. From here onward, unless specified, we will use this new definition of $g$ for massive Schwinger calculations. 

\subsubsection{Schwinger atom} \label{sec: schwinger atom}
We consider a setup composed of a single classical charge $Q$ fixed at the origin $x\!=\!0$. This source has to be balanced by an oppositely charged soliton in the scalar field so that the total electric energy is finite. The choice of initial conditions is therefore a delicate matter and we choose it to be the known analytical form for the kink solution of the unperturbed SG model (Eq.\,\ref{eq:SG_singlefluxon}) that has a total charge of $2\pi g$. For the system to be charge-neutral, we choose $Q=-2\pi g$. At the boundaries of the finite computational domain, radiative BCs are applied to allow the mesons that are created during transient dynamics to leave the system without being artificially reflected back into the domain and affecting the true dynamics.

We first study the case where the initial position of the SG kink is $x_0\!=\!0$, exactly on top of the classical source, and has a starting velocity $u\!>\!0$. 
Due to the finite electric field generated in the region between the classical source and the soliton (Fig.\,\ref{fig:schwinger_atom_x0_0_no_g^2}a), the soliton is constantly pulled towards the source. As a result, a soliton with an initial velocity will oscillate around the source (Fig.\,\ref{fig:schwinger_atom_x0_0_no_g^2}b) and together they form a 1D ``atom" which we refer to as the {\it Schwinger atom}. If we turn off this mediation of attractive interaction by removing the term $g^2\varphi$ from Eq.\,\ref{eq:normalized_skg}, we return to the pure SG equation on the left-hand side with a step function source $-gQ\Theta(x)$ on the right-hand side. The resulting dynamics of a soliton governed by this modified equation is shown in Fig.\,\ref{fig:schwinger_atom_x0_0_no_g^2}d, where instead of oscillating around the source, the charged soliton escapes towards $x\!\rightarrow\!-\infty$ at a constant velocity. It is helpful here to establish an analogy to fluxon dynamics in a JTL, which for the conditions stated here, has  half of its length subject to a bias current (Fig.\,\ref{fig:schwinger_atom_x0_0_no_g^2}c). In that scenario, the dynamics in Fig.\,\ref{fig:schwinger_atom_x0_0_no_g^2}d is equivalent to that of a vortex initially at $x=0$ and moving into the biased region. In this region, the vortex experiences the Lorentz force from the bias current and is pulled towards negative $x$, causing it to slow down and switch its direction. Eventually the vortex exits the biased region and travels towards $x\!\rightarrow\!-\infty$.

\begin{figure*}[t]
    \centering
    \includegraphics[scale=0.135]{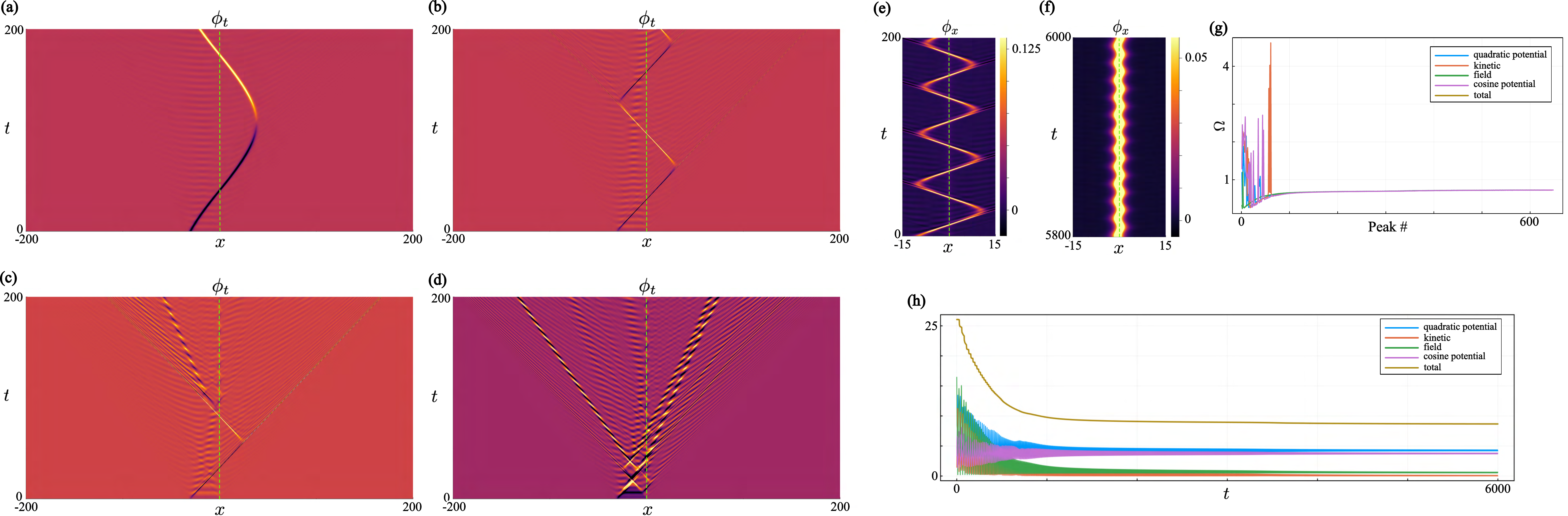}
    \caption{Interaction between a fixed classical charge and an oppositely charged soliton. The classical charge is placed at $x=0$, while the soliton is initially at a distance away from the source and is moving towards it. (a), (b), (c), and (d) show the dynamics of the charge current $\phi_t(x,t)$ at early times $0\leq t \leq 200$ for $g=0.1, \ 0.3,\ 0.35,$ and $0.4$, respectively. In these figures the initial location of the soliton is $x_0=-30$, and vertical green dashed line indicates the fixed location of the classical source. (e) plots the early-time dynamics of $\phi_x$ for $g=0.3$ and $x_0=-10$, and (f) plots the late-time dynamics of the same field. (g) shows the frequencies in the oscillations of each term in the Hamiltonian as a function of the period number. (h) plots the dependence on time of the energies within a finite region $-16\leq x \leq 16$.}
    \label{fig:schwinger_atom_solitonaway}
\end{figure*}

Both the initial spatial range and frequency of the soliton oscillation depend on $g$, as can be seen in Figs.\,\ref{fig:schwinger_atom_x0_0}a-d. The larger $g$ is, the faster the oscillation is and the shorter the round trip that the soliton makes in each period. This oscillation damps out at long times. This is shown in Fig.\,\ref{fig:schwinger_atom_x0_0}e-f where we compare the oscillations at early times $0\leq t\leq 200$ and after a relatively long time, when $5800 \leq t \leq 6000$. We also see that amplitudes in the oscillations of the energy terms decrease accordingly (Fig.\,\ref{fig:schwinger_atom_x0_0}g). Note that although each term in the Hamiltonian oscillates as the soliton winds around the source, the sum of these terms (the total energy) is constant -- DEC-QED is able to preserve this conservation law throughout long-time simulations. In Fig.\,\ref{fig:schwinger_atom_x0_0} we also plot the energies within a finite domain $(-16\leq x \leq 16)$ around the atom and observe a decay in the total energy stored within this domain. This is because as the soliton oscillates around the source, it also releases energy in the form of radiation. The radiation leakage from the atom slows down over time and ends as \mbox{$t\rightarrow\infty$} when the soliton comes to a full stop on top of the source. Note that this state of affairs can only be captured by the correct implementation of radiation BCs at the boundary of the computational domain. Finally, the evolution in the frequencies of the oscillations in each energy term is shown in Fig.\,\ref{fig:schwinger_atom_x0_0}i. At any time, the energy terms oscillate at the same frequency and this frequency evolves over time. In this setup, where the soliton is initially placed right where the source is, the frequency at asymptotic time depends exclusively on $g$ in a nonlinear manner.

We next consider a situation where the soliton is initially at a distance away from the source and is moving towards it. We again choose the initial condition to be a SG soliton with initial position $x_0$ and velocity $u=0.55$. In Figs.\,\ref{fig:schwinger_atom_solitonaway}a-d the results for the current $\phi_t(x,t)$ at early times $0\leq t \leq 200$ are shown for $g=0.1, \ 0.3,\ 0.35,$ and $0.4$, respectively. In these figures, the soliton is initially located at $x_0 = -30$. As the coupling strength $g$ increases, the shape and trajectory of the SG soliton become increasingly distorted and it emits radiation more rapidly. The soliton accelerates as it approaches the external charge, as reflected in the changing slope of its trajectory — most clearly seen in Fig.\,\ref{fig:schwinger_atom_solitonaway}a. However, its speed remains bounded by the speed of light, and in Fig.\,\ref{fig:schwinger_atom_solitonaway}b, we observe the soliton approaching this relativistic limit. In Figs.\,\ref{fig:schwinger_atom_solitonaway}b and \ref{fig:schwinger_atom_solitonaway}c, we also see that as the soliton reaches its maximum distance from the source and begins to reverse direction due to the attractive interaction, part of its charge separates and continues to propagate away. Throughout its motion, the soliton gradually loses its integrity, continuously shedding radiation in the form of mesons that travel outward from the source-soliton system.

The initial position of the soliton with respect to the source also plays a role, because the distance between them sets the initial energy stored in the electric field. In Fig.\,\ref{fig:schwinger_atom_solitonaway}e, the early dynamics are plotted for $g=0.3$ and $x_0=-10$. This is the same value for $g$ as in Fig.\,\ref{fig:schwinger_atom_solitonaway}b, but with a different initial location of the soliton. The closer the soliton starts to the source, the shorter its round-trip trajectory around the atom, and the less radiation is emitted to dissipate the energy initially stored. This process of meson emission and dispersion continues until the soliton has released all its excess energy and settles into the configuration needed to effectively screen the source. Fig\,\ref{fig:schwinger_atom_solitonaway}f shows the long time dynamics of the same setup as in Fig\,\ref{fig:schwinger_atom_solitonaway}e (where the early dynamics is shown). The oscillations of the soliton are now very localized around the source. The long-time behavior of the soliton-source system here is similar to what was observed earlier in Fig.\,\ref{fig:schwinger_atom_x0_0}f with $x_0=0$. This indicates that, for $g$ in a certain range, regardless of the initial condition on the SG soliton it will always eventually be trapped by the fixed source to form an atom. The evolution of the frequencies of the energy terms is plotted in Fig.\ref{fig:schwinger_atom_solitonaway}g and, similar to the previous setup, they approach the same asymptotic frequency value. After initially losing energy to mesons that travel to infinity, the energy of this atom is also stabilized at long times (see Fig.\,\ref{fig:schwinger_atom_solitonaway}h). 

\begin{figure}[t]
    \centering
    \includegraphics[scale=0.105]{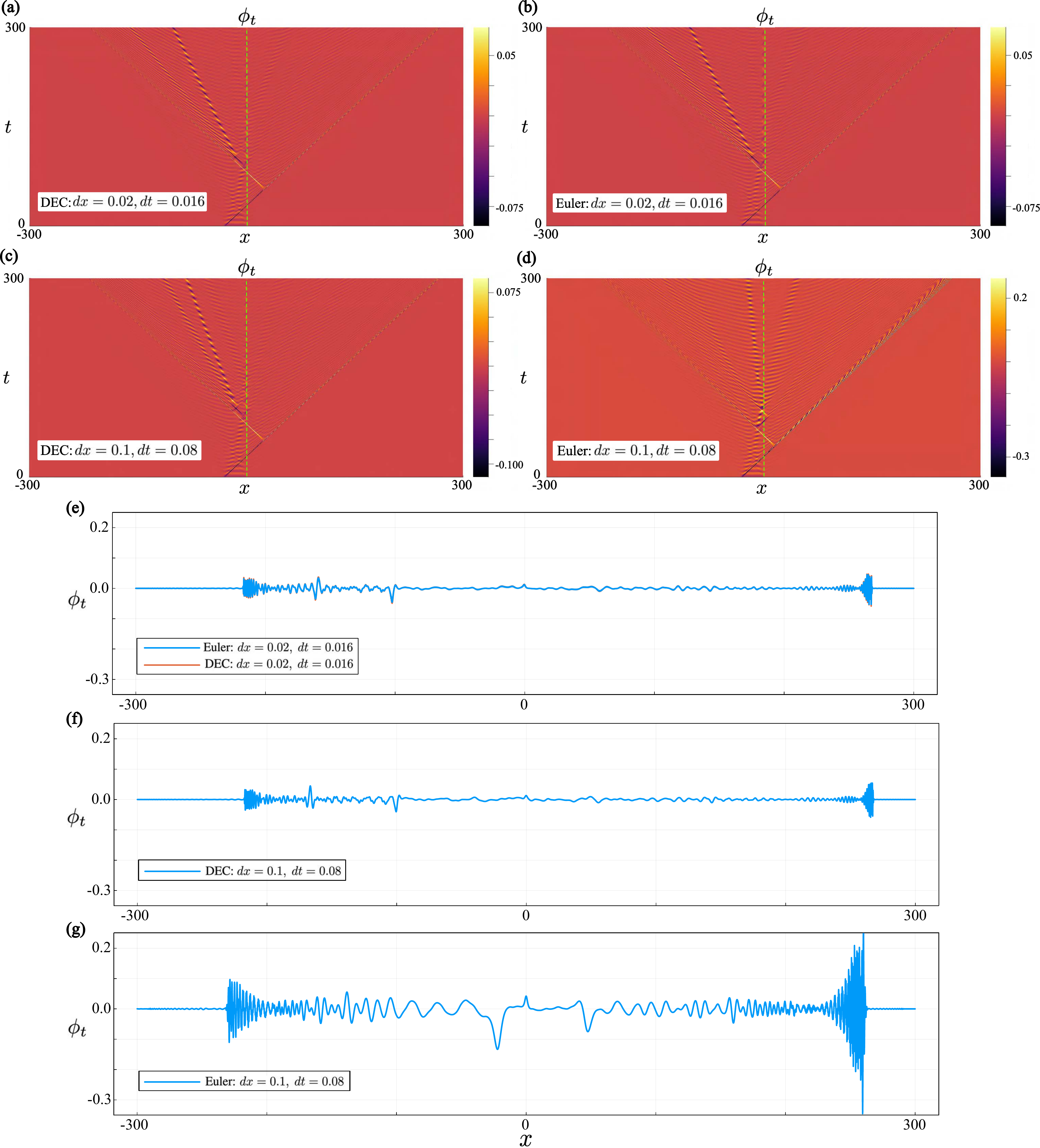}
    \caption{Comparisons between DEC-QED and the Euler method in simulating the out-of-equilibrium dynamics of the massive Schwinger process when a soliton interacts with a fixed classical source. The classical charge is placed at $x\!=\!0$, while the soliton is initially at $x_0\!=\!-30$, away from the source and is moving towards it with velocity $u\!=\!0.55$. The 1+1D distribution of $\phi_t$ for $t\leq T_m$ is computed with  $dx\!=\!0.02,\ dt\!=\!0.016$ using DEC and Euler's method are shown in (a) and (b) respectively. The same field computed with $dx\!=\!0.1,\ dt\!=\!0.08$ using the two methods are shown in (c) and (d), respectively. $\phi_t(x)$ at the time slice $t=T_m$ obtained from both methods with $dx\!=\!0.02,\ dt\!=\!0.016$ are plotted in (e), while the results computed with $dx\!=\!0.1,\ dt\!=\!0.08$ are shown in (f) for DEC and in (g) for Euler.}
    \label{fig:schwinger_atom_dec_vs_euler}
\end{figure}

Figs.~\ref{fig:schwinger_atom_solitonaway}(c-d) show that with a sufficiently large effective mass $g$ and large initial energy, governed by the soliton’s initial displacement and velocity, the soliton undergoes rapid destabilization, fragmenting into radiative modes. Capturing such far-from-equilibrium dynamics poses numerical challenges, as errors are more prone to accumulate for dynamics far away from localized solitary wave solutions. We take this opportunity to benchmark the DEC-QED integrator.

We show the evolution of $\phi_t$ over $0\!\leq\!t\leq\!T_m\!=\!300$ computed with discretization steps $dx\!=\!0.02,\ dt\!=\!0.016$ using DEC-QED (Fig.\,\ref{fig:schwinger_atom_dec_vs_euler}a) and Euler (Figs.\,\ref{fig:schwinger_atom_dec_vs_euler}b). We see that with such a finely discretized grid, the two methods produce practically the same dynamics within the simulated time $t\!\leq\! T_m$. This is also evident in Fig.\ref{fig:schwinger_atom_dec_vs_euler}e, where the details of $\phi_t$ at the last time slice $t\!=\!T_m$ obtained using the two methods are compared in the same plot. However, when the calculations are done with coarser discretization steps, we see that while DEC remains fairly accurate (Figs.\ref{fig:schwinger_atom_dec_vs_euler}c, \ref{fig:schwinger_atom_dec_vs_euler}f), Euler's method quickly diverges from the correct dynamics (Figs.\ref{fig:schwinger_atom_dec_vs_euler}d, \ref{fig:schwinger_atom_dec_vs_euler}g).

\subsubsection{Schwinger positronium}\label{sec:positronium}

\begin{figure*}[t]
    \centering
    \includegraphics[scale=0.18]{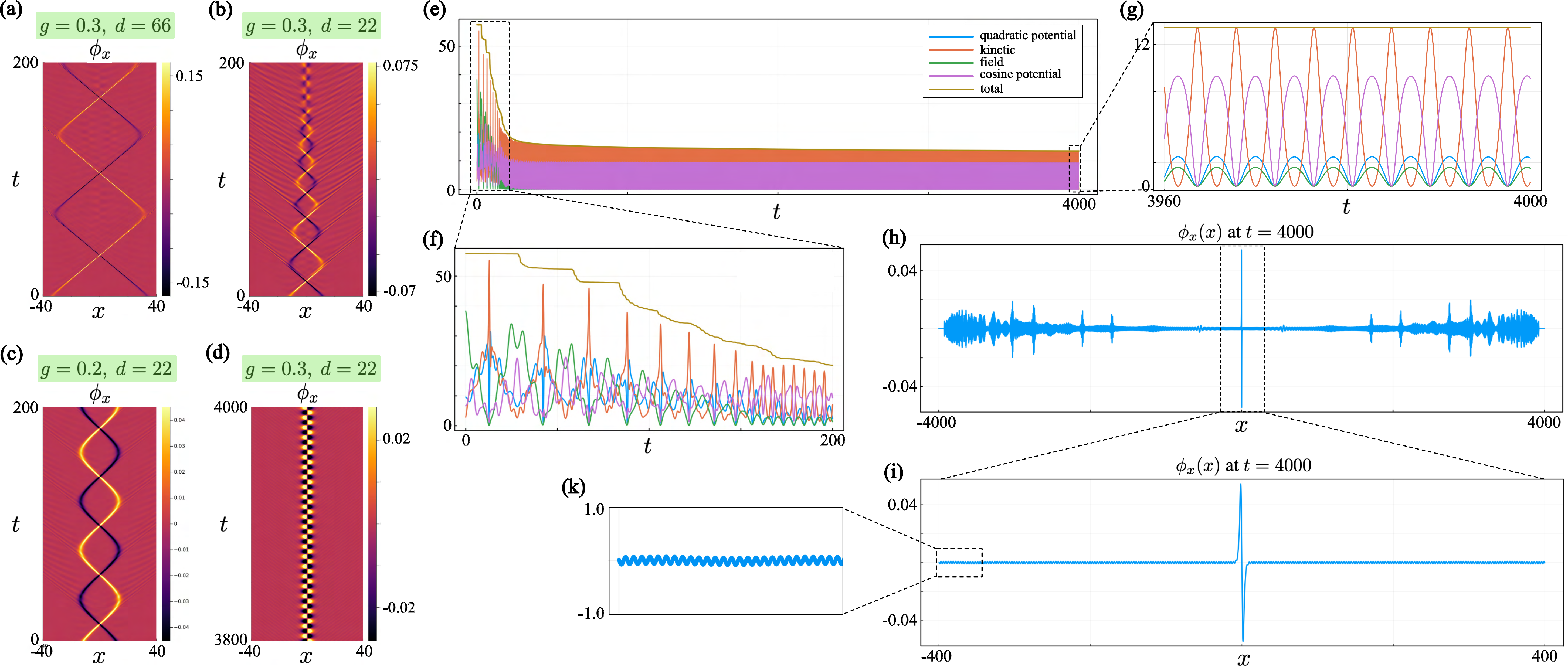}
    \caption{Dynamics of a Schwinger positronium composed of a stable pair of soliton and antisoliton that oscillate around each other. The pair is initially at a distance $d$ from each other and are moving towards each other, each with speed $u=0.55$. (a) plots the field $\phi_x(x,t)$ at early times ($0\leq t \leq 200$) for $g=0.3,\ d=66$. (b) shows $\phi_x(x,t)$ at early times for $g=0.3,\ d=22$. (c) shows $\phi_x(x,t)$ at early times for $g=0.2,\ d=22$. (d) plots the value of $\phi_x(x,t)$ at late times ($3800\leq t \leq 4000$) for $g=0.3,\ d=22$. (e) presents evolution over time of the energies stored inside a finite domain $-16\leq x\leq 16$ with $g=0.3,\ d = 22$. (f) zooms into the early evolution of the energies shown in (e), while (g) zooms into the asymptotic times. (h) plots the field $\phi_x(x)$ everywhere in space at $t=4000$. (i) also plots $\phi_x(x)$ at $t=4000$ with a closed-up view into the range $-400\leq x \leq 400$ around the positronium. (k) shows the closed-up view of the radiation that leaks from the positronium at $t=4000$.}
    \label{fig:schwinger_positronium1}
\end{figure*}

An additional hypothesis of interest is the potential existence of a stable Schwinger ``positronium," which we investigate next. 

In the framework of 1+1D field theory, positronium can be understood as a stable bound state of a soliton–antisoliton pair repeatedly passing through each other as they oscillate around a point $x$. In the physical 3+1D QED, positronium is not a bound state but rather a two-body resonance with a well-studied finite lifetime~\cite{karshenboim2004precision, badertscher2007}.

The evolution from two separated solitons to a bound pair is strictly prohibited by the unperturbed SG model. This means that when two solitons in the pure SG model collide, they simply pass through each other. The solitary wave solutions in that theory are solitons according to the stricter definition of the term~\footnote{In the discussion so far we have not made a distinction between a soliton solution and a solitary wave solution. The former is a narrower class of solutions than the latter, that require stricter asymptotic conditions in time. For more information see Ref.~\cite{ablowitz1981solitons}}. The inclusion of the dynamical mass term $g^2\varphi$, however, breaks the integrability of the SG model and may facilitate the formation of a bound positronium from two initially separated solitons. 

In Figs.\,\ref{fig:schwinger_positronium1}a-c we present the early-time space-time dynamics of the charge density $\phi_x(x,t)$ that corresponds to a soliton-antisoliton pair for different values for $g$ and the initial separation $d$. The initial condition for $\varphi$ in all three cases have been chosen as the analytically known fluxon-antifluxon solution in Eq.\ref{eq:vav_pair}. Comparing Figs.\,\ref{fig:schwinger_positronium1}a and \ref{fig:schwinger_positronium1}b, one can see that smaller $d$ leads to faster decay in the oscillation of the pair. The reason is clear -- a smaller initial separation reduces the initial energy stored in the electric field, which is confined to the space between the two solitons, thereby limiting the energy available to sustain the oscillations that emit radiation. Comparing Figs.\,\ref{fig:schwinger_positronium1}b with Figs.\,\ref{fig:schwinger_positronium1}c, one sees that smaller $g$ also results in slower decay. This occurs because the soliton accelerates when propagating in a large-$g$ background, as we also observed in the case of the atom with a fixed charge. The increased acceleration leads to enhanced radiation emission from the soliton, causing the positronium to lose energy more rapidly.

In all of the three cases presented in Figs.\,\ref{fig:schwinger_positronium1}a-c, although the oscillations decay over time, the pair is not completely annihilated but is reduced to a steady state of persistent oscillation. We show an example in Fig.\,\ref{fig:schwinger_positronium1}d, where the late-time dynamics for $g=0.3, \ d=22$ (the same parameters as Figs.\,\ref{fig:schwinger_positronium1}b) is plotted. The oscillations of the pair are stabilized as the decay rate becomes progressively small. The binding of an initially separated SG soliton-antisoliton pair into a breather may remind us of the vortex-antivortex interaction in a lossy junction that was discussed in Sec\,\ref{sec:vav_annhiliation}. However, unlike in the lossy junction case, where the breather decays until it vanishes entirely due to the effect of the resistive term $\partial_t\varphi$, here the solitons only decay until the system reaches a steady state of self-sustained oscillation. In this massive Schwinger process, the breather seen in Fig.\,\ref{fig:schwinger_positronium1}d is an oscillating electric dipole, and the internal oscillation of the breather at asymptotic time is facilitated by the attractive force between its two oppositely charged halves. 

The evolution of energy within the finite domain $-16 \leq x \leq 16$, shown in Fig.~\ref{fig:schwinger_positronium1}e, offers further insight into these observations. It can be seen that, following an initial rapid decay, the total energy within this region stabilizes at a steady-state value at late times, indicating that energy leakage from the atom eventually ceases. As shown in Fig.~\ref{fig:schwinger_positronium1}g, which provides a close-up view of the energy dynamics in the steady state, the kinetic energy oscillates $\pi$ out of phase with the other energy components. This phase relationship suggests a coherent exchange of energy between the kinetic term, the potential energy of the positronium, and the energy stored in the electric field. 

In Fig.~\ref{fig:schwinger_positronium1}h, which displays the field $\phi_x$ across the entire spatial domain at $t = 4000$, it is evident that most pair creation occurs early in the evolution, as the soliton–antisoliton system stabilizes. These pairs -- mesons -- subsequently propagate away from the positronium. As time progresses, both the rate of pair creation and the group velocity of the emitted mesons diminish. At asymptotically late times, the amplitude of $\phi_x$ in the vicinity of the positronium becomes vanishingly small (see Figs.,\ref{fig:schwinger_positronium1}i–k), indicating that the energy loss from the positronium effectively ceases.

The possibility of a mechanism for radiative relaxation towards a true bound state of a positronium is very unusual and its rigorous establishment requires further asymptotic analysis, which we leave to future work. An interesting hypothesis is that, as opposed to the physical 3+1D case where positronium is a resonance with a finite lifetime, in 1+1D the numerical simulations presented above indicate that a true bound state may exist. As we shall see next, this however is limited to a certain regime of $g$ and not the case generally.

\begin{figure}[t]
    \centering
    \includegraphics[scale=0.2]{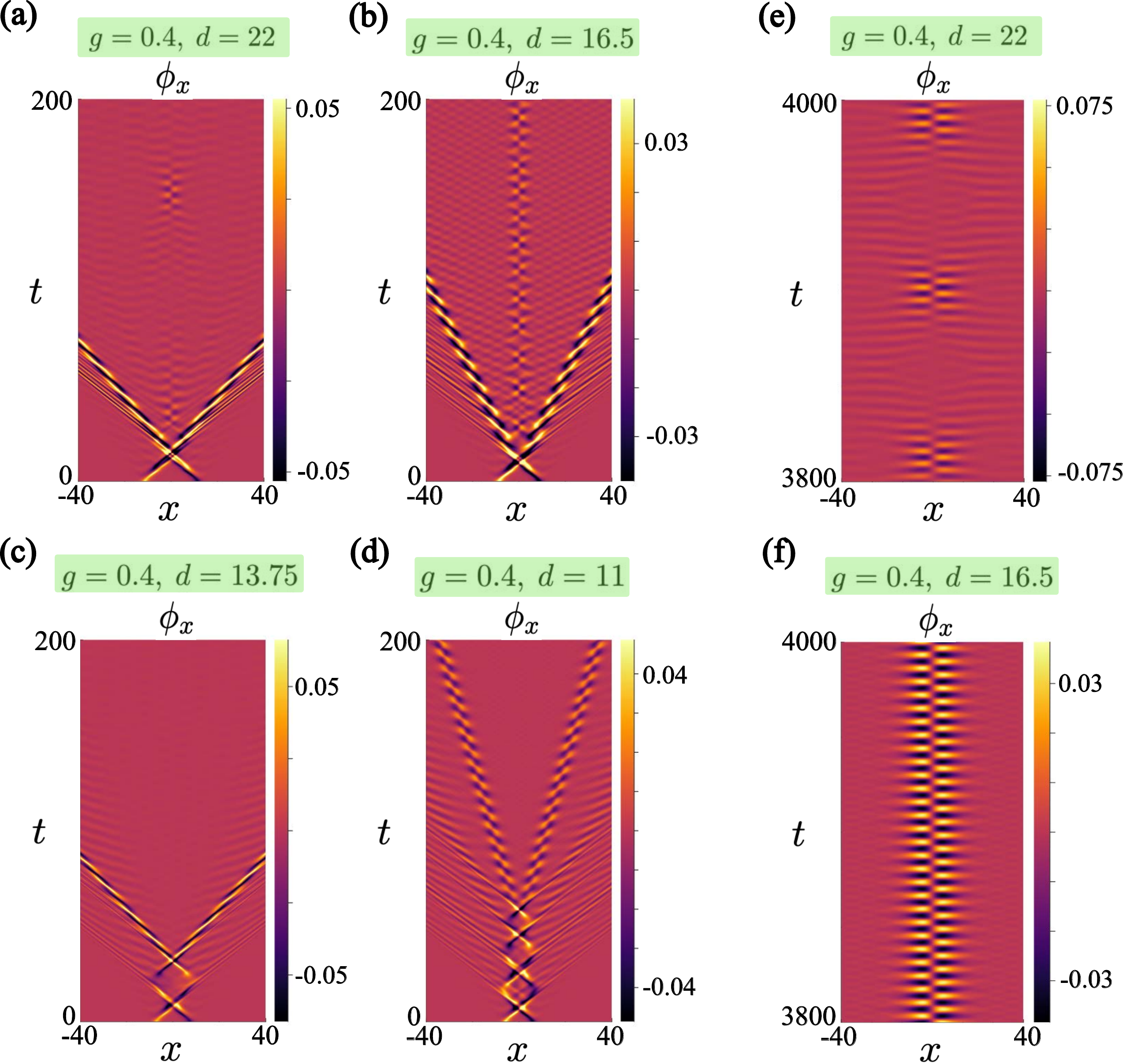}
    \caption{Possible dynamics of a system with large effective mass ($g=0.4$) and is initially composed of a soliton-antisoliton pair moving towards each other at the speed $u=0.55$. The early dynamics of the field $\phi_x$ are shown at (a), (b), (c), and (d) for initial separations of the pair $d=22,\ d=16.5, \ d=13.75, \ $and $d=11$, respectively. Late-time dynamics of $\phi_x$ are shown at (e) and (f) for $d=22$ and $d=16.5$, respectively.}
    \label{fig:unstablepair}
\end{figure}

Up to this point, we have considered only the parameter regimes in which a steady-state positronium is formed. However, if $g$ is large enough, depending on the initial conditions, a soliton-antisoliton system may scatter without forming a stable atom. Fig.~\ref{fig:unstablepair} illustrates this point by showcasing various dynamical scenarios, all initiated with a soliton–antisoliton pair. In Figs.~\ref{fig:unstablepair}a–d, where early-time dynamics is explored for $g = 0.4$ and $u = 0.55$ across different initial separations $d$, the pair is seen to immediately transition into breather-like configurations upon release. A larger $g$ leads to faster meson production, and splitting the initial soliton into an internally oscillating breather helps to suppress the creation of mesons. As each breather acts as an oscillating electric dipole, the pair — being aligned dipoles — mutually attract. Owing to this mutual attraction, the breather pair initially oscillates around each other before eventually dispersing, and it is observed that a smaller initial separation $d$ results in a longer period of oscillation prior to scattering.

In scattering off one another, the breathers emit mesons, and if the electric field holds enough energy — when $d$ is large — stable structures can form and persist near the origin. Figures~\ref{fig:unstablepair}e and \ref{fig:unstablepair}f present the late-time dynamics for the initial conditions used in Figs.~\ref{fig:unstablepair}a and \ref{fig:unstablepair}b, respectively. In Fig.~\ref{fig:unstablepair}e, we observe a steady beating pattern around $x = 0$, where meson density periodically disperses and refocuses. This behavior stems from two secondary breathers, formed from the residual charge left by the original breathers, which oscillate internally and orbit each other. Another scenario, shown in Fig.~\ref{fig:unstablepair}f, involves the formation of a stable positronium-like state after the initial breathers disperse. Here, a single breather remains at $x = 0$, oscillating indefinitely without radiative losses. 
Another possible outcome is a final state in which all the initial energy radiates away to $x \to \pm \infty$. This occurs in a two-breather system lacking sufficient initial energy in the electric field — specifically, when the initial separation $d$ is too small, as shown in Figs.~\ref{fig:unstablepair}c–\ref{fig:unstablepair}d. Although the breather pair oscillates longer for smaller $d$, they eventually disperse. The resulting electric field from the emitted mesons is too weak to generate or sustain further mesons within the region between the departing breathers. 
To validate the different late-time configurations observed in our numerical study of soliton-antisoliton dynamics, we also analyze the evolution of energy terms for each of these cases and the results are discussed in Appendix \ref{append:unstablepair_energy}.

\begin{figure}[t]
    \centering
    \includegraphics[scale=0.2]{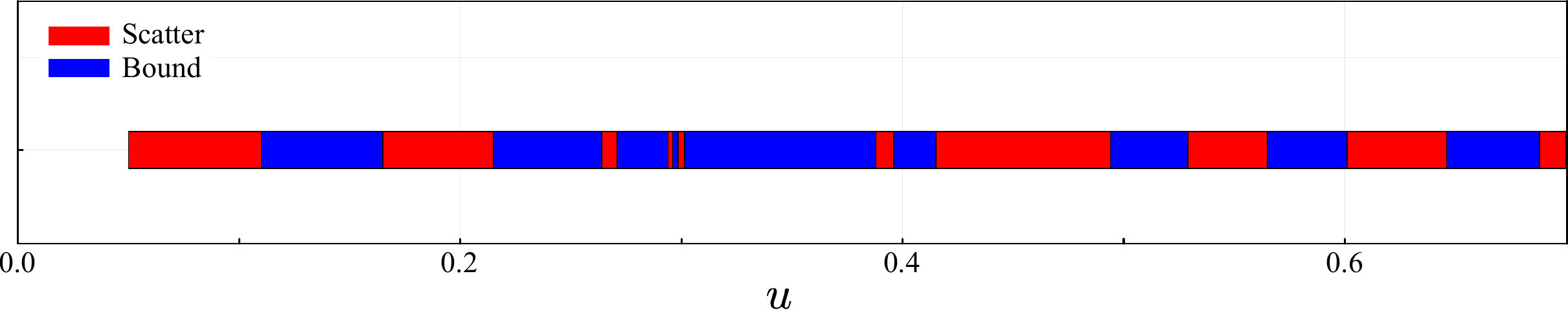}
    \caption{Fractalization of the initial velocity into intervals that correspond to either scattered solitons or a bound positronium in the long-time limit. Here, $g\!=\!0.32,$ and $d/u\!=\!20$.}
    \label{fig:bion_scatter_fractal}
\end{figure}

\subsubsection{Long-time phase diagram of 1+1D Positronium}

Our numerical experiments indicate that for sufficiently large $g$, the fate of an electron–positron pair — whether they scatter or form a stable bound state — depends sensitively on the initial conditions. In Fig.~\ref{fig:bion_scatter_fractal}, we map the final state as a function of the initial velocity $u$ ($0.05 \leq u \leq 0.7$). Interestingly, there is no single critical velocity separating the scattering and bound states; instead, we find a series of disjoint intervals where the system alternates between these outcomes. This “fractal” dependence on the incident velocity closely resembles the complex behavior seen in $Z_2$ soliton dynamics within $\varphi^4$ theory~\cite{anninos1991fractal, campbell1983resonance}.

By comparing the dynamics in Fig.\,\ref{fig:unstablepair}c where $d\!=\!13.75$ with those in Fig.\,\ref{fig:unstablepair}b where $d\!=\!16.5$, we can predict a phase transition at $d$ between these two values where the transition is from a system that sustains a localized atom at steady state to one that does not. Along the same line of thought, by comparing the dynamics in Fig.\,\ref{fig:unstablepair}a (where $g\!=\!0.4$) with those in Fig.\,\ref{fig:schwinger_positronium1}b (where $g\!=\!0.3$) - two dynamics that have the exact same initial conditions for the SG soliton pair but with different values for $g$ - we can predict a phase transition at $g\!=\!g_c$ somewhere between these two values that separates a phase ($g\!<\!g_c$) in which the initial SG soliton pair turns into a stable positronium with a phase ($g\!>\!g_c$) in which the initial solitons turn into breathers. To quantitatively identify in parameter space $(u,d,g)$ the boundaries  of the possible phases that an initial pair of solitons can evolve into at steady state, further detailed analysis will need to be carried out in the future.

\section{Discussion}

In this article, we examine the long-time behavior of a generalization of the Sine-Gordon equation that captures the semiclassical dynamics of the four-current of the Dirac field of fermions in the presence of gauge fields in 1+1 dimensions. Capturing emergent coherent structures -- such as long-lived, particle-like solitary waves -- requires numerical methods that remain accurate deep into the long-time regime. We present a space-time coarse-graining framework designed for this purpose, targeting the dynamics of generalized Sine-Gordon equations. Through systematic benchmarking against analytical solutions and established numerical methods, we validate the reliability of the framework across a range of physically relevant scenarios, before applying it to the dynamics of charge carriers interacting with a localized central charge -- the 1+1D Hydrogen atom.  

Firstly, the results demonstrate that the proposed coarse-graining method accurately captures the formation, propagation, and interaction of localized structures over asymptotically long timescales, with significantly improved numerical stability and reduced dispersion-induced artifacts. In particular, the method resolves regimes where conventional approaches fail to preserve the integrity of solitary wave solutions or suffer from cumulative errors. These capabilities establish the framework as a robust tool for investigating non-perturbative dynamics in semiclassical field theories, including those exhibiting soliton-like behavior and radiative decay.

Secondly, our results motivate further investigation into the analog simulation of relativistic 1+1D QED -- the massive Schwinger model -- using Josephson Transmission Lines (JTLs). While the present study operates strictly within the confines of the semiclassical regime of the quantum Sine-Gordon model, earlier foundational work~\cite{dashen1974, dashen1975}, has established that classical field dynamics can serve as a meaningful starting point for probing quantum phenomena beyond the reach of perturbative techniques, such as those we investigate in this work. In this context, it is essential for future work to clarify which aspects of the massive Schwinger model can be faithfully captured through analog implementations in JTL architectures.

Thirdly, extending the framework developed here to the associated quantum field theory holds the potential to yield novel insights into non-perturbative phenomena that underpin the foundations of quantum theory — namely, the stability and structure of atoms. Atomic spectra and radiative transition rates are traditionally understood through perturbative expansions in the gauge-field coupling and the renormalization framework~\cite{weinberg2005qft}. Even the stability of the Hydrogen atom’s ground state has only been rigorously established within perturbative QED through the ingenious use of inequalities~\cite{lieb2005stability}. A non-perturbative numerical approach capable of accessing such properties -- even in the reduced dimensional setting of 1+1D -- would represent a significant advance, offering a unique window into atomic structure that remains out of reach for perturbative approaches to quantum gauge theory.

\section{Acknowledgments}
We gratefully acknowledge discussions with Thomas Maldonado. This research is supported by the U.S. Department of Energy, Office of Basic Energy Sciences, Division of Materials Sciences and Engineering, under Award No. DESC0016011. The simulations presented in this article were performed on computational resources managed and supported by Princeton Research Computing, a consortium of groups including the Princeton Institute for Computational Science and Engineering (PICSciE) and the Office of Information Technology's High-Performance Computing Center and Visualization Laboratory at Princeton University.

\appendix
\section{Derivation of the SG equation from the gauge-invariant EHDS equations}\label{append:SG_from_EHDS}
\begin{figure}
    \centering
    \includegraphics[scale=0.8]{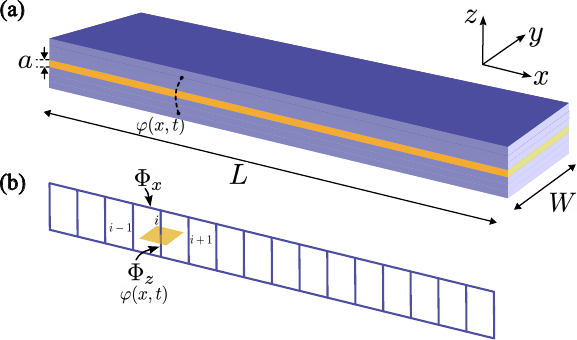}
    \caption{(a) Schematic of a long Josephson junction. The thickness of the effective insulating region is labeled $a$. The long dimension is $L\!\gg\!\lambda_J$ while the other dimension is $W\!\ll\!\lambda_J$. (b) The discretization of the long Josephson junction using a quasi-1D grid, in which the Josephson phase $\varphi(x)$ is encoded on the single layer of $z-$oriented edges that encompass the thickness of the junction along this dimension. The orange patch indicates the face dual to the primal $i^{th}$ $z$-edge.}
    \label{fig:longjj_schem}
\end{figure}

In this appendix, we show that the sine-Gordon equation can be derived from an electrohydrodynamics description of superconductors (EHDS)~\cite{dec-qed, maldonado2024pressure} 
applied to the long Josephson junction. This will confirm that although the SG equation is a 1D effective field equation in which the spatial variation along the thickness of the junction is not considered, it can still be rigorously traced back to a first principles formulation. 

Consider a long Josephson junction composed of two superconducting islands that are separated by a thin ($\lesssim$ few $nm$) layer of either insulator or normal metal, as illustrated in Fig.\,\ref{fig:longjj_schem}. The long dimension of the junction is along $x$ and has length $L\!\gg\!\lambda_J$, where $\lambda_J$ is the Josephson penetration depth. The superconductor-insulator-superconductor variation in material properties is along $z$, while the remaining dimension $W$ of the junction along $y$ is assumed to be much less than $\lambda_J$ so that there is effectively no field variation inside the junction in this direction. If the separation between the superconducting island is thin, the superconducting condensate can leak into this region and, if the decay tails from two islands touch, a supercurrent can be established in the junction~\cite{JosephsonReview_1965}. Let $\psi$ be the macroscopic order parameter of the collective superconducting condensate fluid that lives in the entire junction and is made of Cooper pairs (with mass $m\!=\! 2m_e$ and charge $q\!=\!2e$, where $m_e$ and $e$ are the electron mass and charge respectively). Using the Madelung representation $\psi=\sqrt{\rho} e^{i\theta}$, where $\rho$ is the local density of condensate and $\theta$ is the gauge-dependent superconducting phase, the EHDS equation governing the dynamics of the condensate in the presence of the electromagnetic field $(V,{\bf A})$ is given by
\begin{align}\label{eq:Aprime_waveeq_rho}
 {\bf\nabla}\!\times\!{\bf\nabla}\!\times\!\bm{\mathcal{A}} + \mu_0 &\epsilon\frac{\partial^2\bm{\mathcal{A}}}{\partial t^2} + \frac{\mu_0 q^2}{m}\rho\bm{\mathcal{A}}  -\frac{\mu_0\epsilon q}{2m}\frac{\partial}{\partial t}\nabla\big|\bm{\mathcal{A}}\big|^2 \nonumber \\
  & + \frac{\mu_0\epsilon\hbar^2}{2mq}\frac{\partial}{\partial t}\nabla\bigg[\frac{\nabla^2(\sqrt{\rho})}{\sqrt{\rho}}\bigg] =  \mu_0{\bf J}_{src},
\end{align}
where \mbox{$\bm{\mathcal{A}} = {\bf A} - \frac{\hbar}{q}\nabla\theta$} is the gauge-invariant hybridized field that lives everywhere throughout space. When applied to a short Josephson junction with a sufficiently thin separation between the superconducting islands, it was rigorously shown in Ref.\,\cite{dec-qed} that the Josephson current-phase relation is produced from the last two nonlinear terms on the left hand side of Eq.\,\ref{eq:Aprime_waveeq_rho} canceling each other. For a long Josephson junction, the same cancellation holds if the spatial variations of the field $\bm{\mathcal{A}}$ along $x$ and $y$ appear at much larger length scales than along $z$. In the case of Josephson fluxons, the fluxons are assumed to be translationally invariant throughout the $y$ direction. The sizes of the fluxons along $x$ are determined by the Swihart velocity and the speed of the fluxons. These conditions typically result in fluxon sizes on the order of 100\,nm -few\,$\mu$m. Therefore, the condition for the mentioned cancellation of two nonlinear terms is well satisfied. 

To proceed, we discretize the long junction using a quasi-1D grid composed of an array of $xz$-facing rectangular cells aligned along $x$ (Fig.\,\ref{fig:longjj_schem}b). The $z$-oriented edges start from deep in the bulk of one superconducting island and end at the other island.
This minimal discretization also takes into account the fact that the field variation over the thickness of the junction along $y$ is negligible and can be captured by a single layer of edges. We define the edge fields $\Phi_{z}\!=\!\int_{e_z}\!\!{\bf d\ell}\!\cdot\!\bm{\mathcal{A}}$ which live on the edges along $z$ and $\Phi_{x}^+\!=\!\int_{e_x^+}\!\!{\bf d\ell}\!\cdot\!\bm{\mathcal{A}}$, $\Phi_{x}^-\!=\!\int_{e_x^-}\!\!{\bf d\ell}\!\cdot\!\bm{\mathcal{A}}$ which live on the upper and lower horizontal edges along $x$, respectively. 
Upon integrating both sides of Eq.\,\ref{eq:Aprime_waveeq_rho} over the dual face $e_z^\dagger(i)$ of the $z-$edge $e_z(i)$ and applying DEC to the curl-curl operator we obtain
\begin{widetext}
    \begin{align}\label{eq:ehds_jjdiscrete}
    \frac{-\Phi_z(i\!-\!1)\! +\! 2\Phi_z(i)\!-\!\Phi_z(i\!+\!1)\!+\!\Phi_x^+(i)\!-\!\Phi_x^-(i)\!-\!\Phi_x^+(i-1)\!+\!\Phi_x^-(i-1)}{\Delta x^2} + \mu_0 \epsilon\frac{\partial^2\Phi_z(i)}{\partial t^2} - \mu_0J_s^z(i) = \mu_0J_{src}^z(i),
\end{align}
\end{widetext}
where $J_s^z(i)$ and $J_{src}^z(i)$ are the supercurrent and the externally applied current flowing from one superconducting island to the other along $z$. Since the upper and lower horizontal edges are deep in the superconductor and because of the mirror symmetry between two islands, we have $\Phi_x(i)^+\!=\!\Phi_x^-(i)$. This allows for the simplification of the first term in Eq.\,\ref{eq:ehds_jjdiscrete}. By introducing the normalized flux variable $\varphi=-2\pi\Phi/\Phi_0$, where $\Phi_0$ is the flux quantum, Eq.\,\ref{eq:ehds_jjdiscrete} becomes 
\begin{align}\label{eq:sg_discrete}
    &\frac{-\varphi(i\!-\!1)\! +\! 2\varphi(i)\!-\!\varphi(i\!+\!1)}{\Delta x^2} + \mu_0 \epsilon\frac{\partial^2\varphi(i)}{\partial t^2}\nonumber\\
    &+ \mu_0J_0\sin\varphi(i) = \mu_0J_{src}^z(i),
\end{align}
where we have used the sinusoidal dependence on $\varphi$ to express the supercurrent. By rewriting the first term of Eq.\,\ref{eq:sg_discrete} above into $-\partial_x^2\varphi$, we obtain the sine-Gordon equation.

\section{Bosonization of the massive Schwinger model}\label{append:bosonization}

As introduced in Section~\ref{ref:massiveSchwinger}, the massive Schwinger model is described by the  Lagrangian
\begin{equation}
\label{Eq: MSM Lagrangian}
    \mathcal{L} = \Bar{\psi}(i\gamma^\mu D_\mu - m) \psi - \frac{1}{4} F_{\mu\nu} F^{\mu\nu}
\end{equation}
in the absence of the background $\theta$-angle, where $\psi \equiv \big( \begin{smallmatrix} \psi_{R}\\ \psi_{L} \end{smallmatrix} \big)$ is the Dirac fermion field in 1+1D, $D_{\mu} = \partial_{\mu} - i e A_{\mu}$ is the gauge covariant derivative, and $F_{\mu\nu} = \partial_{\mu} A_{\nu} - \partial_{\nu} A_{\mu}$ is the electromagnetic tensor. 

For bookkeeping purposes, we define the field theory on the finite domain $x \in [-L/2, L/2]$ and consider the theory on $\mathbb{R}$ as the $L \rightarrow \infty$ limit. Provided that the $\psi$ field can be expanded in modes labeled by integers (e.g. the momentum modes for periodic boundary conditions), the $n$-fermion Hilbert space constructed from the fermionic ladder operators $c_{k}^{\eta}, c_{k}^{\eta \dagger}$ is spanned by the following bosonic creation operators acting on the $n$-fermion ground state:
\begin{equation}
\label{Eq: bdagger def}
    \begin{split}
        b_{q}^{\eta \dagger}
        \equiv
        \frac{1}{\sqrt{q}} \sum_{k\in\mathbb{Z}} u_{k}^{\eta} \cdot
        c_{k+q}^{\eta \dagger} c_{k}^{\eta}
        ,\qquad
        \textrm{for}\;q=1,2,3\cdots
    \end{split}
\end{equation}
where $\eta$ labels the fermion species while $u^{\eta}_{k}$ is a species-dependent phase factor ($\abs{u^{\eta}_{k}}=1$) determined by the boundary conditions. For periodic boundary conditions, $\eta \in \{R, L\}$ labels the chiral fermion components and $u^{\eta}_{k} = i$. However, for other boundary conditions, $\eta$ does not indicate fermion chirality in general. For example, with closed boundary conditions $\eta$ would label and even and odd modes instead. 

More precisely, employing the partition function analysis by Haldane~\cite{Haldane_1981}, one can prove the following operator identity:
\begin{equation}
\label{Eq: bosonization dictionary}
\begin{split}
&
\psi^{\eta}(x)
=
U^{\eta} g^{\eta}(n^{\mu}; x) e^{ -i \cdot 2\sqrt{\pi} \phi^{\eta}(x) }
\end{split}
\end{equation}
where $\phi^{\eta}(x)$ is the real scalar field constructed from the ladder operators $b_{q}^{\eta},b_{q}^{\eta \dagger}$, the Klein factor $U^{\eta}$ is the unitary operator that removes the highest-energy $\eta$-fermion from any (free-theory) $n^{\eta}$-fermion ground state, and $g^{\eta}(n^{\eta};x)$ is a c-number function for fixed fermion number $n^{\eta}$ whose analytical expression can be determined by evaluating the ground-state expectation value of $U^{\eta \dagger} \psi^{\eta}(x)$.

In 1+1D, the gauge field can be integrated out from Eq.\,\ref{Eq: MSM Lagrangian} and expressed in terms of the real scalar field
\begin{equation}
\label{Eq: Electric field in terms of phi}
    \begin{split}
    \varphi(t,x)
    \equiv
    \sum_{\eta} \phi^{\eta}(t,x)
    =
    \frac{e}{\sqrt{\pi}} E(t,x)
    \end{split}
\end{equation}
In the $L \rightarrow \infty$ limit, Eq.\,\ref{Eq: bosonization dictionary} and Eq.\,\ref{Eq: Electric field in terms of phi} allow us to rewrite Eq.\,\ref{Eq: MSM Lagrangian} as the following Lagrangian density in terms of $\varphi(t,x)$ alone:
\begin{equation}
    \begin{split}
        \mathcal{L}
        =&
        \frac{1}{2} \partial_{\mu} \varphi(t,x) \partial^{\mu} \varphi(t,x)
        -
        \frac{e^{2}}{2\pi} \varphi^{2}(t,x)\\
        &
        +
        \frac{c}{\pi \epsilon} \cos\big( 2\sqrt{\pi}\varphi(t,x) \big)
    \end{split}
\end{equation}
where the order-$1$ constant $c$ is a function of Euler's constant (e.g. $c = e^{\gamma}$ for periodic boundary conditions), while $\epsilon \equiv \frac{L}{\pi\Lambda}$ with $\Lambda$ the cutoff imposed on the summation in Eq.\,\ref{Eq: bdagger def}.

In particular, the bosonization dictionary suggests that the mass term in the fermionic theory gives us the following cosine potential in the bosonic theory
\begin{equation}
m \Bar{\psi} \psi
\to
-
\frac{c}{\pi \epsilon} \cos(2 \sqrt{\pi} \varphi)
\end{equation}
If the cosine is properly normal ordered with respect to the renormalized boson modes, the factor of $\frac{1}{\epsilon}$ will be canceled and the renormalized Lagrangian takes the form
\begin{equation}
    \mathcal{L} = \frac{1}{2} \partial_\mu \varphi \partial^\mu \varphi + \frac{\kappa}{2} \cos (2 \sqrt{\pi} \varphi ) + \frac{g}{2} \varphi \epsilon^{\mu\nu} F_{\mu\nu} - \frac{1}{4} F_{\mu\nu} F^{\mu\nu}
\end{equation}
One obtains a $\varphi^{2}$ term in the bosonic Lagrangian due to the gauge interaction, which gives rise to a free Schwinger boson mass of $g$.  The new coefficients $g$, $\kappa$ are obtained after applying the bosonization dictionary and normal ordering with respect to the renormalized boson modes.  They are defined as
\begin{equation}
    g \equiv \frac{e}{\sqrt{\pi}}
\end{equation}
\begin{equation}
    \kappa \equiv \frac{mee^\gamma}{\pi^{3/2}} \Bigg( \frac{e^\gamma m}{e/\sqrt{\pi}} + \sqrt{1 + \frac{(e^\gamma m)^2}{e^2/\pi}} \Bigg)
\end{equation}
In the limit of small $\frac{m}{e}$, we obtain the usual expression for $\kappa \equiv \frac{e^\gamma m e}{\pi^{3/2}}$.

\section{Boundary conditions in DEC}\label{append:BCs}
\begin{figure}[h]
    \centering
    \includegraphics[scale=0.25]{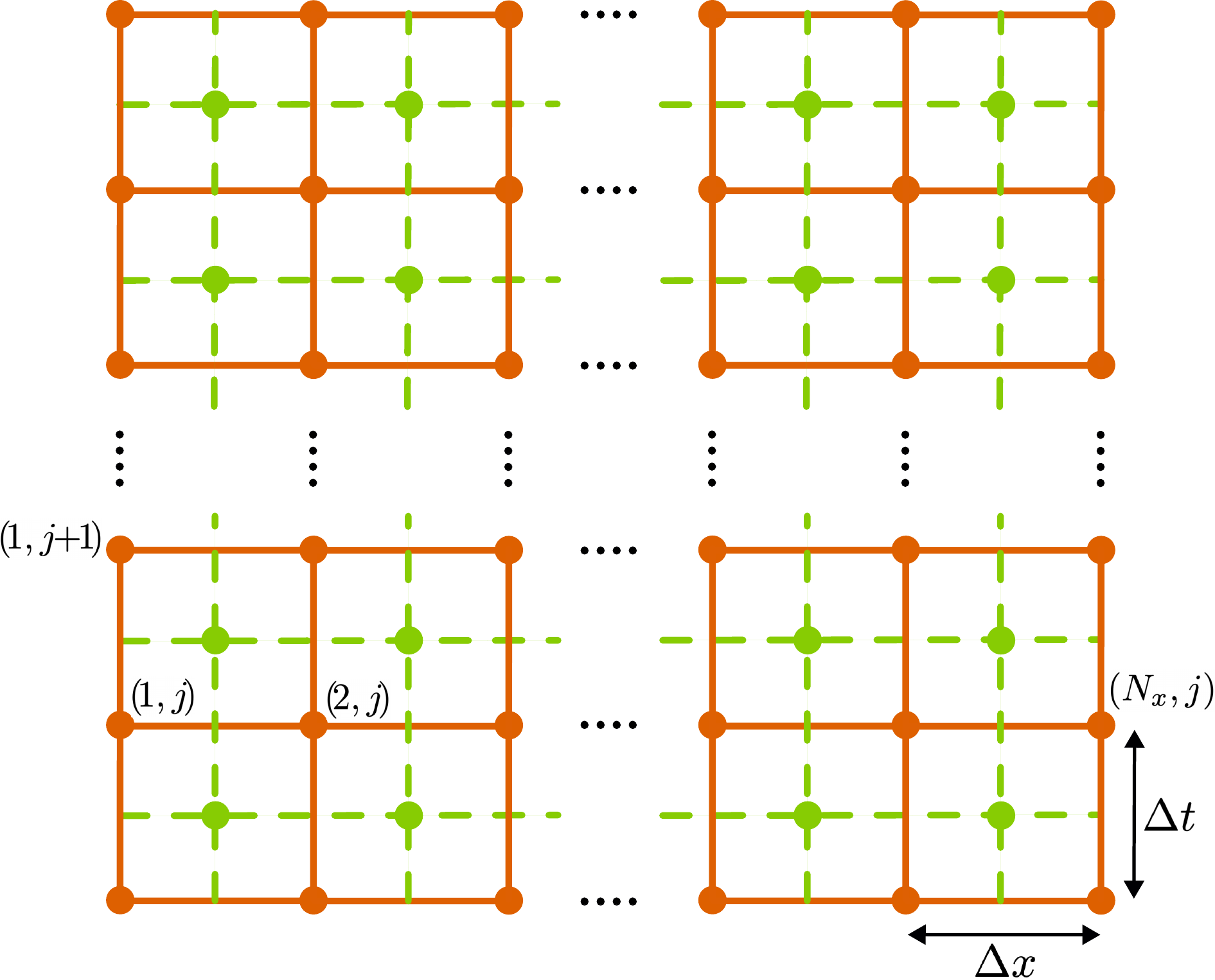}
    \caption{An example spacetime grid used in DEC-QED simulation.}
    \label{fig:grid_boundary}
\end{figure}

In this appendix, we discuss how boundary conditions are implemented in DEC and on a 1+1D spacetime grid. We focus on boundary conditions that are typically met in long Josephson junctions and in the Schwinger process.

{\it Closed Dirichlet boundary condition}: this is a BC imposed on the value of the scalar field $\varphi$ at the boundary (for example: $\varphi(0,t)=f(t)$, $\varphi(L,t)=h(t)$, where $x=0,L$ are the spatial boundary coordinates). This boundary condition can be straightforwardly imposed at the boundary vertices.

{\it Ordinary Neumann boundary condition}: examples of this type of boundary condition were presented in Eqs.\,\ref{eq:longjj_BC1}, \ref{eq:longjj_BC2}, and \ref{eq:boundarypulse}, in which the conditions on $\partial_x\varphi$ at the boundary incorporate how external magnetic field and injected boundary currents affect the internal dynamics of the junction. These BCs, which set the value for $\phi_x$ at the boundary, can be directly implemented on the boundary edges $e_x[(1,j),(2,j)]$ and $e_x[(N_x-1,j),(N_x,j)]$  at every time step $j$ in the spacetime grid (Fig.\,\ref{fig:grid_boundary}). Similarly, boundary conditions imposed solely on the time derivative of the scalar field, $\partial_t\varphi$, can be implemented on the field $\phi_t$ on the boundary edges $e_t[(1,j),(1,j+1)]$ and $e_t[(N_x,j),(N_x,j+1)]$.

{\it Outgoing boundary condition}: this type of boundary conditions is typically inhomogeneous, i.e. it depends both on the spatial and temporal derivatives of the field as well as on the specific wave equation in consideration. Here we discuss a general approach for outgoing boundary conditions which is effective for weakly nonlinear wave equations of the type
\begin{equation}\label{eq:general_wave_eq}
    \varphi_{tt} -\varphi_{xx} + U(x,t)\varphi = 0, 
\end{equation}
with $U(x,t)\ll 1$. By performing a Fourier transform \mbox{$\varphi(x,t)\!=\! 1/\!(4\pi^2)\!\int\!\!\int\! e^{i(kx+\omega t)}\hat{\varphi}(k,\omega)dkd\omega$} to Eq.\,\ref{eq:general_wave_eq} and collecting the terms in the integrand we obtain 
\begin{align}
    k^2 &= \omega^2 - \frac{1}{4\pi^2}\!\int\!\int\!\hat{U}(k',\omega')e^{i(k'x+\omega't)}dk'd\omega' \nonumber\\
    &= \omega^2 - U(x,t).
\end{align}
This yields the dispersion
\begin{align}\label{eq:dispersion_series}
    ik=\pm i\omega\sqrt{1-\frac{U(x,t)}{\omega^2}} \approx \pm i\omega\left[1+\frac{U(x,t)}{2(i\omega)^2} + ...\right]
\end{align}
The series expansion in Eq.\,\ref{eq:dispersion_series} allows for order-by-order improvements in the open boundary condition such that the reflection of waves exiting the computational domain is suppressed. Through the mappings $ik\rightarrow\partial_x$ and $i\omega\rightarrow \partial_t$, the zeroth and first order outgoing boundary conditions at the left boundary is given by
\begin{align}
    \partial_x\varphi &= \partial_t\varphi, \label{eq:0th_outgoingbc} \\
    \partial_{xt}\varphi &= \partial_{tt}\varphi + U\varphi/2. \label{eq:1st_outgoingbc}
\end{align}
Similar outgoing conditions for the right boundary are obtained by flipping the sign on the left-hand side of Eqs.\,\ref{eq:0th_outgoingbc}-\ref{eq:1st_outgoingbc}. Higher order boundary conditions can be readily obtained by keeping more terms in the expansion in Eq.\,\ref{eq:dispersion_series}.

The DEC implementation of the BC in Eq.\,\ref{eq:0th_outgoingbc} can be done by setting
\begin{equation}
    \phi_t[(1,j),(1,j+1)] = \left(\Delta t/\Delta x\right)\phi_x[(1,j),(2,j)],
\end{equation}
while the BC given in Eq.\,\ref{eq:1st_outgoingbc} is imposed by setting
\begin{align}
    \phi_t&[(1,j),(1,j+1)] =  \phi_t[(1,j\!-\!1),(1,j)] \nonumber\\
    &+ \left(\Delta t/\Delta x\right)\left\{\phi_x[(1,j),(2,j)]-\phi_x[(1,j\!-\!1),(2,j\!-\!1)]\right\} \nonumber\\
    &-\Delta t^2 U(1,j)\varphi(1,j)/2.
\end{align}

\section{Profiling simulation times}\label{append:runtimes}
\begin{figure}[t]
    \centering
    \includegraphics[scale=0.35]{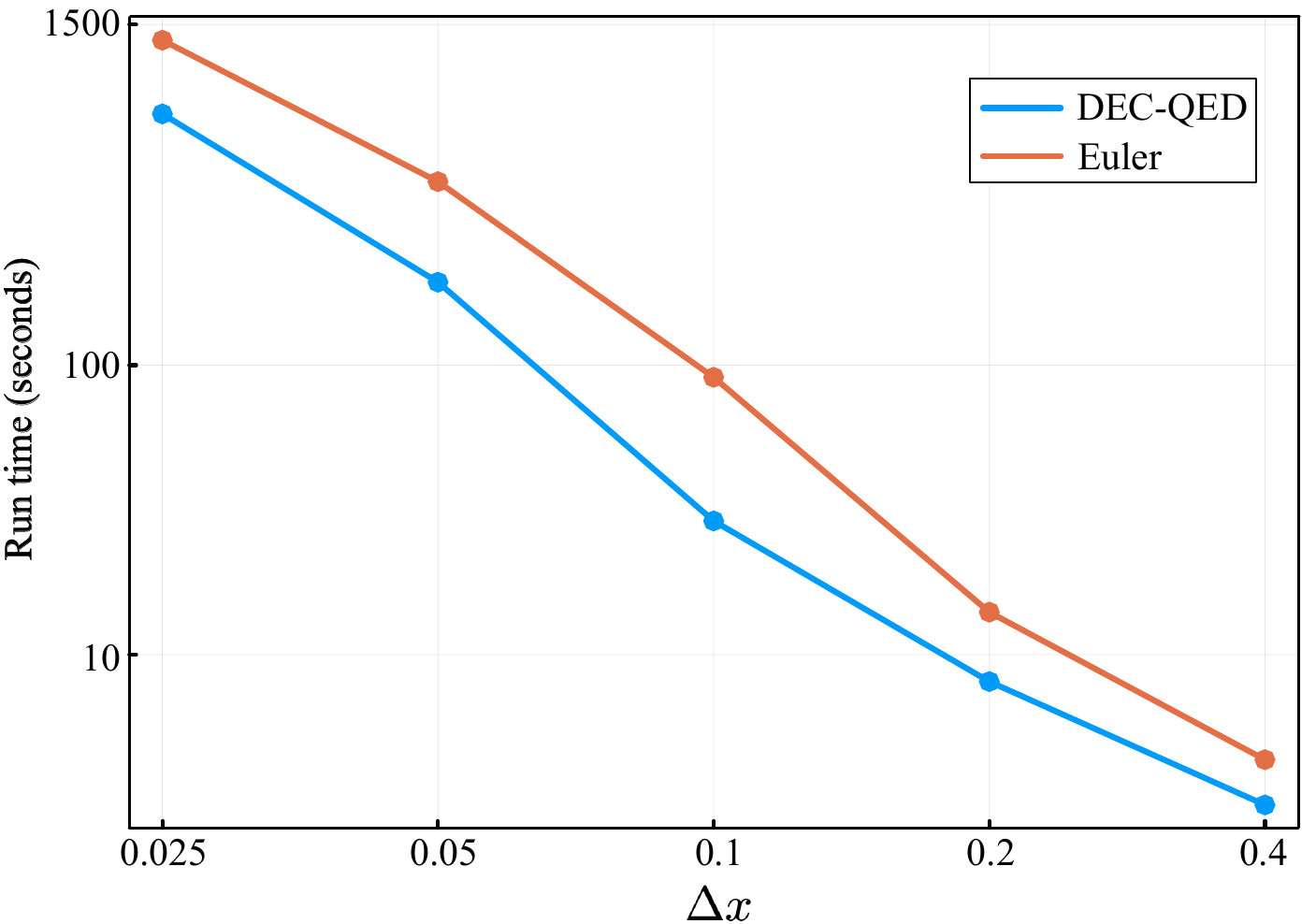}
    \caption{Comparisons on the run times of DEC-QED and Euler methods in simulating a fluxon trapped inside an unperturbed Josephson junction. For all simulations here, $T_m=50000, \Delta t=0.8\Delta x$, $u=0.55$, and $L=100$.}
    \label{fig:runtime_DEC_euler}
\end{figure}

A comparison on the runtimes of DEC-QED and the Euler method is presented in Fig.\,\ref{fig:runtime_DEC_euler}. The problem used for the comparison is the dynamics of a sine-Gordon fluxon trapped inside an unperturbed Josephson junction of length $L=100$. The simulated fluxon moves inside the junction at speed $u=0.55$ and for a duration of $T_m=50000$. For all the spatial step size $\Delta x$ tested, we choose $\Delta t=0.8\Delta x$. Since the two methods have the same complexity, the number of computations scales linearly with the number of gridpoints in the spacetime grid. However, because DEC-QED deals with the first-order PDE of the edge fields, it is faster than solving the second-order noninear wave equation for the scalar field $\varphi$ using the traditional Euler method. As a result, DEC-QED is slightly faster than the Euler method regardless of the step size used in the simulation. One should also note that if the sine term is removed and the dynamcis become linear, the two methods will have the same convergence rate for the derived fields $\phi_x$ and $\phi_t$. By nature of it being a local and time-propagating method DEC-QED also requires that the Courant condition is respected in order for the evolution to be stable. In a nonlinear and out-of-equilibrium setting, however, DEC-QED remains stable for a considerably longer simulations compared to standard Euler's method - as shown in Fig.\,\ref{fig:verylongtime_laststep_DEC_euler_CN} and Fig.\,\ref{fig:schwinger_atom_dec_vs_euler}.

\section{Modification of vortex dynamics due to biased boundary conditions}\label{append:verylongtime_biasedBC}
\begin{figure*}[t]
    \centering
    \includegraphics[scale=0.41]{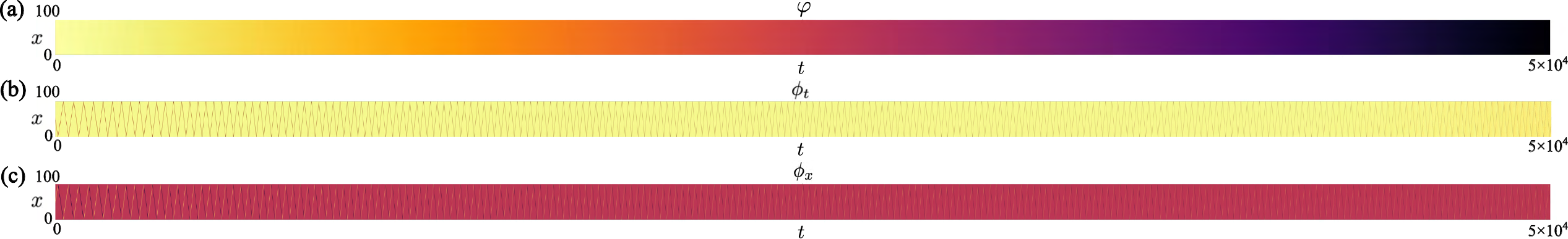}
    \caption{Long-time dynamics of a single fluxon trapped inside a Josephson junction. The junction length is $L=100$, the total time of the dynamics is $T_m=50000$, the initial velocity of the fluxon is $u=0.55$, $\eta=0.002$ and $\xi=0.006$. (a), (b) and (c) show the value of  $\varphi$, $\phi_t$, and $\phi_t$ respectively.}
    \label{fig:verylongtime_bareJJ_biasedboundary}
\end{figure*}

Figs.\,\ref{fig:verylongtime_bareJJ_biasedboundary}a, \ref{fig:verylongtime_bareJJ_biasedboundary}b, and \ref{fig:verylongtime_bareJJ_biasedboundary}c show the results for $\varphi$, $\phi_t$, and $\phi_x$ with $\eta=0.002$ and $\xi=0.006$.
We observe a gradual speeding up of the soliton over time.  This is because every time the soliton collides with a boundary where $\eta$ and $\xi$ are non-zero, it is in contact with an externally injected current and a magnetic field that either adds to or decreases its kinetic energy depending on the relative orientation of the current of the (anti)vortex with these boundary effects. For the parameters chosen for the simulations in Figs.\,\ref{fig:verylongtime_bareJJ_biasedboundary}a-c, after each round-trip (i.e. after colliding with the boundary on both ends) the soliton obtains a small additional energy and speeds up.  With each subsequent round trip, the velocity of the (anti)vortex approaches that of the boundary current.  As this happens, the (anti)vortex speeds up by less and less with each collision.  In total, the dynamics shown in Fig.\,\ref{fig:verylongtime_bareJJ_biasedboundary}a-c has 453 collisions with the boundaries, covering a total distance slightly greater than 18875 times its original size. To compare with the vortex dynamics in Figs.\,\ref{fig:verylongtime_laststep_DEC_euler_CN} in the main text, where we have the same initial conditions and parameters except that $\eta=\xi=0$, the fluxon only collides 278 times with the boundary.

\section{Vortex in narrow constrictions}\label{append:constrictions}
\begin{figure}[t]
    \centering
    \includegraphics[scale=0.1]{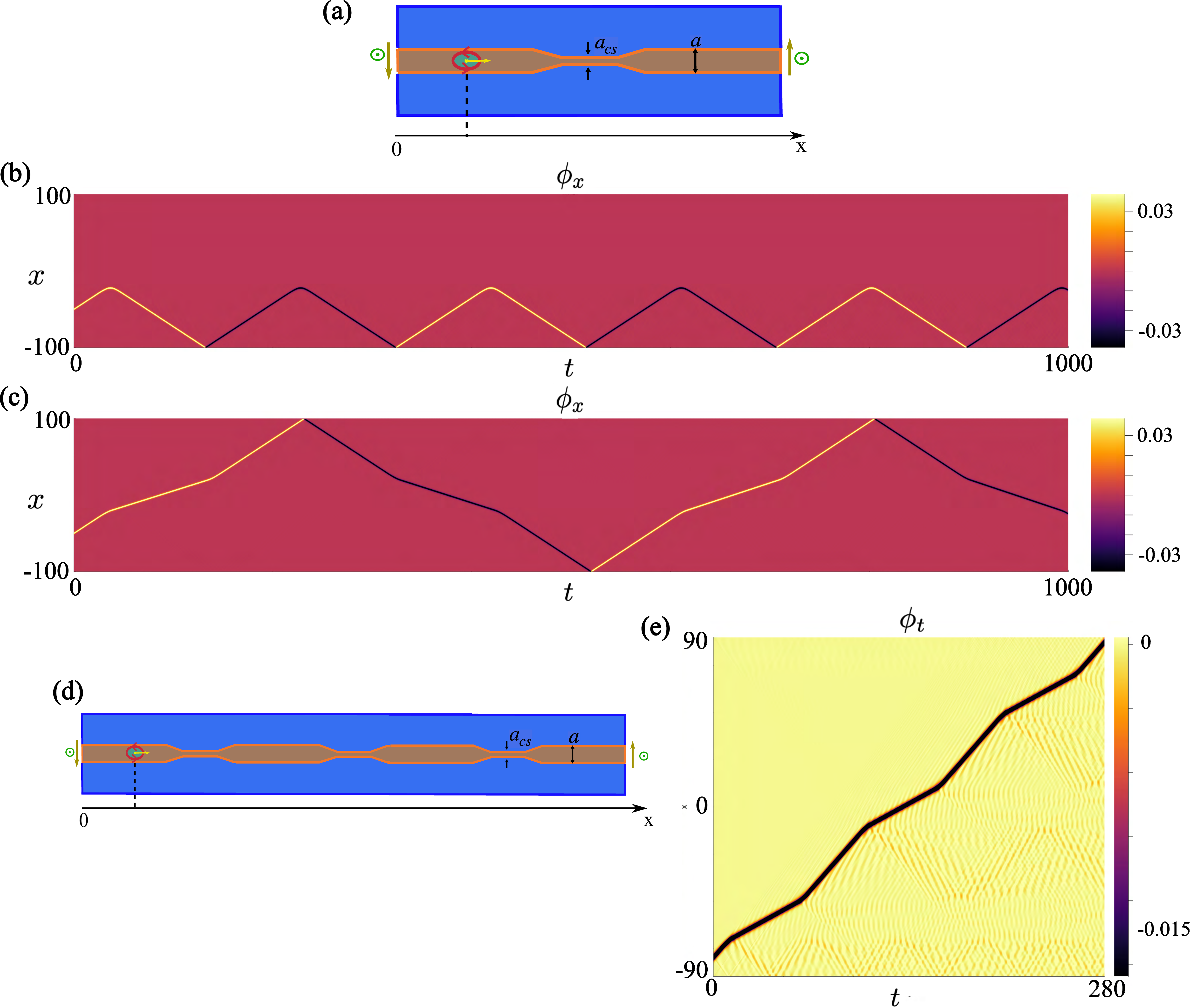}
    \caption{(a) Schematic of a vortex moving inside a long Josephson junction that contains a constriction region. (b) The $\phi_x$ field when a fluxon travels inside a junction having a finite region of constriction. The critical current inside the constriction is ten times the critical current in the normal region. (c) The $\phi_x$ field when the fluxon travels in a constricted junction, but when the critical current inside the constriction is three times the critical current in the normal region. (d) Schematic of a long junction containing three consecutive constrictions. (d) $\phi_t$ field when the fluxon travels a full length of the triple-constriction junction.}
    \label{fig:constriction}
\end{figure}
Typically, in a long Josephson junction, it is difficult to maintain uniform thickness and material properties throughout the entire length of the junction. The microshorts discussed in Section \ref{subsec:microshort}  are extreme cases, where the thickness of the insulating region is near zero and the superconducting islands are essentially in contact with each other. In general, the thickness of the insulating region can vary along the junction, and the sine-Gordon equation can be modified accordingly to capture this behavior
\begin{equation}
    \partial^2_t\varphi -\partial^2_x\varphi + \alpha\partial_t\varphi + \mu(x)\sin\varphi = -\beta
\end{equation}
where the critical current $\mu(x)$ is now a function of $x$. Using DEC-QED, we consider the evolution of a fluxon in a junction that has a narrow constriction inside (see Fig.\,\ref{fig:constriction}a), where the insulating region is thinner than the rest of the junction, leading to a stronger critical current there. In general, the critical current depends exponentially on the thickness of the insulating region. In the limit of thin junctions, the dependence is approximately inversely proportional to the thickness of the junction, $a$, ~\cite{dec-qed}, i.e. \mbox{$\mu\propto 1/\sinh(\xi a)\sim 1/a$}. In Fig.\,\ref{fig:constriction}b-c the dynamics of a fluxon moving in a constricted junction is shown. The junction length is $L=200$, the length of the constriction is $\ell = 40$, and the lengths of the two tapered regions that connect the constriction with the rest of the junction are $b=10$. 
In Fig.\,\ref{fig:constriction}b, the critical current inside the constriction is $\mu_{cs} = 10$, while the critical current at the normal regions are unity. The fluxon has an initial velocity of $u=0.85$ and travels without resistive loss. Similar to the microshort case, this narrow constriction draws energy from the fluxon, preventing it from entering and repeling it from the constriction. The movement of this fluxon is therefore limited to bouncing between the junction boundary and the edge of the constriction.

If the critical current in the constriction is sufficiently relaxed, which is the situation in Fig.\,\ref{fig:constriction}c where $\mu_{cs}=3$, the fluxon can enter the narrow channel. When the fluxon is inside this potential barrier, part of the kinetic energy is transferred to potential energy and the soliton slows down, as can be seen by the change in the slope of the trace in Fig.\,\ref{fig:constriction}c. Note that the higher critical current in a narrow junction means the Swihart velocity is also lower there.
When the soliton has just entered the constriction, however, it can still have a velocity higher than the maximum velocity allowed in this region. As the soliton is abruptly slowed down to below the Swihart velocity in the constriction, it also releases energy in the form of radiation. These are similar to Cherenkov radiations that appear in nonlinear optical systems that have high-order dispersions such as in fiber optics~\cite{akhmediev1995cherenkov}. This effect is evidently seen in Fig.\,\ref{fig:constriction}e, where the dynamics of a soliton moving through a series of three constrictions is shown. The fringes caused by radiation coming out of the soliton are most intense at the bending corners of the trajectory, where the soliton enters a narrow constriction and has to slow down. In this example, after the soliton exits the last constriction, roughly $\approx 3\%$ of its initial kinetic energy has been transferred to radiations.

\section{Creation of solitons using applied external pulse}\label{sec:pulse}

\begin{figure}[t]
    \centering
    \includegraphics[scale=0.115]{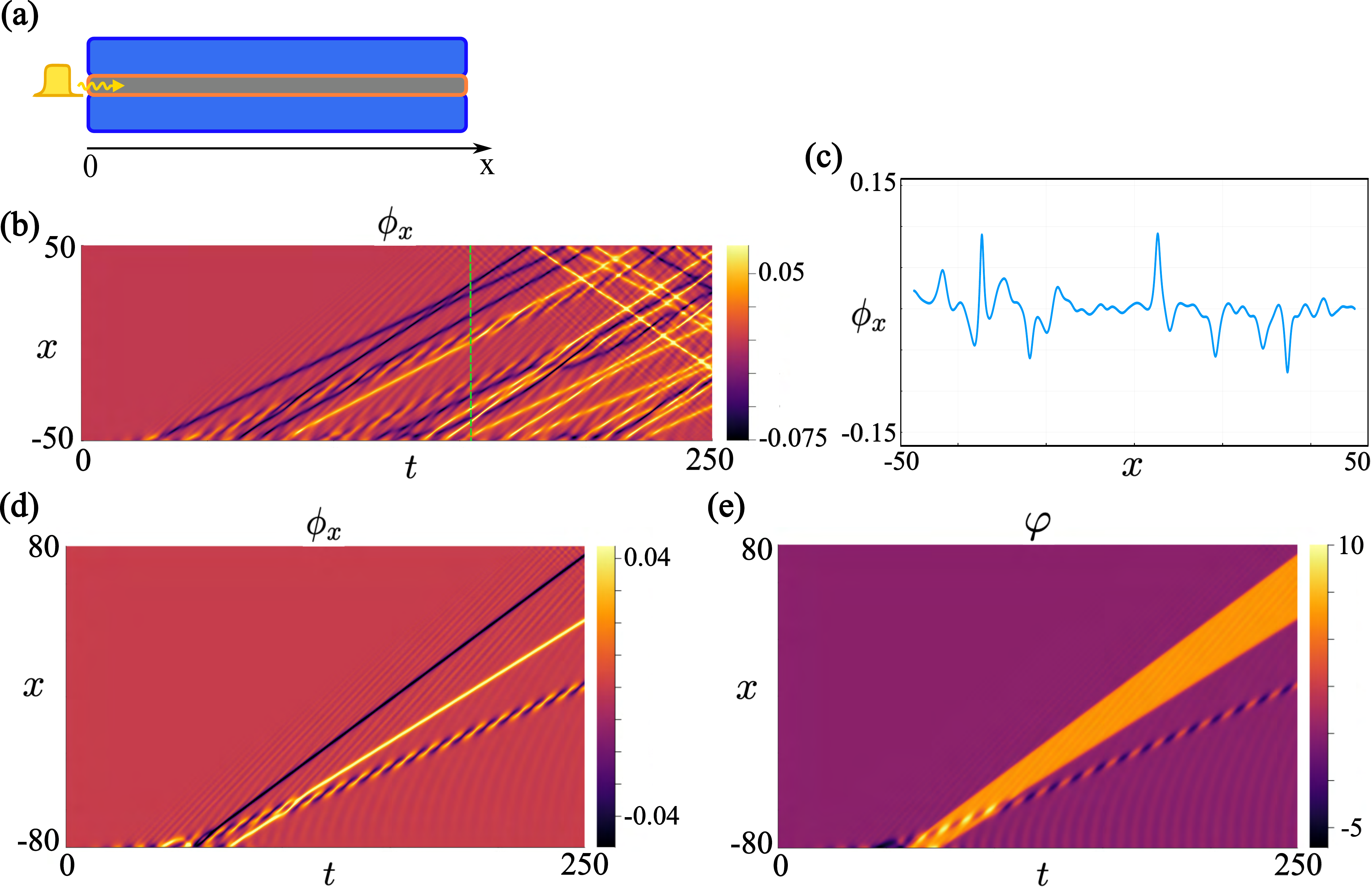}
    \caption{(a) Schematic of a Josephson junction excited by an external pulse at the boundary. (b) An example of dynamics inside a junction constantly excited by a square pulse with Gaussian rise and fall edges. The total duration of the pulse is $T_m=250$, with frequency $\omega=0.8$. The rise time and fall time are $\sigma_{rise}=\sigma_{fall}=10$. Between the rising and falling period, the pulse amplitude is held at $A=1.5$. A time slice at $t=160$ of the dynamics shown in (b) (labeled by the green dashed line) is plotted in (c).
    (d) and (e) plot the $\phi_x$ and $\varphi$ fields inside a junction excited by a tuned pulse such that there is exactly one fluxon, one antifluxon, and one breather created in the system. The junction has length $L=160$. The parameters for the pulse are: $\sigma_{rise}=\sigma_{fall} = 10, \omega = 0.6, A = 1.4,$ and the total duration of the pulse is $T_p\approx 70$.}
    \label{fig:pulse}
\end{figure}

So far we have discussed a wide variety of possible situations involving solitons embedded in Josephson junctions and demonstrated how DEC-QED is able to reliably capture the evolution of these systems. These simulations all begin with solitons as initial conditions. However, in a realistic experimental setup, these fluxons or breathers have to be created by a well-controlled application of external fields~\cite{kivshar1991solitoncreation,sakai1987solitoncreation}. 
Therefore, it is also important to be able to simulate how these solitons can emerge from the background via external excitation. In this section, we discuss how DEC-QED is a suitable tool for this. 

We consider a long junction subjected to an external magnetic pulse at one boundary edge of the junction. The mechanism of generating solitons by applying a boundary pulse is similar to those studied in the physics of nonlinear supratransmission\,\cite{geniet2002energy, khomeriki2009fluxon, schlawin2017terahertz, desantis2022generation, desantis2022supratransmission}.
In our current setting, the effect of the pulse to the junction is modeled through a boundary condition of the form
\begin{equation}\label{eq:boundarypulse}
    \partial_x\varphi(x,t) = H(t)\sin\omega t,
\end{equation}
where $\omega$ is the frequency of the pulse, and $H(t)$ is the amplitude. We model the amplitude by a square envelope for $H(t)$ that is smoothed by Gaussian edges to account for the finite rise time $\sigma_{rise}$ and fall time $\sigma_{fall}$ of the pulse. 

When a junction experiences a finite pulse at the boundary, radiations, fluxons, and antifluxons  of any velocity $u\leq 1$ can occur on that boundary and travel into the junction. This is demonstrated in Fig.\,\ref{fig:pulse}b, where we choose a junction whose length is $L=160$. The pulse has a total duration of $T_m=250$, with frequency $\omega=0.8$. The rise time and fall time are $\sigma_{rise}=\sigma_{fall}=10$, and the pulse amplitude after the rise time and before termination is $A=1.5$. In the 2D $x-t$ heatmap in Fig.\,\ref{fig:pulse}b one can see very clearly that the generated excitations are a series of fluxons, antifluxons, and breathers, all moving at different speeds and interfering with each other, plus some background radiations. It is also worthwhile to note that instead of looking at this holistic view of the full dynamics, one can plot the field at a specific time slice (as done in Fig.\,\ref{fig:pulse} where $\phi_x$ at $t=160$ is shown), and it appears as if there is no structure to the dynamics generated by the pulse which is actually composed of mostly solitons. 

Given the efficiency and accuracy of the numerical scheme, one can easily sweep a large parameter space to tune the pulse to produce the desired results. To demonstrate, in Fig.\,\ref{fig:pulse}c and \ref{fig:pulse}d we plot the $\phi_x$ and $\varphi$ fields for a setup in which the external pulse creates exactly one fluxon, one anti-fluxion, and one breather. To find the correct pulse, we had sampled the parameter space of $(A,\omega,\sigma_{rise},\sigma_{fall}, T_p)$ in the following ranges and steps: $1\!\leq\! A\!\leq\! 2$ with step $\Delta A\!=\!0.1$, $5\!\leq\!\sigma_{rise}\!=\!\sigma_{fall}\!\leq\! 15$ with step $\Delta\sigma\! =\! 1$, $0.5\leq\omega\leq 0.9$ with step $\Delta\omega\!=\!0.1$, and $40\!\leq\! T_p\!\leq\! 80$ with $\Delta T_p\!=\! 2$. This sampling amounts to 12700 coordinates in the parameter space. We choose a junction length of $L=160$ that is discretized by grid spacings $\Delta x=0.02$, while the simulated time is $T_m=250$ with $\Delta t = 0.016$.
For each set of parameters for the pulse, the simulation is therefore done with a spacetime mesh of $125$ million grid points. The total time it takes for the sweep to return the desired pulse is $5.5$ hours.

\section{Massive capacitor discharge}\label{append:capdischarge_massive}

\begin{figure}[t]
    \centering
    \includegraphics[scale=0.108]{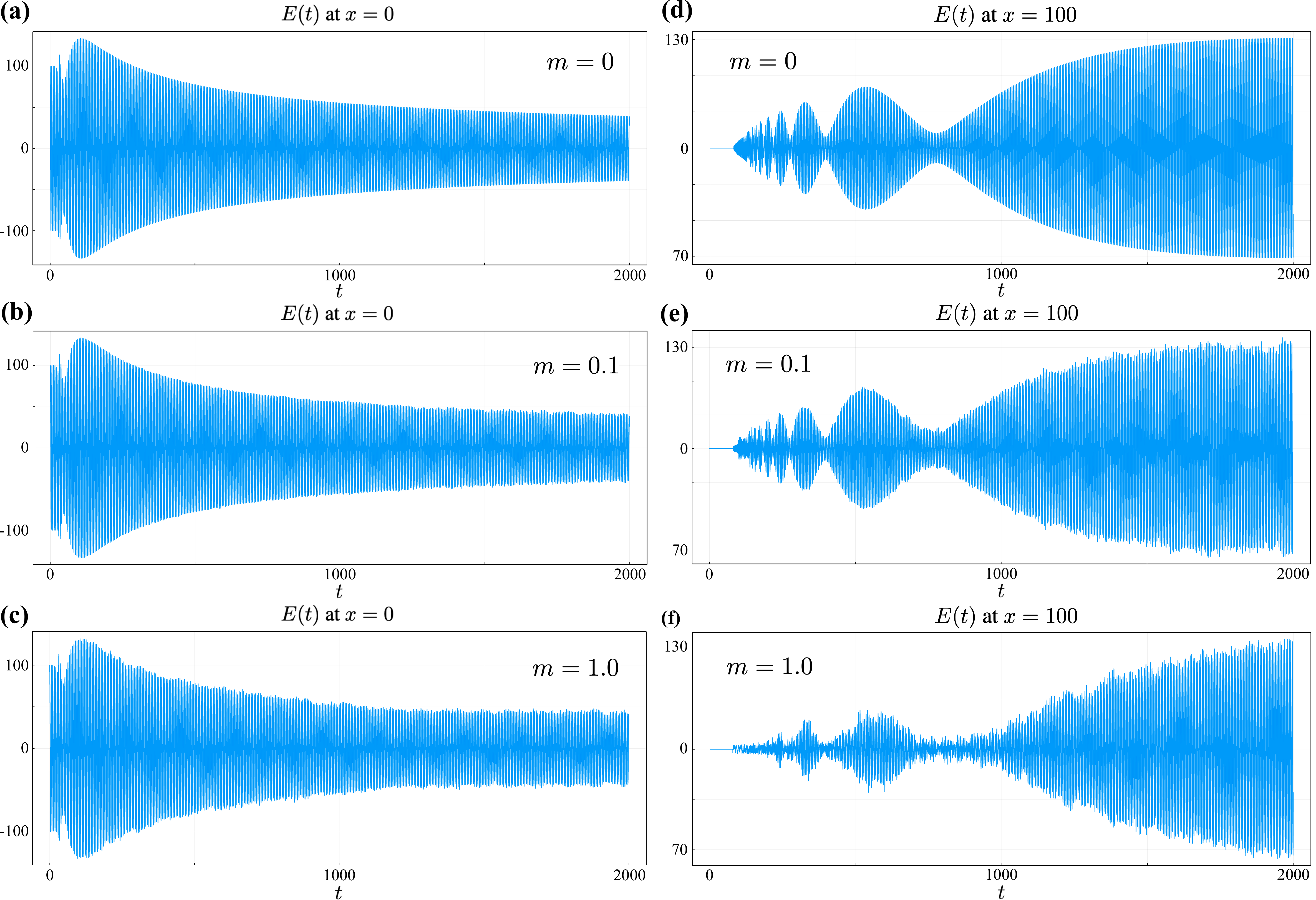}
    \caption{Time evolution of the electric field due to a sharged capacitor and the back reaction of the fermionic fluid. The capacitor is charged at $Q=100$, with the plates located at $\pm L/2=\pm20$, and $g=1.2$. The dynamics is captured up to $T_m=2000$. The electric field $E$ at $x\!=\!0$ is shown in (a), (b), and (c) for when the mass of the fermions are  $m=0,\ m=0.1,\ $ and $m=1.0$, respectively. In a similar order, the evolution of $E$ at $x\!=\!100$ when $m=0,\ m=0.1,\ $ and $m=1.0$ are shown in (d), (e) and (f).}
    \label{fig:capdischarge_massive}
\end{figure}

In Section \ref{ref:masslessSchwinger} we studied the backreaction dynamics of massless fermions under the influence of a charged capacitor. In this appendix, we present the results for the capacitor discharge dynamics when the fermions are massive. Figs.\,\ref{fig:capdischarge_massive}a-c provide direct comparisons of the electric field at the center of the capacitor when the fermion mass is $m=0,\ 0.1,$ and $1.0$, respectively. Figs.\,\ref{fig:capdischarge_massive}d-f present the electric field at a finite distance away from the capacitor. As evidently shown in Figs.\,\ref{fig:capdischarge_massive}, a finite mass $m$ leads to fast oscillations in the field due to the presence of the term $\sin\varphi$ in the equation of motion, with larger $m$ resulting in oscillations with higher amplitudes. 

\section{Different long-time scenarios of the Schwinger soliton-antisoliton pair}\label{append:unstablepair_energy}

\begin{figure*}[t]
    \centering
    \includegraphics[scale=0.22]{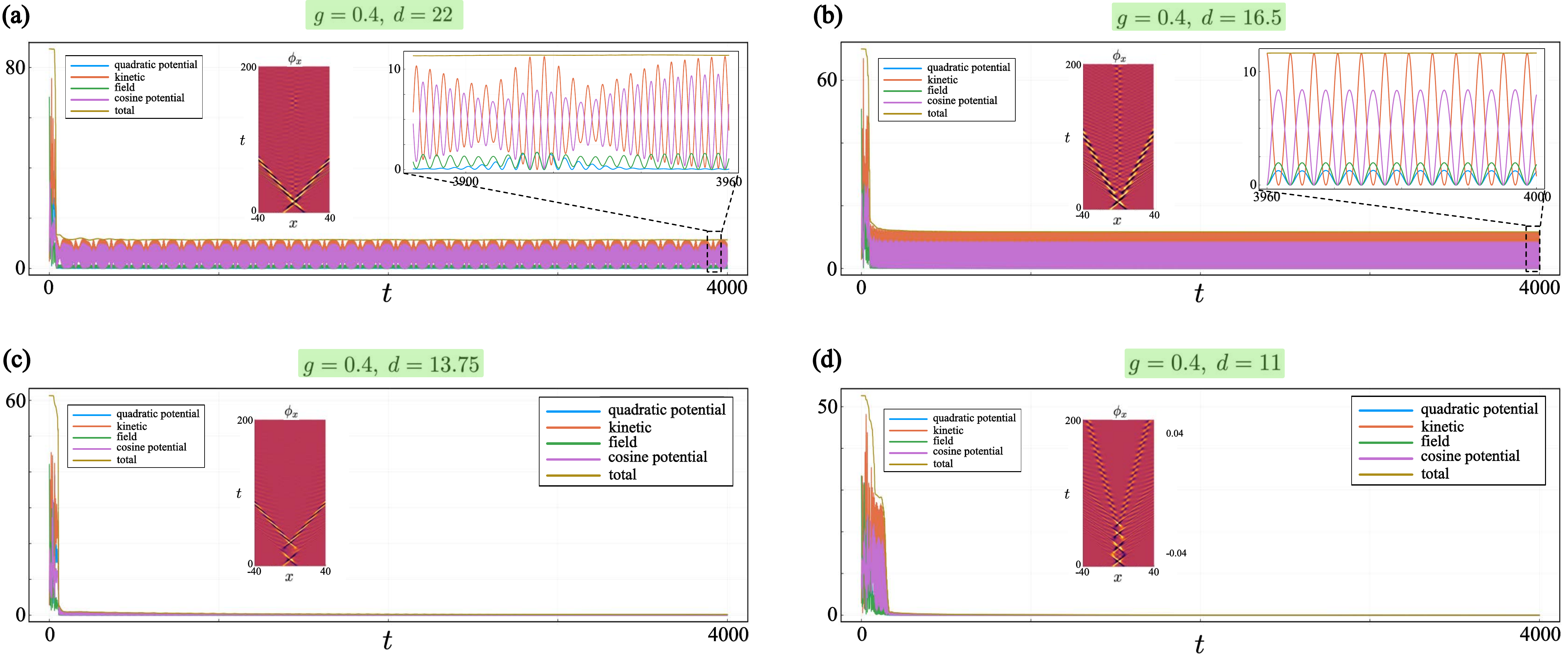}
    \caption{Possible dynamics of a system with large effective mass ($g=0.4$) and is initially composed of a soliton-antisoliton pair moving towards each other at the speed $u=0.55$. The evolution over time of the energies stored inside a finite domain $-16\leq x\leq 16$ are shown in (a), (b), (c) and (d) for $d=22,\ d=16.5, \ d=13.75, \ $and $d=11$, in that order.}
    \label{fig:unstablepair-energy}
\end{figure*}

In Section \ref{sec:positronium} we discussed the various long-time scenarios observed in our numerical investigation of the Schwinger model when the initial condition is a soliton-antisoliton pair carrying opposite unit charges. In addition to an oscillating positronium, other possible long-time configurations are shown in Fig.\,\ref{fig:unstablepair}. In this Figs.\,\ref{fig:unstablepair-energy}a-d, we present the results on the evolution of the energies in each of those scenarios. The early-time dynamics of $\phi_x$ in Figs.\,\ref{fig:unstablepair}a-d are included as insets in Figs.\,\ref{fig:unstablepair-energy}a-d correspondingly for convenience.

When the long-time configuration around the center is a pair of secondary breathers oscillating around each other (resulting in a steady beating pattern with the meson density periodically dispersing and refocusing), the energy evolution also exhibits the beating behavior (see Fig.~\ref{fig:unstablepair-energy}a). The close-up inset reveals that, in the steady state, the energy associated with the quadratic potential term, $(\partial_x \varphi)^2$ (blue curve), undergoes periodic collapse and revival as its dynamics intermittently synchronize and desynchronize with the other energy components. At steady state, the background meson oscillations near the system’s center of mass also settle into a standing wave pattern. This indicates the absence of propagation in these oscillations, meaning that no energy escapes from the finite region surrounding the core.
The energy dynamics in Fig.~\ref{fig:unstablepair-energy}b corresponds to scenario shown in Fig.\,\ref{fig:unstablepair}b in which the late-time dynamics is a single breather remaining at $x\!=\!0$; the total energy in the region around the origin reaches a steady state, with energy cycling continuously at a fixed frequency among kinetic, potential, and electric field contributions. Lastly, in Figs.\,\ref{fig:unstablepair-energy}c-d we show the evolution of the energy terms over time when the initial solitons turn into breathers and scatter off to infinity without generating further mesons in the region between them. The total energy within the finite region around $x = 0$ drops to zero and remains there once the breathers have left.

\bibliography{refs}

\end{document}